\documentclass[preprint,11pt]{aastex}

\shorttitle{VLA Protostars}
\shortauthors{Tobin et al.}

\newcommand{\msun}{\mbox{$M_{\sun}$}}

\newcommand{\kms}{\mbox{km s$^{-1}$}}
\newcommand{\spitzer}{{\it Spitzer}}
\newcommand{\herschel}{{\it Herschel}}
\newcommand{\microjy}{\mbox{$\mu$Jy}}
\begin{document}

\title{The VLA Nascent Disk and Multiplicity Survey of Perseus Protostars (VANDAM). II. Multiplicity of Protostars in the Perseus Molecular Cloud}
\author{John J. Tobin\altaffilmark{1,10}, Leslie W. Looney\altaffilmark{2},
Zhi-Yun Li\altaffilmark{3}, Claire J. Chandler\altaffilmark{4}, Michael M. Dunham\altaffilmark{5}, 
Dominique Segura-Cox\altaffilmark{2}, Sarah I. Sadavoy\altaffilmark{6},
Carl Melis\altaffilmark{7}, Robert J. Harris\altaffilmark{1}, Kaitlin Kratter\altaffilmark{8}, Laura Perez\altaffilmark{4}}

\altaffiltext{1}{Leiden Observatory, Leiden University, P.O. Box 9513, 2300-RA Leiden, The Netherlands; tobin@strw.leidenuniv.nl}
\altaffiltext{2}{Department of Astronomy, University of Illinois, Urbana, IL 61801}
\altaffiltext{3}{Department of Astronomy, University of Virginia, Charlottesville, VA 22903}
\altaffiltext{4}{National Radio Astronomy Observatory, P.O. Box O, Socorro, NM 87801}
\altaffiltext{5}{Harvard-Smithsonian Center for Astrophysics, 60 Garden St, MS 78, Cambridge, MA 02138}
\altaffiltext{6}{Max-Planck-Institut f\"ur Astronomie, D-69117 Heidelberg, Germany}
\altaffiltext{7}{Center for Astrophysics and Space Sciences, University of California, San Diego, CA 92093}
\altaffiltext{8}{University of Arizona, Steward Observatory, Tucson, AZ 85721}
\altaffiltext{10}{Veni Fellow}

\begin{abstract}

We present a multiplicity study of all known protostars (94) in the Perseus molecular cloud
from a Karl G. Jansky Very Large Array (VLA) survey at Ka-band (8~mm and 1~cm) and C-band (4~cm and 6.6~cm).
The observed sample has a bolometric luminosity range between 0.1~L$_{\sun}$ and $\sim$33~L$_{\sun}$, with a median of
0.7~L$_{\sun}$. This multiplicity study is based on the Ka-band data, 
having a best resolution of $\sim$0\farcs065 (15~AU) 
and separations out to $\sim$43\arcsec\ (10000~AU) can be probed. The overall 
multiplicity fraction (MF) is found to be of 0.40$\pm$0.06 and the companion star
fraction (CSF) is 0.71$\pm$0.06. The MF and CSF of the Class 0 protostars are
0.57$\pm$0.09 and 1.2$\pm$0.2, and the MF and CSF of Class I protostars
are both 0.23$\pm$0.08. The distribution of companion separations
appears bi-modal, with a peak at $\sim$75 AU and another peak 
at $\sim$3000~AU. Turbulent fragmentation is likely the dominant mechanism on
$>$1000~AU scales and disk fragmentation is likely
to be the dominant mechanism on $<$200 AU scales.
Toward three Class 0 sources we find companions separated by $<$30 AU. These systems
have the smallest separations of currently known Class 0 protostellar binary systems. Moreover,
these close systems are embedded within larger (50~AU to 400~AU) structures and may be
candidates for ongoing disk fragmentation.

\end{abstract}


\section{Introduction}

A significant fraction of stars are found in multiple systems.
The frequency of multiplicity is a strong function of spectral type (or stellar mass): most
O and B stars are multiples \citep[e.g.,][]{sana2011}, as are about half of all solar-type (G)
stars \citep{raghavan2010}, around one third of M stars \citep{lada2006}, and 20 - 25\% of brown dwarfs or
very low mass stars \citep{allen2007}, also see the review by \citet{duchene2013}. 
Thus, multiple stellar systems are a common outcome of the star formation process and our physical
understanding of star formation must account for the formation of multiple 
systems \citep[e.g.,][]{mathieu1994,tohline2002,reipurth2014}.

Multiple systems are expected to form early in the star formation process when there is 
a large mass reservoir available. Multiple systems may form through several possible processes (and combinations thereof): 
1) turbulent fragmentation of the molecular cloud \citep[e.g.,][]{padoan2002,offner2010,bate2012},
 2) the thermal fragmentation of strongly perturbed, rotating, and infalling core 
\citep[e.g.,][]{inutsuka1992,bb1993,boss2013,boss2014}, and/or
3) the fragmentation of a gravitationally unstable circumstellar disk
\citep[e.g.,][]{adams1989,bonnell1994a,bonnell1994b,machida2008,stamatellos2009}. Fragmentation
due to scenarios 1) and 2) will lead to multiple systems that are 
initially separated by several hundred to 1000s of AU; direct observational evidence for this
process taking place may have been observed in \citet{pineda2015}. On the other hand, scenario 3)
will form companions with initial separations of 100s of AU or less and there are
several examples for which this process may have taken place \citep[e.g.,][]{rodriguez1998,takakuwa2012,tobin2013}.
Furthermore, dynamical interactions in an initially close triple system 
(presumably formed by one of the route mentioned above) can eject one member 
into a wide orbit, providing an alternate mechanism for the production of wide 
systems \citep{reipurth2010,reipurth2012}.

The distribution of companion separations in multiple systems can reflect their likely formation mechanism.
The characteristic separation for solar-type multiples is 45 AU \citep{raghavan2010}, but 5.3 AU for low-mass
(0.5 \msun\ to 0.1 \msun) stars \citep{fischer1992}. However, the main-sequence field multiple systems have been shaped 
by dynamical evolution \citep[e.g., three-body interactions, interactions
with cluster members]{marks2012}. Therefore the present distribution
of separations in field multiples is likely substantially different
from their initial separation distribution. This makes it difficult to infer
the likely formation routes of multiple systems from the observations of field stars alone, and to
gain a better understanding of multiple star formation, characterizing the multiplicity properties of young,
forming stars is crucial.

Young stars are typically divided into four observational
classes, Class 0, I, II, and III \citep[e.g.,][]{dunham2014}.
Class 0 protostars are considered the youngest 
and most deeply embedded within dense envelopes of gas and dust 
\citep{andre1993}, Class I protostars are still surrounded by envelopes but are less
embedded than Class 0s, Class II sources have no (or very tenuous) envelopes and are comprised of a
dusty disk around a pre-main sequence star, and Class III sources are pre-main sequence stars without
substantial disk emission, but may have debris disks. Note that the Class 0 
definition is not independent of Class I because Class 0 is based on 
submillimeter luminosity and the
Class I, II, and III definitions are based on the near to mid-infrared spectral slope \citep{lada1987}.
Alternatively, bolometric temperature ($T_{bol}$) is also used and has boundaries
defined for all Classes \citep{myers1993}; however, the observational classification may not
necessarily reflect the true evolutionary state of the YSO due to 
extinction and inclination effects \citep{launhardt2013,dunham2014}. The expected lifetime in the Class 0 phase is expected
to be $\sim$160 kyr and the combined Class 0 and Class I phase is 
expected to last $\sim$ 500 kyr \citep{dunham2014}, assuming a 2 Myr lifetime of the Class II phase.

Multiplicity is becoming well-characterized for pre-main sequence stars (Class II and III sources)
with radial velocity
and high-contrast imaging techniques \cite[e.g.,][]{kraus2011,kraus2008,reipurth2007,kraus2012}.
The typical statistics derived from multiplicity studies are the multiplicity fraction (MF)
and companion star fraction (CSF). These measures can be thought of as the probability of
a given system having companions and the average number of companions
per system, respectively. The MF is defined by 
\begin{equation}
MF = \frac{B+T+Q+...}{S+B+T+Q+...}
\end{equation}
and the CSF is defined by
\begin{equation}
CSF = \frac{B+2T+3Q+...}{S+B+T+Q+...}
\end{equation}
where S, B, T, and Q stand for the number of single, binary, triple, and quadruple systems respectively. 
The overall CSF of Class II and Class III objects in Taurus is $\sim$0.7 \citep{kraus2011} with the
fractions of single systems only being about 0.25 to 0.33. In contrast to the more distributed population
in Taurus, the Orion Nebula Cluster only has a CSF of $\sim$0.08 between 67.5 AU to 675 AU, about 2.5$\times$ lower
than Taurus for the same range of separations \citep{reipurth2007}. This reduction
in companions is thought to result from dynamical interactions in the dense cluster environment that strip wider
companions.

The multiplicity of Class I protostars, on the other hand, has not been as well-characterized due to their embedded nature.
\citet{connelley2008} conducted a near-infrared survey of Class I sources in several star forming regions, finding
companions toward 27 of 136 targets with separations between 200 AU and 2000 AU. At separations between 50 AU and 200 AU,
\citet{connelley2009} and \citet{duchene2007} 
found 15 companions out of 88 targets. Thus, Class I protostars have CSF of $\sim$0.36 from
between 50 and 2000 AU and the distribution of separations is rather flat 
between 100 AU and 2000 AU, with an increase at $\sim$3000 AU \citep{connelley2008}. 
Thus, multiplicity is commonly observed toward Class I and II sources,
but the distribution of separations does not appear universal for all star forming regions. 

The Class 0 sources remain poorly characterized in terms of multiplicity. 
This is because these sources are even more deeply embedded than Class I protostars, and
their multiplicity can typically only be examined at wavelengths $\ga$10 \micron.
Since the protostar is obscured by a thick envelope, 
emission at $\lambda$ $<$ 10 \micron\ is typically from 
scattered light and/or shock-heated material in the outflow \citep[e.g.,][]{tobin2007, seale2008, tobin2010a};
Thus, the multiplicity of Class 0 sources has been characterized 
principally with interferometers at millimeter and centimeter 
wavelengths due to limited resolution in the mid and far-infrared.
\citet{looney2000} conducted a 2.7 mm survey of 11 nearby Class 0 protostars, finding 
that all the protostellar sources were in multiple systems with
separations between 140~AU and 8000~AU. Most, however, were found with separations $>$ 400~AU 
\citep[corrected for the updated distance to Perseus, d $\sim$ 230 pc][]{hirota2008,hirota2011}. 
\citet{maury2010} observed 5 systems at high angular resolution ($\sim$0\farcs4), only finding single sources.
They then claimed that there was no evidence for multiplicity in Class 0 sources on scales between 150 AU and 400 AU 
(also corrected for the updated distance to Perseus), based on their non-detections combined with
the results from \citet{looney2000}.

More recently, \citet{chen2013} used archival millimeter/submillimeter data taken toward Class 0
protostars to characterize multiplicity toward 33 systems. On scales between 50~AU and 5000~AU, 
\citet{chen2013} found an MF of 0.64 and a CSF of 0.91, with
most companions being separated by more than 1000 AU. The main limitation of that study was 
that it was not conducted in a uniform manner in terms of sensitivity or resolution. 
This is because the data were drawn from archival
observations toward various star forming clouds at various distances. The survey had
spatial resolutions that ranged between 30 AU and 1800 AU with a median of 600 AU. 
However, they found 3 multiple Class 0 systems
with separations between 150 AU and 430 AU, with a total of 5 sources between 50~AU and 430~AU). 
This survey was a large step forward in the characterization of wide companions toward Class 0 
protostars, but was limited in addressing close multiplicity.

Nonetheless there has been some progress in characterizing multiplicity on scales $<$ 400 AU. 
\citet{tobin2013} found two Class 0/I sources (out of a sample of 3) with companions separated by
100 AU. Moreover, \citet{tobin2015} found a companion toward the Class 0 system NGC 1333 IRAS2A separated by 142 AU,
perhaps the driving source of a secondary east-west outflow observed in this system. 
Also, \citet{tobin2015b} found a companion toward L1448 IRS3B separated by
$\sim$210~AU. Thus, statistics have been building up for Class 0 sources at 
smaller separations, but in a slow, piecemeal fashion.

To make a large stride in the characterization of protostellar multiplicity in the Class 0 and I phases, 
a survey with uniformly high sensitivity and high-resolution ($<$ 50 AU) is necessary. This is one of the driving goals
of the VLA Nascent Disk and Multiplicity (VANDAM) survey, undertaken with the Karl G. Jansky Very Large
Array (VLA). The VANDAM survey was conducted at Ka-band (8 mm and 1 cm), where the observations
are sensitive to emission from both thermal dust and free-free jets. Furthermore, complementary
observations were taken in C-band (4 cm and 6.4 cm) to characterize the spectral slope 
of the free-free emission. In this survey, we have observed all known protostars in the Perseus
molecular cloud (82 Class 0 and I sources plus 12 Class II sources ) in order to characterize the multiplicity of Class 0 and Class I protostars
with as little sample bias as possible.

This paper is focused on the multiplicity results of the VANDAM survey derived 
from the Ka-band data only. Several following papers will
focus on the resolved disk candidates \citep{seguracox2016}, polarization results, C-band radio spectra, and full survey results. The sample
is described in Section 2, the observations, instrument setup, and data reduction are described
in Section 3, the multiplicity results are described in Section 4, the results are discussed
in Section 5, and the summary and conclusions are given in Section 6.

\section{The Sample}
The VANDAM sample leverages the large body of work that has already
been done to identify and characterize the protostellar content within Perseus. 
The sample of sources we observed and the pointing
centers are given in Table 1. Our sample is primarily based on the catalog
published by \citet{enoch2009}, which considered all the available \textit{Spitzer Space Telescope} \citep{werner2004}
photometry and Bolocam\footnote{1.3 mm continuum instrument on the Caltech Submillimeter Observatory} 
data taken toward Perseus. \citet{enoch2009} lists 66 protostars within Perseus, 27 of which 
they classify as Class 0 and 39 are Class I based on $T_{bol}$. 
However, rather than a stringent transition from Class 0 to Class I, we refer to 
sources with 60 K $\le$ $T_{bol}$ $\le$ 90 K as Class 0/I objects because the measured $T_{bol}$ has a 
dependence on viewing angle that can make Class 0 sources appear as Class I and 
vice-versa \citep{launhardt2013, dunham2014}. In addition, Per-emb-44 (SVS13) is also denoted
a Class 0/I because its continuum and outflow properties 
are more consistent with Class 0 objects \citep{looney2000,plunkett2013}. We have also updated the $L_{bol}$ and $T_{bol}$
for the sources published in \citet{sadavoy2014} that include \textit{Herschel} 
photometry. Thus, from the sources listed in \citet{enoch2009}, we classify 27 as Class 0 sources, 8 Class 0/I sources, and 
31 Class I sources in our sample.

While the \citet{enoch2009} survey still represents the best near to far-infrared characterization published thus far, 
there have recently been candidate first hydrostatic cores (FHSCs) and 
Very Low Luminosity Objects (VeLLOs) identified by millimeter 
interferometry \citep{hirano1999,enoch2010,chen2010,schnee2012,pineda2011}
that were not detected in the infrared. Moreover, some of the \citet{enoch2009} sources that 
were listed as one source were known to comprise multiple millimeter continuum
sources \citep[e.g.,][]{looney2000} and the low-resolution of \textit{Spitzer} at 24~\micron\ and 70~\micron\
prohibited these sources from being identified as discrete objects 
by \citet{enoch2009}; all these sources, 11 in total,
were added to the sample. Many of these Class 0 sources 
known from millimeter observations, but not detected clearly by \textit{Spitzer},
 were resolved in the far-infrared by \textit{Herschel} 
\citep{pezzuto2012, sadavoy2014}.
Many of these sources are highly obscured at  24 \micron\ and may be analogous to the PACS Bright Red 
Sources (PBRS) discovered in Orion by \citet{stutz2013}.
We note that the 70 \micron\ emission from the source 
Per-emb-37 has a $\sim$6\arcsec\ position shift relative to the 24 \micron\ position.
This is due to the source being faint at 24 \micron\ and 
a nearby Class II source being much brighter at 24 \micron\ and shorter wavelengths.
Per-emb-37 was also identified by \citet{sadavoy2014} as 
a Class 0, while \citet{enoch2009} classified it as Class I due to the mis-association of 70~\micron\ emission
with the shorter wavelength emission.

We also examined the \textit{Herschel} 70 \micron, 100 \micron, and 
250 \micron\ maps of the Perseus region in an attempt to identify additional
sources that may have been missed by the \textit{Spitzer} survey. We found 17 sources that
were bright at 70 \micron\ and 100 \micron\ and these were added to the sample; however, these
sources are classified as either Class II or flat spectrum 
sources (borderline between Class I and Class II) \citep{evans2009}.
In total, our sample is comprised of the \citet{enoch2009} sample plus 28 sources additional sources.

Thus, our sample contains all currently known/published Class 0/I protostars in the Perseus region
and flat spectrum/Class II sources that are bright in the far-infrared. 
It is possible that there are some undiscovered protostars in Perseus, given that classifications 
for the entire cloud using \textit{Herschel} photometry remain unpublished. 
However, our efforts to identify bright sources in the \textit{Herschel} data
did not turn up a significant number of new Class 0 or Class I sources. Therefore, the sample 
presented in Table 1 is as complete as possible, given the current knowledge of 
the protostar population in Perseus. The sample includes a total of 94 targeted sources, 
37 of which are Class 0 protostars (FHSCs and VeLLOs included), 8 are Class 0/I 
protostars, 37 are Class I protostars (flat spectrum included), and 12 are Class II sources, see Table 1.
The sources included in \citet{enoch2009} are denoted by Per-emb-XX and the additional
young stellar objects that did not have more common names are denoted EDJ2009-XXX \citep{evans2009},
 where X refers to a number. The sources not included in either of those 
catalogs are referred to by their most common name.

The range of luminosities sampled is between $\sim$0.1~L${\sun}$ and $\sim$33~L${\sun}$, with a median luminosity
of 0.7~L${\sun}$. The median luminosities of the Class 0 and Class I
sources are 0.9~L${\sun}$ and 0.7~L${\sun}$, respectively. This range of
luminosities is consistent with the typical distribution of protostellar luminosities observed
in Orion and the Gould Belt Clouds \citep{dunham2014,dunham2015}. Therefore, our sample is comprised of
a reasonably representative sample of protostellar objects.

\section{Observations and Analysis}

\subsection{Observational Setup and Procedure}
We conducted observations with the VLA in B-configuration 
between 2013 September 28 to 2013 November 20 and in 
A-configuration during 2014 February 24 to 2014 May 31 and 
2015 June 19 to 2015 September 21. 
The B-configuration (also referred to as B-array) has
a maximum baseline (antenna separation) of 11.1 km and at 8 mm provides a resolution of
$\sim$0\farcs2 (46 AU). The A-configuration (A-array) has a maximum baseline of 36.4 km, providing
a resolution of $\sim$0\farcs065 (15 AU).

For each source in Table 1 we observed a single pointing toward the coordinates
listed. However, the sources Per-emb-21, Per-emb-42, IRAS4B$^{\prime}$, and SVS13C were located less than 15\arcsec\
from another source and one pointing was sufficient. Then the sources, 
EDJ2009-233, SVS3, EDJ2009-173, and EDJ2009-235 were serendipitously detected
within the primary beams of adjacent target sources and we report their detections as well. Observations 
have been obtained for the entire sample in B-configuration (except for EDJ2009-268),
and observations in A-configuration have been conducted for all sources detected
in B-configuration. 

The Ka-band observations were conducted in 8 GHz continuum mode using 
3-bit samplers with one 4 GHz baseband centered at 36.9 GHz and another centered at 28.5 GHz.
The full 8 GHz of bandwidth was divided into 128 MHz spectral windows, 
each having 64 channels that were 2 MHz wide, and we recorded full polarization
products. The B-configuration scheduling blocks (SB) were 
3.5 hours in length, observing three sources per SB. Each SB started with
observations of the absolute flux density calibrator (3C48), followed 
by observations of the bandpass and polarization leakage calibrator (3C84).
The observations were conducted with fast-switching, observing the 
complex gain calibrator (J0336+3218) for $\sim$25 seconds and then the source for $\sim$75 seconds.
The pointing solutions were updated every 50 minutes and each 
source received $\sim$30 minutes of on-source integration in each scheduling block. 
Each scheduling block ended with an observation of 3C138
to calibrate the linear polarization angle; thus, the VANDAM dataset in 
B-configuration has all the necessary calibrations taken to examine
the polarization toward these protostars at 8 mm and 1 cm. 
See \citet{cox2015} 
for details on the polarization calibration and results toward NGC 1333 IRAS 4A.

The A configuration Ka-band data were observed with the same spectral setup, 
but with scheduling blocks that were
1.5 hr, 2.5 hr, or 2.75 hr in length. The shorter scheduling
blocks were necessary due to the limited windows for observing Perseus 
during the A-configuration. The 1.5 hr scheduling blocks 
observed only 1 source and 2 sources were observed in the 2.5 hr and 2.75 hr 
blocks. Each A-array SB started in the same manner as the B-array SBs
and we also achieved a similar on-source time. The difference was that we 
did not observe 3C138 at the end of the SBs and 
rely instead on 3C48 for polarization angle calibration.

\subsection{Data Reduction}

The VANDAM survey data were all reduced using the Common Astronomy Software 
Applications (CASA\footnote{http://casa.nrao.edu}) package \citep{mcmullin2007}.
The data taken in 2013 and 2014 were reduced using version 1.2.2 
of the VLA pipeline in CASA version 4.1.0
and the data taken in 2015 were reduced using version 1.3.1 of the pipeline in CASA version 4.2.2.
The two versions of the pipeline are found to produce consistent results for our data.
The VLA pipeline applies flags generated by the online system, 
as well as at the edge-channels of the spectral windows where sensitivity is reduced. 
The pipeline then applies standard calibration procedures for
the delays, bandpass calibration, absolute flux calibration, and the time-dependent 
gain and phase calibration. We inspected the resulting calibration tables to 
ensure proper calibration and that bad/uncalibrated
data are not included in the final data products. We first verified the absolute 
flux density calibration accuracy by editing the gain table used as input to
the \textit{fluxscale} task. We flagged the calibrator solutions that were at 
significantly different elevations and those with substantial time
variation. We then re-ran \textit{fluxscale} with the edited calibration table
and compared the flux densities calculated for the gain calibrator and
bandpass calibration with those applied during the pipeline script. If the 
values agreed within 10\% we accepted the flux density scale 
as-is, when they did not agree (only one SB), we flagged the known bad antennas and reran the pipeline.
The overall uncertainty in the flux density scale is estimated to be $\sim$10\%.

Following the flux density calibration check, we inspected the final gain 
versus time table and flagged gain solutions that were discrepant from the general
trends versus time. We also inspected the gain versus frequency tables to ensure 
that specific spectral windows did not have abnormally large scatter. Lastly
we inspected the phase versus time tables to identify periods of unusually 
large phase scatter or phase jumps. Following the gain table flagging, we ran
the \textit{applycal} task with the \textit{mode=flagonly} option enabled, 
which flags the on-source data with no corresponding calibration data,
based on the flagged gain tables.

We note that our reduction method only applies flagging \textit{a posteriori} 
and the gain solutions are computed with some bad data. However,
there is a large amount of redundancy in the computation of the closure phase 
and gain solutions because the VLA has 27 antennas. To determine
the effect of \textit{a posteriori} flagging versus \textit{a priori} 
flagging, we imaged a dataset in which flagging was applied after pipeline
calibration and then applied the same flags to a raw measurement set 
before running the pipeline. The source structure and root-mean-squared 
(rms) noise in the resultant maps were statistically indistinguishable.
Therefore, we have used the \textit{a posteriori} flagging method exclusively. The good agreement between these two methods
is attributable to the redundancy in the data with so many antennas.
Following the application of gain table flags, we split out each 
source into an individual measurement set, averaging all 64 channels 
in each spectral window to 1 channel for the Ka-band data.

With the measurement sets for each source, we generated 
naturally-weighted dirty maps of the full Ka-band 
(9 mm effective wavelength) and each 4 GHz baseband individually (8.1 mm and
1.05 cm effective wavelengths, respectively); multi-frequency synthesis
imaging mode was used in all cases. We defined regions to deconvolve using the \textit{clean}
algorithm by drawing CASA regions around the peak source emission in each dirty map and then performed non-interactive
cleaning down to $\sim$3$\times$ the rms noise using natural weighting. We then examined the 
cleaned images for additional source emission that was apparent after 
cleaning the strong sources. If additional source emission was detected, we repeated 
the above steps with additional clean masks. We also imaged the data using
Briggs weighting with robust parameters of 0.5, 0.25, and 0; the robust parameter 
adjusts the relative weighting of the short and long baselines in the deconvolution process. 

A robust parameter of 2 is comparable to natural weighting and typically 
produces an image with the lowest noise, but
also lower resolution; a robust parameter of -2 is comparable to 
uniform weighting and has the best resolution
but with a higher noise level because there are fewer data at the 
longer baselines relative to short baselines.
Intermediate values of the robust parameter enable the image 
resolution and sensitivity to be adjusted to 
find an optimal balance. This methodology 
was applied to both the A and B configuration data. 
For the data with both A and B configuration
observations, we produced a merged measurement set using the \textit{concat} 
task and performed the same imaging steps as noted above.
The largest angular scales 
that can be recovered in A and B-configurations
are $\sim$0\farcs8 (180 AU) and $\sim$2\farcs6 (600 AU), respectively; these numbers apply to
natural weighted maps, and the maps made using Briggs weighting (with robust between 0 and 0.5) will have 
smaller largest angular scales\footnote{https://science.nrao.edu/facilities/vla/docs/manuals/oss2016A/performance/resolution}.

\subsection{Data Analysis}

To find companion sources, we visually inspected the images from each 
Ka-band baseband (8 mm and 1 cm) and the full bandwidth (9 mm) for multiple sources.
We define a multiple system as the detection of multiple discrete continuum sources, detected at 8 mm with
a S/N $\ga$6 or at 9 mm with S/N $\ga$ 5; however, if there is a previous detection
in the near-IR, we allowed sources to have S/N of 4. We also examined images from each
robustness level given that some companions only became apparent with robust levels $<$0.5; 
this is because the sources may be blended with natural weighting and only resolved at the 
higher resolution provided by images with a lower robust parameter.
The flux densities of the sources (multiple and single) were measured using the 
CASA task \textit{imfit}, and the peak flux densities are measured directly from the 
images. Most sources are within the inner 20\arcsec\ of the primary beam,
so the correction is $<$15\%. The integrated and peak flux densities reported for all sources have
the primary beam correction applied.

For single sources, the flux densities were measured 
from the B-configuration image generated with natural weighting, 
given that those data would be most sensitive to the largest scale
of emission. The flux densities of the multiples separated by $<$500 AU were
measured from the A+B configuration images generated with natural weighting; the multiples
separated by $<$50 AU have their flux densities measured from the A-configuration data alone.
The spectral indices of the integrated and peak intensities were calculated from the 8 mm and 1 cm
flux density measurements and the spectral index error results from the 
standard error propagation \citep{chiang2012}. All the detected sources and companions have detections
at both 8 mm and 1 cm.

The separations of multiple systems are determined by simultaneously fitting multiple Gaussian components and
calculating the distance between Gaussian central positions. The measured flux densities
of the single and multiple sources are given in Table 2. The separations of apparent
companion sources are given in Tables 3, 4, and 5 and are further discussed in the following sections.

\section{Results}

The VANDAM survey data provide an unprecedented characterization of
protostellar multiplicity in terms of sample size, angular/linear resolution, and sensitivity.
The current results probe previously uncharted regions of protostellar 
companion separations, with a complete sample probing scales
down to $\sim$15 AU. We identify multiple sources out to separations of 
43\arcsec\ ($\sim$10000 AU; $\sim$0.05 pc). This upper limit
to multiple system separation is not physically motivated, but this scale is at the half-power point of
the VLA primary beam at 8 mm. However, this scale is also comparable to the typical radius of protostellar
envelopes \citep[0.05 pc;][]{bm1989} and the break point at 0.04 pc between clustering 
(on larger scales) and multiplicity (on smaller scales) in 
the Taurus molecular cloud \citep{larson1995}. Moreover, on scales $\ga$20\arcsec\ multiplicity in Perseus has been 
characterized in the infrared and (sub)millimeter \citep[e.g.,][]{looney2000, chen2013}. 
Thus, the main discovery space opened by our survey is on scales less than 1000 AU. 
The nature of the multiple continuum 
sources we detect is discussed further in Section 5.

In total, we have found 26 multiple systems in the Perseus molecular cloud with our
VLA data, assuming that sources out to 10000 AU
constitute a single system; this number changes depending on the range of separations considered. Of these 26
multiple systems, 16 are new detections or reflect the discovery of a new component to an 
existing multiple system. The newly discovered multiple systems are described in Section 4.1. The continuum
properties for all detected sources are given in Table 2, and the multiple systems broken down into classes
are given in Tables 3, 4, and 5.

\subsection{Close Multiples}

\subsubsection{Multiple Systems Separated by $<$ 500 AU}
The VANDAM data dramatically improve our
knowledge of protostellar multiplicity on scales $<$ 500 AU. Toward
the Class 0 sources, in particular, there have only been a few studies with 
small samples having spatial resolution $<$ 500 AU \citep[e.g.,][]{looney2000,maury2010,chen2013}. 
Scales $<$ 500 AU are important because this is the size of largest disks observe toward
Class II sources \citep[e.g.,][]{simon2000}, and
at smaller scales companion sources may form within
gravitationally unstable disks \citep{adams1989}.

\begin{figure}[!ht]

\begin{center}
\includegraphics[scale=0.33]{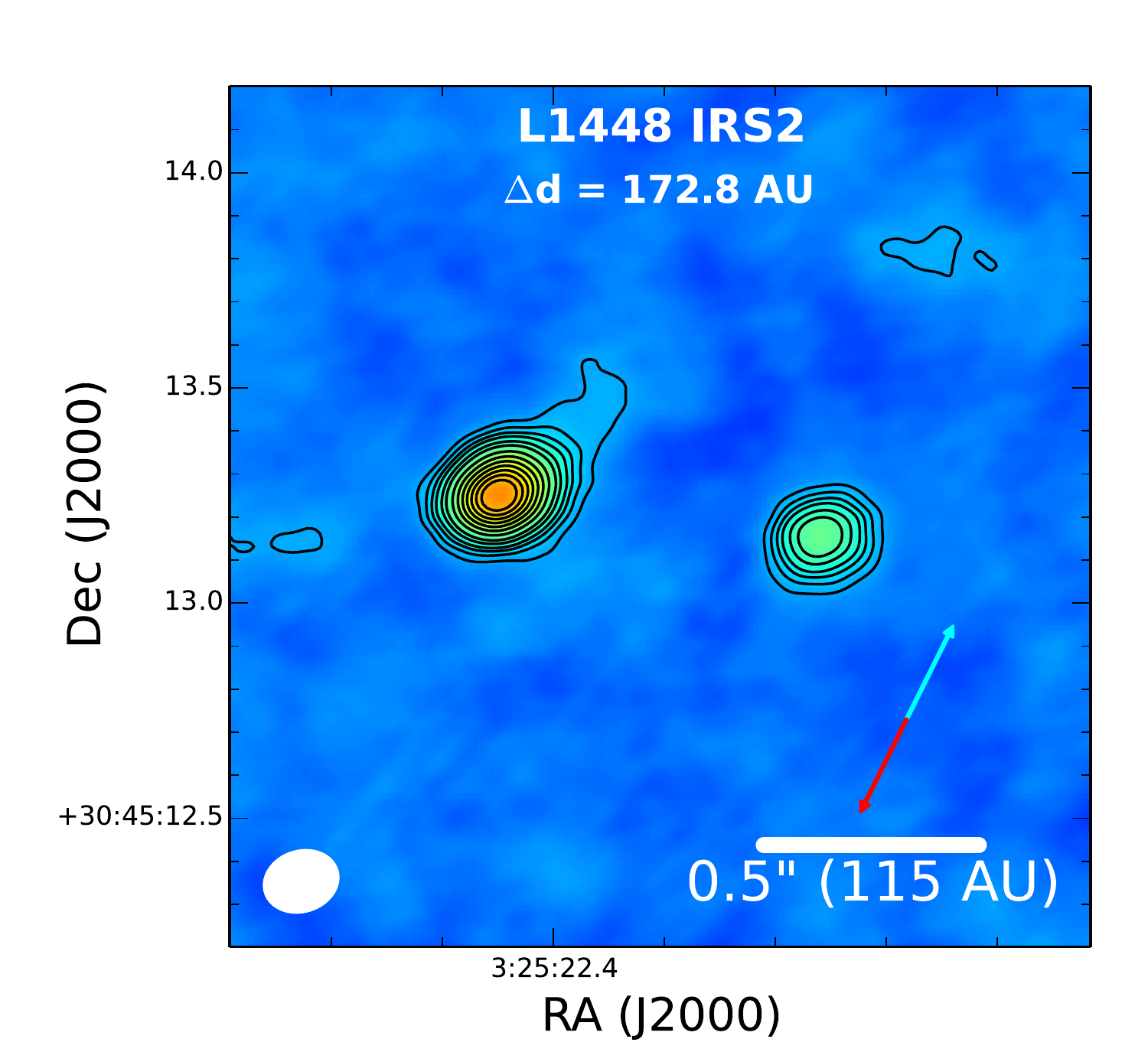}
\includegraphics[scale=0.33]{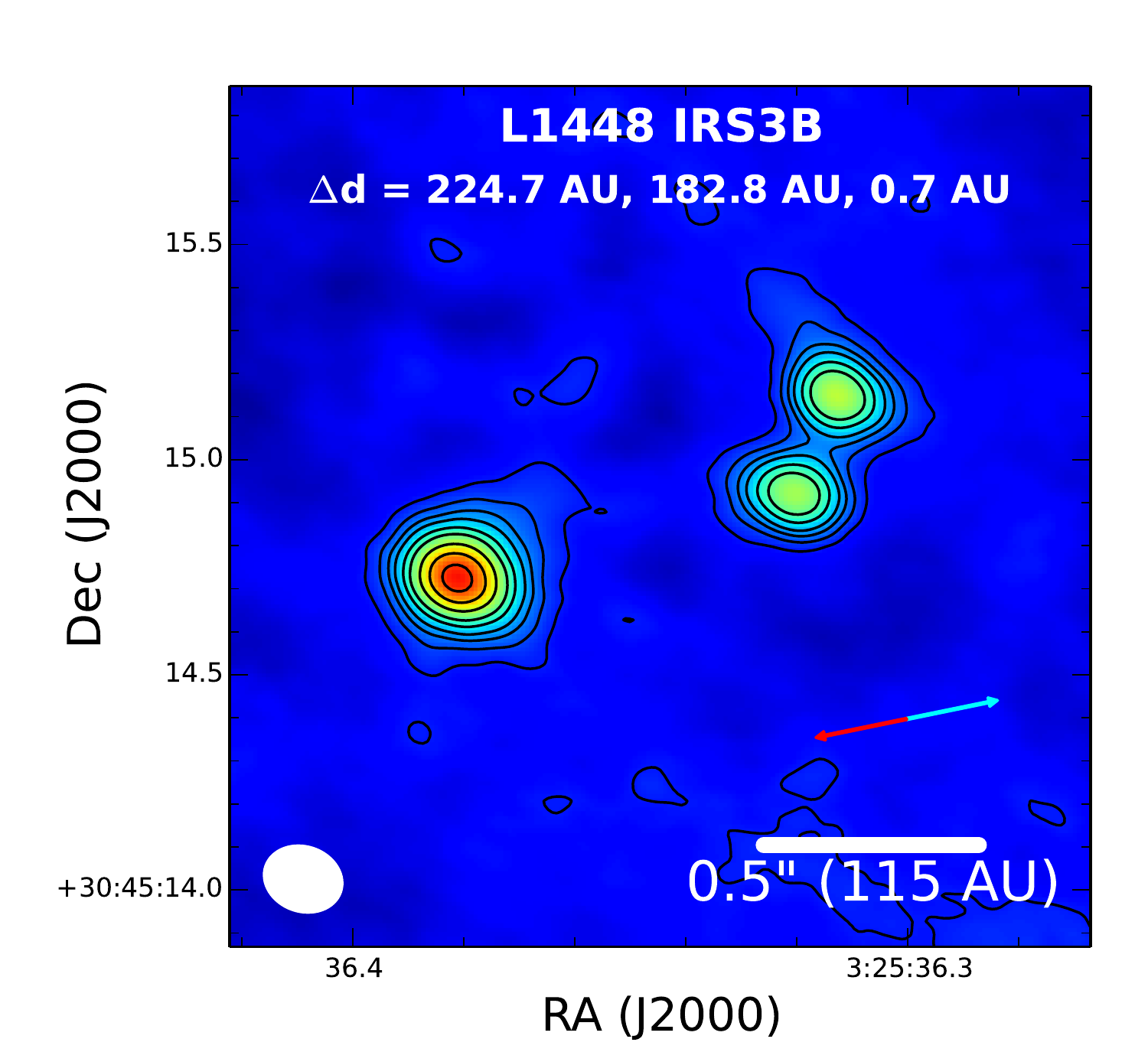}
\includegraphics[scale=0.33]{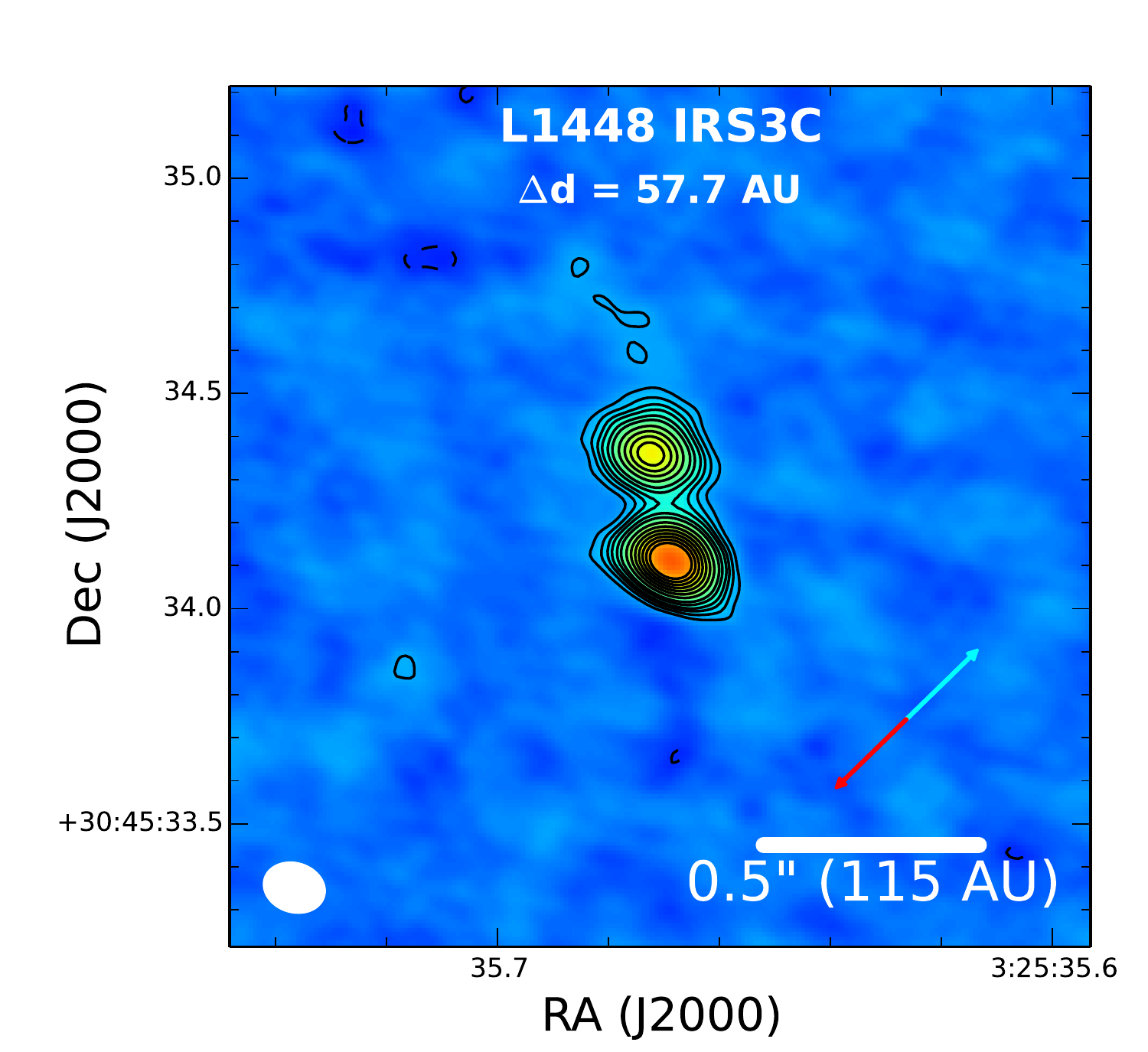}
\includegraphics[scale=0.33]{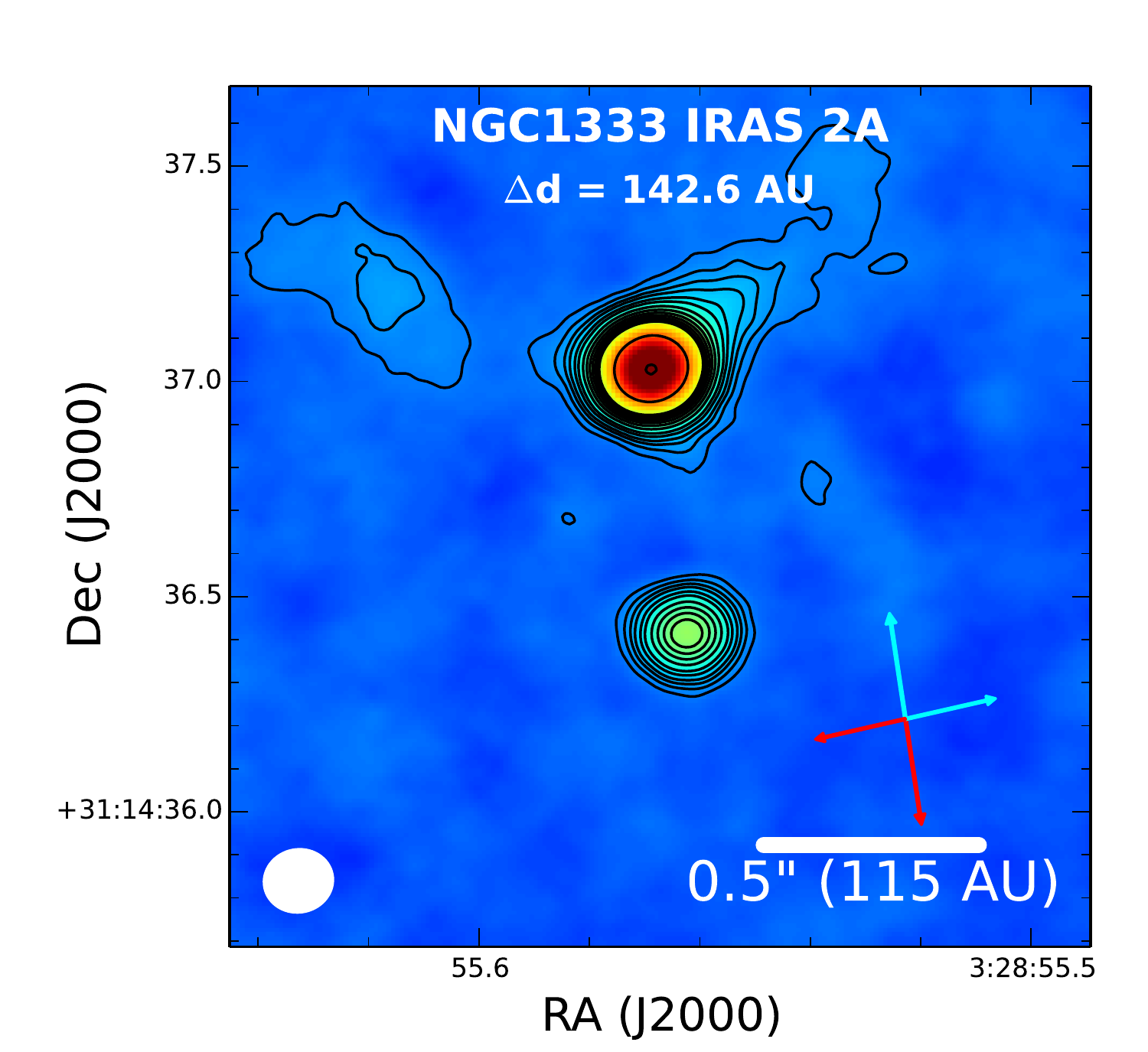}
\includegraphics[scale=0.33]{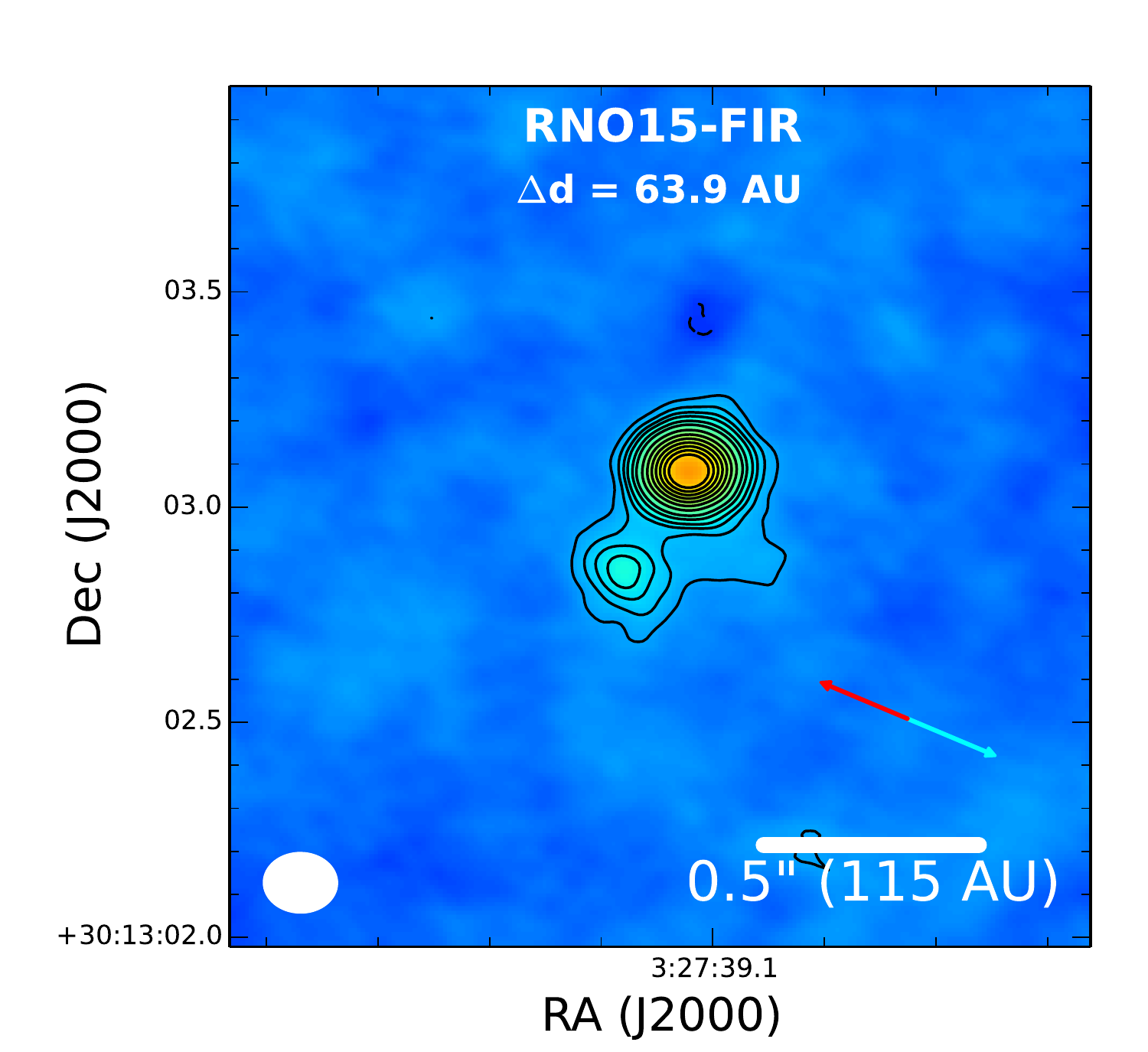}
\includegraphics[scale=0.33]{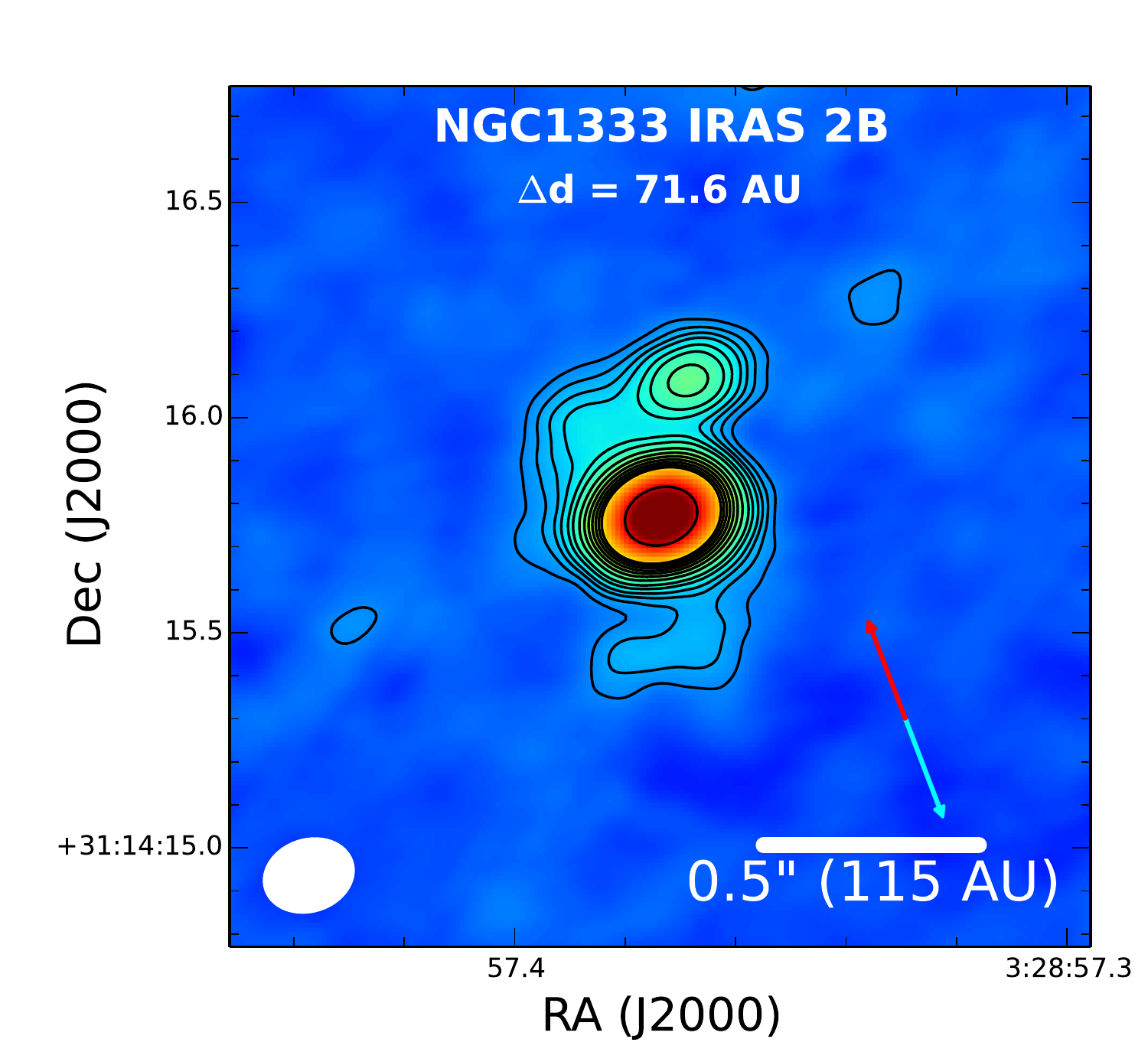}
\includegraphics[scale=0.33]{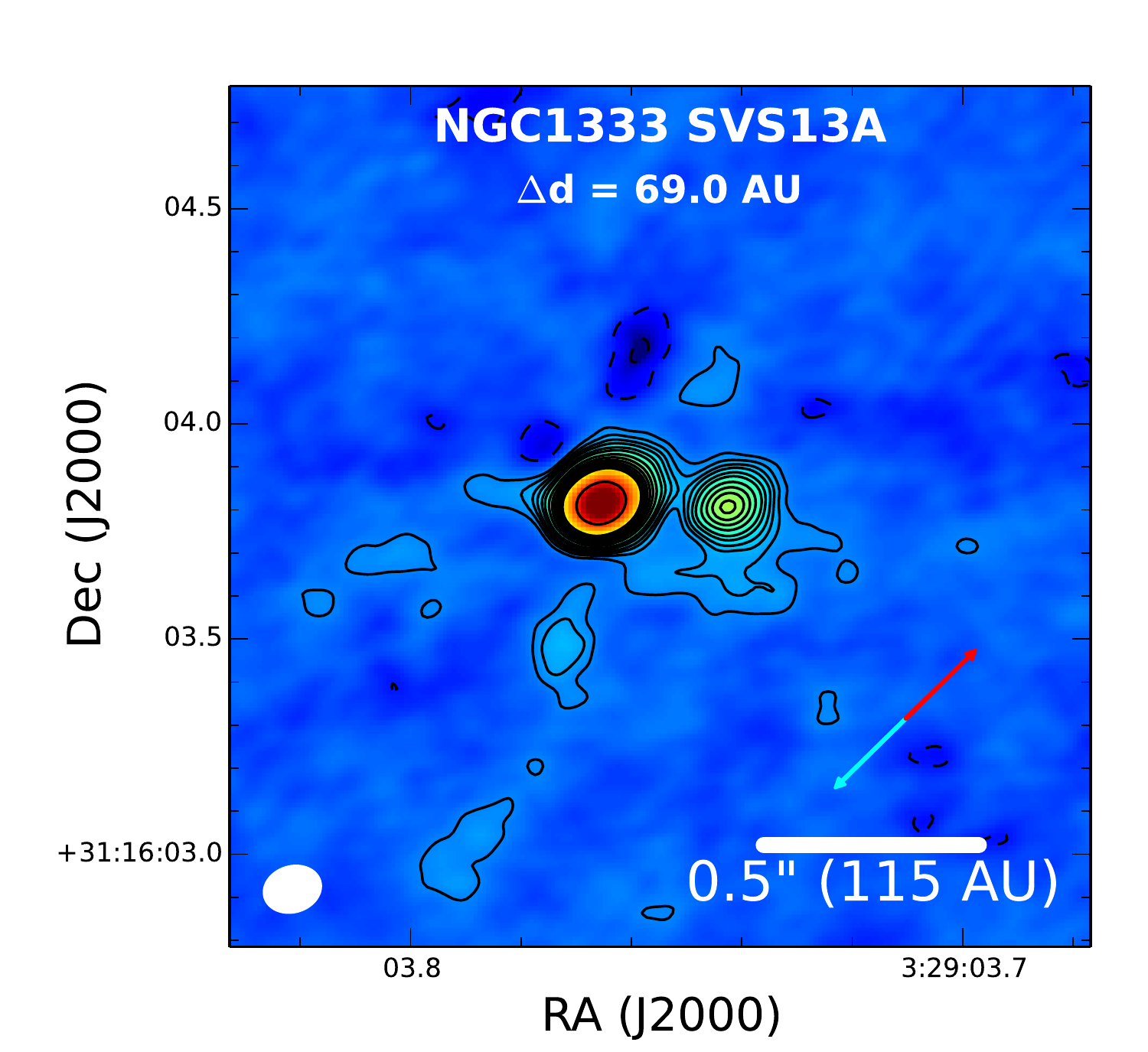}
\includegraphics[scale=0.33]{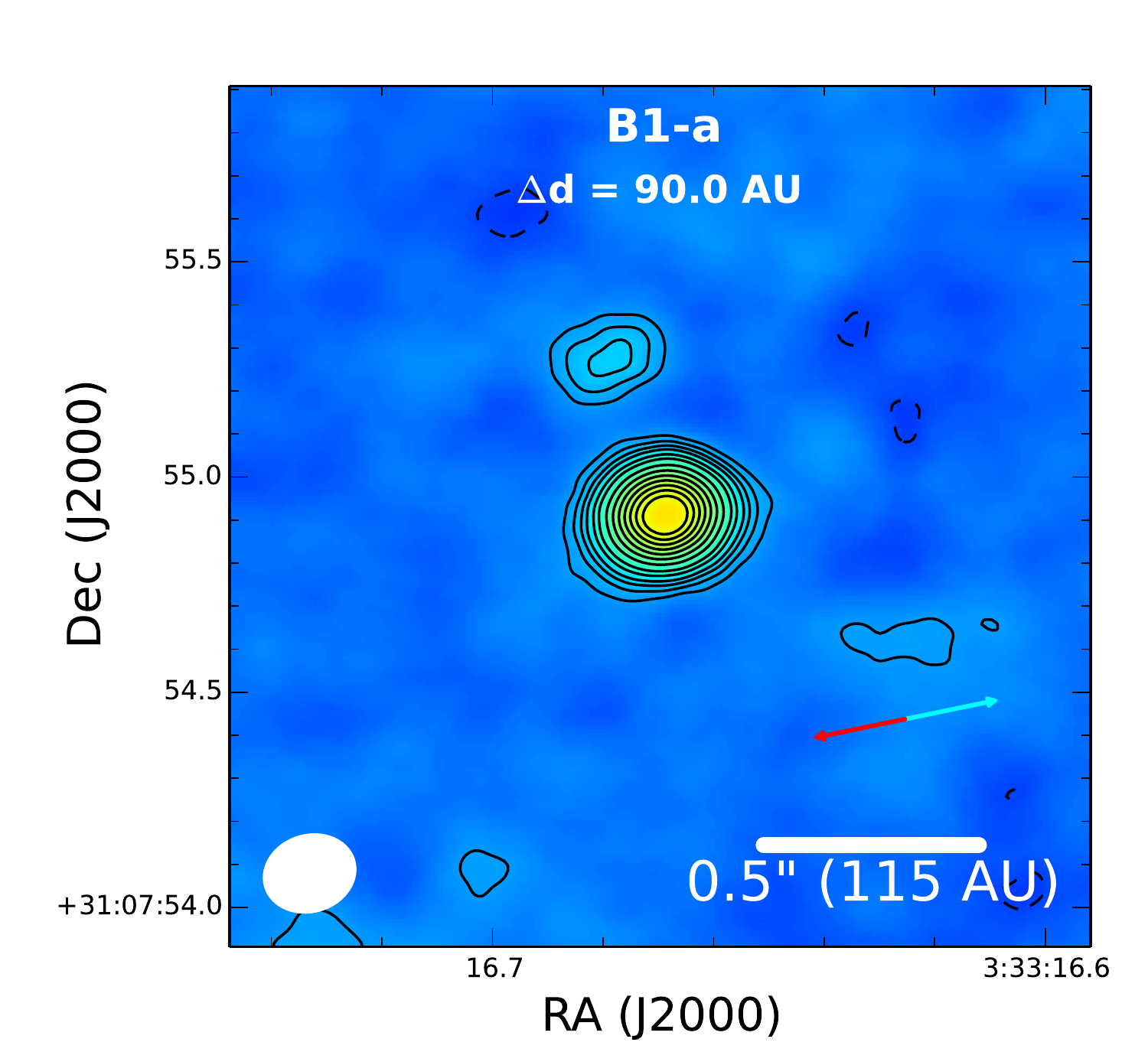}
\includegraphics[scale=0.33]{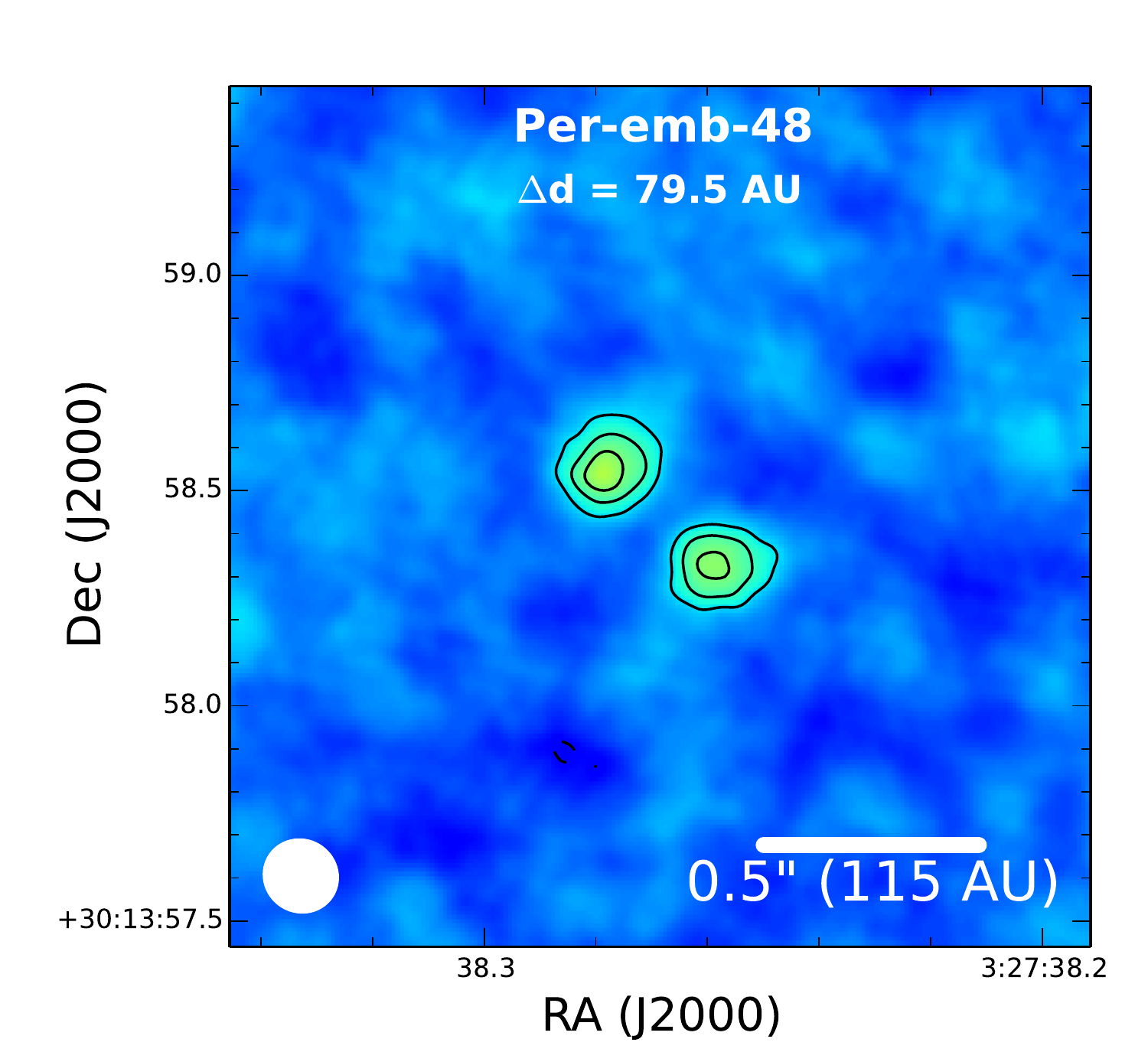}
\includegraphics[scale=0.33]{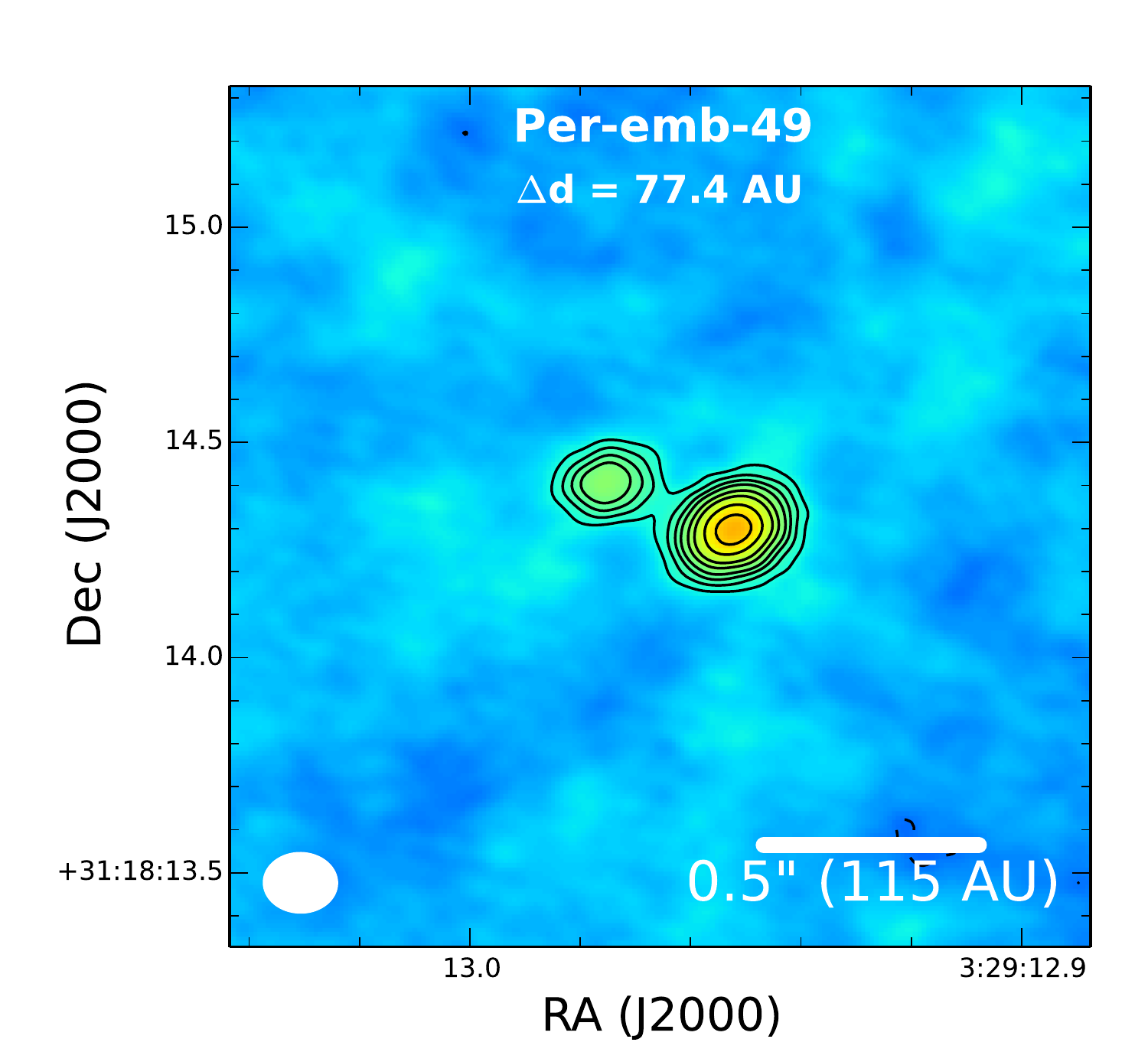}
\includegraphics[scale=0.33]{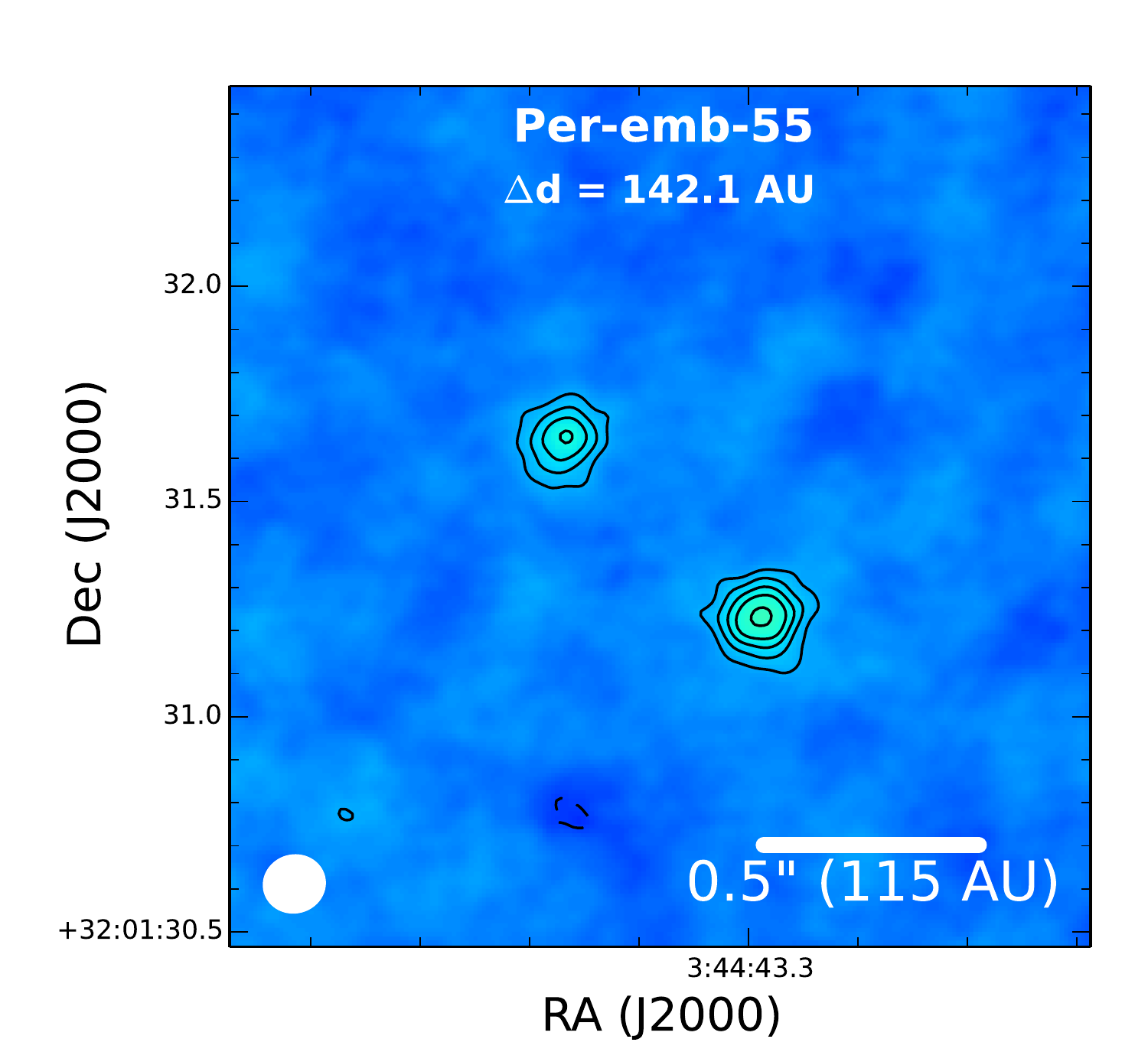}
\includegraphics[scale=0.33]{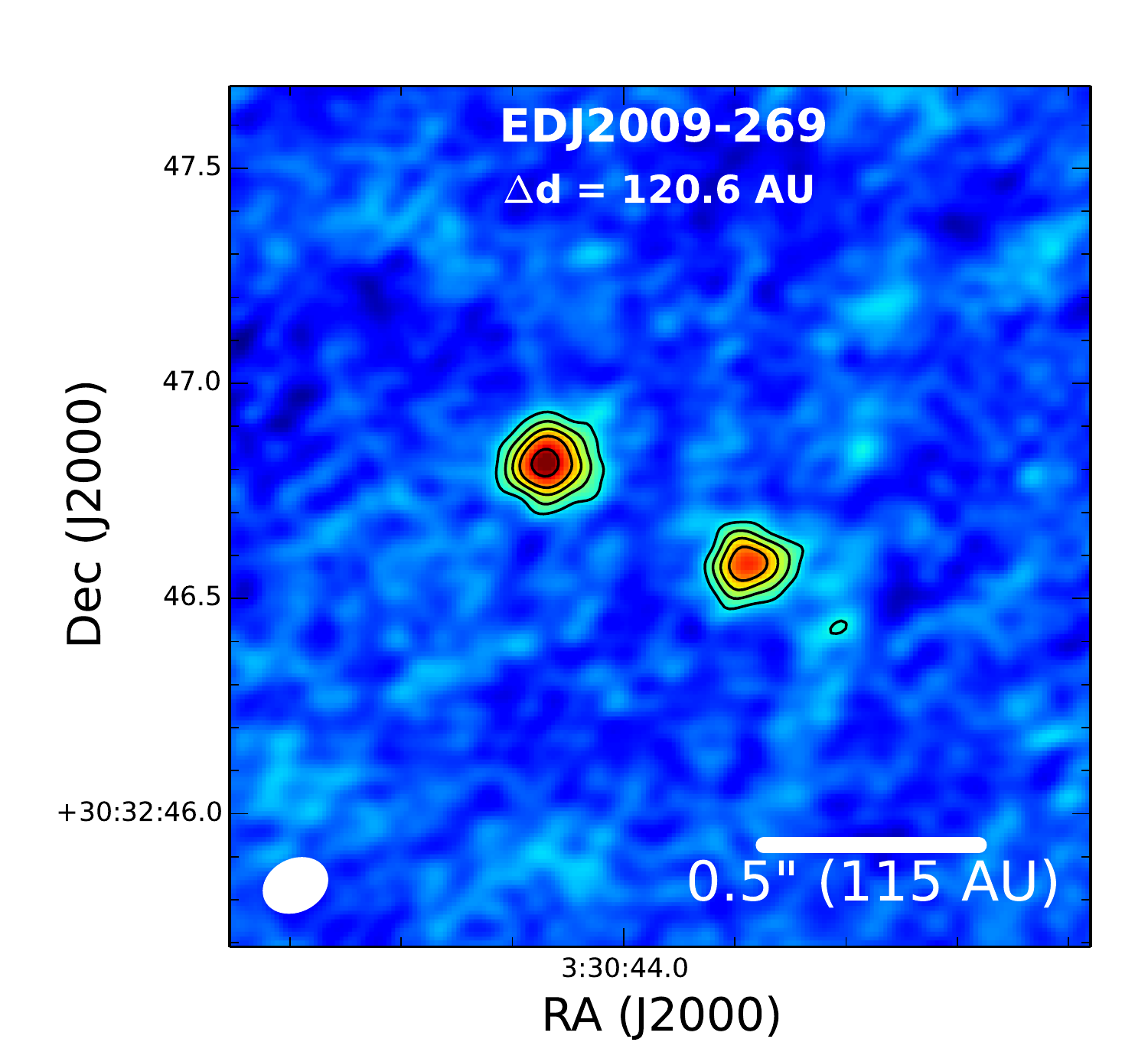}
\end{center}
\caption{Images of multiple systems with separations $<$ 500 AU.
 The images are produced 
from the combined A and B configuration data. Natural weighting is used, along with the full 
Ka-band bandwidth (9 mm effective wavelength;
SVS13A uses Briggs weighting with robust=0.5).
The outflow directions (where available) are indicated by blue and red arrows in the lower
right corner. The outflows are assumed to be driven by the brightest source; the only source with a definitive second
outflow is NGC 1333 IRAS2A \citep{tobin2015}. The contours are [-6,-3, 3, 5, 7, 9, 12, 15, 20, 25, 30, 35,
40, 45, 50, 100, 200, 300, 400, 500, 600]
 $\times$ $\sigma$.
White circles are drawn around low S/N companions that are verified at other wavelengths, see Section 4.1.}
\label{lt500AU}
\end{figure}

We identified 13 new companion sources separated by 30 AU to 500 AU out of 
the 18 total close multiple systems shown in Figure \ref{lt500AU}. Of these new companions, 5 are in Class 0 systems, 6 
are in Class I systems, and 2 are in Class II systems. Prior to the VANDAM survey, only two Class 0 sources
had been known to have companions on $<$ 500 AU,  NGC 1333 IRAS 4A and SVS13A \citep{looney2000,
rodriguez1999,anglada2004}. The companion toward NGC 1333 IRAS2A (Per-emb-27)
 was previously presented in the first VANDAM paper \citep{tobin2015} and we 
include it with the new detections. We note that two Class I multiples (L1448 IRS1 and EDJ2009-183) in Figure \ref{lt500AU} 
have companions that are quite faint. However, we know that these detections are real because the
companions had been previously detected in the near-IR \citet{connelley2008}.

The companions with separations between 30~AU and 500~AU have a 
variety of relative flux densities, the faintest being
$\sim$10 times fainter than the brightest source in the system, see 
Tables 3, 4, and 5. Furthermore, the spectral index 
of the 8 mm and 1 cm emission is positive for all companion sources, but often less than 2,
indicating a combination of dust and free-free emission is responsible for generating
the observed emission. The spectral index ($\alpha$) for
dust emission is expected to be steeply rising with $\alpha$ $\sim$ 2 + $\beta$ (if optically-thin), 
where beta is the dust opacity spectral index. Free-free emission typically has a flatter spectral
index as compared to dust, with a 2 $\ge$ $\alpha$ $\ge$ -0.1 \citep{rodriguez1993}. Non-thermal
synchrotron emission on the other hand typically has $\alpha$ $\sim$ -0.7 \citep{condon1984}. Thus, it 
is unlikely for any companion sources to be background extragalactic objects. See Section 4.6 for more details on the
estimated number of extragalactic background sources.

\subsubsection{Multiple Systems Separated by $<$ 30 AU}

The spatial resolution of 15 AU afforded by our observations
enables us to uncover strong evidence for multiplicity on scales $<$ 30 AU for 3 Class 0 sources.
These three sources are shown in Figures \ref{IRAS03292}, \ref{IRAS03282}, and \ref{Per18}; the top panels show
the emission at multiple resolutions and the bottom panels show the spectral index maps. All three systems are
embedded within a larger structure and the companions are only revealed at the highest resolutions. Furthermore,
the spectral index maps show that both dust and free-free emission are contributing to the source fluxes. These
three close multiple sources have separations between 18.5 AU and 22.3 AU, making them
the most compact multiple protostar systems 
directly detected. Previously, the closest known  
deeply embedded systems detected at millimeter/centimeter wavelengths
were the 45 AU system in L1551 IRS 5 \citep{looney1997,rodriguez1998} and
the 40 AU system in IRAS 16293-2422A \citep{wootten1989}. The implications of these systems will be discussed
further in Section 5.2 and more details of these sources are discussed in Appendix A. 
In addition to these three systems, four others showed evidence for resolved structure
on $<$30~AU scales but did not have enough S/N to be regarded as a multiple system, and these additional sources are also
shown in Appendix B.

\subsection{Multiple Systems Separated by $>$ 500 AU}

We show images of the multiple systems on scales $>$ 500~AU in Appendix C.
Of the wide multiples shown, only Per-emb-37
(see Appendix C) is a new detection, 
though the companion sources are distinct
in \textit{Spitzer} IRAC imaging.
Formally, our level of completeness is a function of separation given
the decreased sensitivity away from field center, but the primary beam response is still 85\% at 20\arcsec\ from
the field center. On scales $>$20\arcsec\ (4600~AU), the multiplicity of protostars has been characterized at
infrared and submillimeter wavelengths. We detect all known wide multiples with separations between 4600~AU and 10000~AU.
The analysis of expected extragalactic background sources given in Section 4.6 suggests that there may be a background source
within the VLA primary beam in a few fields. To check for such sources, 
we have cross-compared our images with infrared imaging from \textit{Spitzer} or \textit{Herschel} \citep{evans2009,sadavoy2014}
to verify that there are associated infrared sources with the wide multiple systems, and that their
colors and flux densities that are inconsistent with being extragalactic objects.

\subsection{Multiplicity Statistics}

In our analysis of multiplicity in Perseus, we only consider the sources detected as multiples
in our data and not those reported from other studies for consistent. See Appendix C for a discussion
of non-detections of previously reported multiples. The detected Class 0 multiple systems are listed in Table 3, the detected Class I multiple 
systems are listed in Table 4, and the Class II multiple systems are listed in Table 5. 

The MF and CSF (see section 1 for definitions) are the key figures of merit for
describing the multiplicity for collections of stars. 
We have calculated these statistics for the VANDAM Perseus Survey: 
for the entire sample, MF = 0.40 $\pm$ 0.06 and CSF = 0.71 $\pm$ 0.06 (S:B:T:Q:5:6=37:17:5:2:2:1),
for the Class 0 sources MF = 0.57 $\pm$ 0.09 and CSF = 1.2 $\pm$ 0.2 (S:B:T:Q:5:6=15:9:5:2:2:1), and
for the Class I sources MF = 0.23 $\pm$ 0.08 and CSF = 0.23 $\pm$ 0.08 (S:B:T:Q=20:6:0:0),
\footnote{Note that the uncertainties throughout the text
are calculated assuming binomial statistics, $\sigma_{CSF}$ = (N$_{comp}$(1-N$_{comp}$/N$_{sys}$)$^{-0.5}$ 
$\times$ 1/N$_{sys}$ where N$_{comp}$ is the number of companions and N$_{sys}$ is the number of systems.
$\sigma_{MF}$ is calculated similarly, but by substituting N$_{mult}$ (number of multiple systems) for N$_{comp}$.
Poisson statistics are not used because the criteria of N$_{comp}$ $>>$ N$_{sys}$ is not met. However, we note
that the variance calculated assuming binomial statistics is only slightly smaller than that of 
Poisson statistics. For the case of CSF $>$ 1.0, $\sigma_{CSF}$ is not a real number and we revert to 
Poisson statistics in this case.}. The statistics are further
enumerated in Table 6 for different ranges of separations 
for the full sample, Class 0 sub-sample, and Class I sub-sample.
Note that the Class 0 systems that have a wide Class I or Class II companion are only considered
in the Class 0 MF and CSF, and the Class 0/I systems are also only considered in the Class 0 statistics.
Furthermore, we only include the Class 0 and Class I systems detected in our survey within these statistics.
Because the smallest separations that we can probe is $\sim$15 AU, the MF and CSF values 
given here and in Table 6 should be considered lower limits.

The Class II sources have MF = 0.33 $\pm$ 0.19 and CSF = 0.33 $\pm$ 0.19 (S:B:T:Q=4:2:0:0), but our
survey only included a small number of Class II sources and these systems are bright
in the far-infrared. Thus, the Class II statistics are biased and too small to draw meaningful conclusions.

The multiplicity of Class~0 protostars was previously examined by \citet{chen2013} and those authors found
MF=0.64 $\pm$ 0.08 and CSF = 0.91 $\pm$ 0.05 for a separation range of 50 AU to 5000 AU. For the Class~0
multiples within this separation range, we find an MF = 0.45 $\pm$ 0.09 and 
CSF = 0.88 $\pm$ 0.06. The results are comparable, and the difference
in the MF could be due to sample bias in \citet{chen2013} and the fact that we do not detect multiplicity toward all Perseus
sources where  \citet{chen2013} reported multiplicity (see Sections 5 and the Appendix for further discussion).

For the Class I sources, \citet{connelley2008} find a MF = 0.35 $\pm$ 0.03 and 
CSF = 0.45 $\pm$ 0.04 (S:B:T:Q=122:51:12:4). \citet{duchene2013} presented a combined analysis of
\citet{connelley2009} and \citet{duchene2007} to derive a CSF of 0.35 $\pm$ 0.05 for Class I sources 
with separations between 50~AU and 2000~AU. For Class I multiples in the same separation range, 
we find both the MF and CSF = 0.28 $\pm$ 0.08; this
is consistent with the results of \citet{duchene2013} within the uncertainties. 
We note that there are two systems comprised of 
a Class 0 and a Class I source within this range of separations
that were included in the Class 0 statistics only.
If we added these sources to the Class I statistics, the MF and CSF 
would be more consistent with the \citet{duchene2013} value.

We find that the overall values of MF and CSF for the Class 0 and Class I sources are not 
significantly different from previous studies, despite our larger and improved sample for several
reasons: 1) many systems already considered multiple in the MF were found in our survey
to have additional closer systems,
2) the number of new multiple systems is balanced by the number of additional systems confirmed to be single, and
3) some systems previously considered to be multiple are not confirmed in our study. 
The MF of Class 0s is lower, likely due to our unbiased sample which detected more single systems. Furthermore,
past studies have often focused on systems that were known to be multiple, 
and samples were biased to the brightest sources at millimeter
wavelengths.

\subsection{Separation Distribution}

Figure \ref{comb-histo} shows the distribution of companion separation for our 
full sample and for the Class 0 and Class I sub-samples, using 
the separations listed in Tables 3, 4, and 5. For systems comprised
of 3 or more members, the distances are all referenced to a single source, usually the most luminous. Thus,
only two separations are considered for a triple system, not all three possible separations.
In the case
of a quadruple (or higher order) comprised of two close multiple systems (e.g., L1448-N, Appendix C) then only the brightest members
in each close multiple system are used to compute the distance to the more widely separated system.
For the full sample, we find a bi-modal distribution with peaks at separations of $\sim$75 AU and $\sim$3000 AU. Between
these peaks there is a valley with only 7 companion sources detected between 
200 AU and 1000 AU. There is also a notable decline in multiplicity at separations $<$ 57.7 AU,
with only three sources having strong evidence of multiplicity. 
We emphasize that the range of spatial scales examined and the numbers of multiple 
systems detected and characterized is currently without precedent, especially
for a sample within the same molecular cloud at a common distance. Much of the improvement
in statistics comes at scales less than 500 AU, where there had been few previous observations. The
largest previous study for Class 0 protostars by \citet{chen2013} had median resolution of 600 AU.  
We do note, however, that the statistical significance of 
the two peaks is marginal in the histograms, but we will statistically compare the cumulative distribution
in the following section.

The size of our sample enables us to examine 
the multiplicity of Class 0 and Class I
systems independently and Figure \ref{comb-histo} also shows 
several key differences between the Class 0 and Class I separation
distributions. First, the Class I systems
have a peak in companion frequency at $\sim$75 AU scales and only a few multiples on scales larger than 100 AU.
The Class 0 systems on the other hand retain the double-peaked distribution seen 
for the full sample. We constructed
cumulative distributions for the two samples (see Figure \ref{cumulat-class}) and 
performed an Anderson-Darling (AD) test\footnote{The Anderson-Darling test is 
similar to the Kolmogorov-Smirnoff (KS) test, but is 
more statistically robust. This because the KS-test uses the maximum deviation to calculate the probability
and is not as sensitive when deviations are at the ends of the distribution or when there are 
small but significant deviations throughout the distribution. https://asaip.psu.edu/Articles/beware-the-kolmogorov-smirnov-test} \citep{scholz1987}, the results of which indicate that  
the probability of the Class 0 and Class I sources being drawn from the same distribution is only 0.16. 
The inclusion of wide multiples comprised of both Class 0 and Class I sources
 with Class 0 would decrease the probability of the
two samples being drawn from the same distribution, but if they were included with the Class I
distribution only, that would make it more likely that the Class 0 and Class I samples were drawn from the same
distribution. Thus, our results are suggestive of differences between the separation distributions
of the Class 0 and Class I protostars but with marginal statistical significance.

\subsection{Constraining the Functional Form of the Separation Distribution}

We compared our dataset to several simple models to determine what the data
can and cannot rule-out in terms of the underlying separation distribution. There are several possible models
that could describe the underlying distribution of separations, 
and we tested a log-flat distribution, 
a model that represents the fields solar-type star separation distribution,
and a model that employs multiple Gaussian functions.

We first compared to a 
log-flat distribution of multiples between 15 and 10000 AU, also known
 as {\"O}pik's Law \citep{opik1924}.
Such a distribution would produce a constant level of 
multiplicity at all separations in a histogram like that of 
Figure \ref{comb-histo}.
The cumulative distribution for a log-flat distribution
of separations is drawn in Figure \ref{cumulat-full} and compared to the data.
The log-flat distribution is always in excess of the observed distribution, except for the largest separations, and
the AD probability for this distribution
is 0.1, so a log-flat distribution of separations is unlikely.

We also considered a model that represents the separation distribution of field
solar-type multiple systems.
The distribution was fit with a Gaussian by \citet{raghavan2010}
with a mean $log(a)$ = 1.7 ($\sim$50 AU) and $\sigma_{loga}$ = 1.52 in units of log(AU). These
are derived from $log(P)$ = 5.03, $\sigma_{logP}$ = 2.28 in units of log(days) assuming a 1.5 $M_{\sun}$ primary mass.
We compare our separation distribution to the \citet{raghavan2010} fit, finding an
AD probability of 0.00015, indicating that the separation distribution of 
solar-type multiples is very unlikely to match that of our protostellar multiples. 
The disagreement provides further evidence that binary systems dynamically 
evolve from their initial separations.

Finally, the double-peaked histogram in Figure \ref{comb-histo} suggests that the separation distribution might
be represented by two Gaussians. We compared the observed distribution to a grid of Gaussian functions
and found that two Gaussians are consistent with the data (probabilities of 0.99 are achieved). 
However, the parameters of the Gaussians are not well-constrained;
a typical fit has the inner peak at $\sim$ 90 AU and the outer peak between 3000~AU and 10000~AU.

\subsection{Extragalactic Background Estimation}

Extragalactic sources that are dominated by synchrotron emission
increase in brightness at longer wavelengths and can become a source of contamination
in sensitive radio surveys.
We have followed the analysis for background objects presented in
\citet{anglada1998} 
to estimate the number of background source that we expect to find
in our survey. Our typical sensitivity was 10 \microjy, thus we estimate the number
of extragalactic background sources at Ka-band 
with a flux density $\ge$30 $\mu$Jy within a 5\arcsec\ (1150 AU) field of view. This
is done by extrapolating the 5 GHz number counts and
assuming a typical spectral index of $\alpha$ = -0.7 for optically-thin synchrotron 
emission \citep{condon1984}. We find that there is a probability of only
3.3 $\times$10$^{-4}$ of finding a background source within a 5\arcsec\ field of view; 
the probability becomes 0.041 for a 60\arcsec\ field of view. This analysis ignores the 
potential contributions of radio emission from submillimeter galaxies, 
where the combination of bright dust and free-free emission associated
with star formation will likely produce flatter spectral indices, making them more detectable. 
For 90 observed fields, we expect to detect $\sim$4 extra-galactic sources. We 
conclusively identify two likely extragalactic sources in our observations, see Tables 1 and 2. They
have negative spectral indices at Ka-band and no corresponding detections at shorter wavelengths. These numbers
are consistent with the expected number of extragalactic sources considering that a portion of the fields
observed overlapping regions of sky.
Thus, it is very unlikely that any close or wide multiples are false detections due
to extragalactic confusion.

\section{Discussion}

The origin of stellar multiplicity has gained significant attention recently
due to the downward revision of solar-type star multiplicity frequency to 0.46 \citep{raghavan2010}
and the finding that the fraction of single M-stars is $\sim$0.63 \citep{lada2006}. Furthermore,
the searches for brown dwarf and planetary mass companions around pre-main sequence stars
\citep[e.g.,][]{white2001,kohler2006,reipurth2007,kraus2008,kraus2011}, have produced large statistical 
samples of multiplicity.
 Nevertheless, connecting these statistics to 
multiple star formation remained uncertain due
to a lack of definitive results on the multiplicity and separation distribution toward
embedded protostars.

The primary routes for the formation of multiple systems are (1) the fragmentation of the core or filament
and (2) disk fragmentation. Core fragmentation can be either thermal (Jeans) fragmentation aided
by rotation and asymmetry \citep[e.g.,][]{bb1993, bonnell1993} or turbulent fragmentation
\citep{padoan2002,padoan2004,offner2010}; these routes tend to produce companions on $\sim$1000 AU scales, but can also
result in companions with ultimate separations $<$ 100 AU via migration \citep{offner2010,bate2012}.
Fragmentation of the protostellar disk via gravitational instability 
can also directly form close companion systems \citep[e.g.,][]{adams1989,bonnell1994a,kratter2010,zhu2012}.

Large simulations of entire star forming molecular clouds have been conducted with enough resolution
to examine fragmentation on the scales from the cloud down to the disks \citep[e.g.,][]{bate2009,bate2012}.
The multiplicity results from such simulations are typically
compared to the field star multiplicity; however, several Gyr
of dynamical evolution in the field population will impact such comparisons
to simulations of younger systems. 
Observations of more deeply embedded multiple systems, such as those presented in this paper,
 will provide a more direct diagnostic to test models of star formation, given that their ages are most likely 
all less than 0.5 Myr \citep{dunham2014}, comparable to the length of time
explored in the simulations.

There has been debate on the origin and frequency of multiplicity 
in the Class 0 protostellar phase, centering
around studies that have small, biased samples of sources. 
\citet{looney2000} examined 11 Class 0 protostellar systems, finding a
preponderance of multiplicity in these systems. However, the sources in the sample 
are among the brightest millimeter sources in the nearby star forming regions and
may not be representative. \citet{maury2010} then examined 5 systems
\citep[including 2 Very Low Luminosity Objects, protostellar sources which have internal
luminosities $<$ 0.1 L$_{\sun}$; ][]{young2004}, not
finding any multiples on scales $\la$ 1600 AU. Their sample, combined with that of 
\citet{looney2000}, led them to conclude that there was no evidence for multiplicity
on scales between 150 AU and 400 AU for Class 0 protostars; the separation of 400 AU reflects the updated
distance to Perseus, which affects the separation of NGC 1333 IRAS4A. Moreover, \cite{maury2010}
went on to tentatively suggest that multiplicity \textit{increased} from the Class 0 to Class I
phase, at least for separations between 150 AU and 400 AU. This would not necessarily be
a true increase in multiplicity but possibly an evolution in separations from initially
wider separations to closer separations \citep[e.g.,][]{offner2010,zhao2013}. Nonetheless,
the robustness of these findings was unclear given the small sample sizes of both
\cite{maury2010} and \cite{looney2000}.

\citet{chen2013} made use of archival SMA data to better characterize
multiplicity in the Class 0 phase using a sample of 33 protostars located in various star forming regions.
For the separation range (50 AU to 5000 AU), \citet{chen2013} showed that the multiplicity fraction for Class 0 protostars is $\sim$0.65.
This is much higher than the $\sim$0.35 for Class I systems \citep{connelley2008}
 and $\sim$0.2 for solar-type field stars \citep{raghavan2010}, indicating that multiplicity is highest in 
the Class 0 phase in this separation range. However, that study did not necessarily rule-out
the conclusion by \citet{maury2010} of multiplicity increasing for separations
between 150 AU and 400 AU. This is because \citet{chen2013} lacked homogeneous
sensitivity and resolution (median resolution of 600 AU), but multiples were reported
by \citet{chen2013} in the range between 150 AU to 400 AU.

The VANDAM survey surmounts these limitations of the previous
studies by observing a large number of protostars (94; 77 detected) in a single star forming region, at
nearly uniform sensitivity (apart from the sensitivity attenuation of the primary beam) and resolution. 
Multiple sources can be resolved with separations
as small as $\sim$0\farcs065 (15 AU). This survey contains the largest
and least biased sample of protostars ever observed with sub-arcsecond resolution.
This survey also boasts the highest ever sensitivity in the 8 mm to 1 cm wavelength range for protostellar multiples. 
Thus, we have been able to characterize protostellar multiplicity with unprecedented statistics.

Although the results from this survey represent enormous progress, there are limitations to how well multiplicity
can be characterized in the context of the protostellar properties. A major limitation is that we
do not know the masses of the protostars (or systems) themselves. We only know the bolometric luminosities
sampled from the near-infrared to submillimeter, which range between $\sim$0.1 L${\sun}$ and $\sim$ 33 L${\sun}$, with 
a median of 0.7 L${\sun}$. The range and distribution of luminosities are typical of the population of known
protostars \citep{dunham2014,dunham2015}. However, it is not trivial to directly translate luminosity to stellar mass
for protostars because the emergent luminosity is dominated by (or has a significant component from) accretion
processes that can be highly variable.

To make estimates of the protostar masses, we can compare to models 
of the protostellar luminosity function with an underlying protostellar mass function,
assuming smooth accretion \citep{offner2011,mckee2010}. Within the context of these models,
most protostars in our sample are expected to be progenitors of K and M-stars.
However, even if those models are reliable, the bolometric luminosities of the components to multiple
systems separated by $\la$ 1500~AU
cannot be determined due to the resolution limitations at mid to far-infrared wavelengths.
 Thus, we cannot say anything about the mass or luminosity ratios of the close 
protostellar binaries themselves. Finally, there is an inherent bias in characterizing
multiplicity at millimeter/centimeter wavelengths, and we may not detect all 
companion sources as evidenced by some of the faint
companion sources detected toward some Class I systems. 
Therefore, our statistics represent lower limits
to the MF, CSF, and the companion frequency as a function 
of separation, see section 5.6 for further discussion.

\subsection{Origin of the Bi-modal Separation Distribution}

The distributions of separations shown in Figure \ref{comb-histo}
represent the most complete snapshot of protostellar multiplicity and also the highest resolution study that
has been compiled in a single star forming region. It is tempting to interpret the distribution of separations
as the initial distribution of separations in multiple systems; however, even 
at these very young ages it is possible that significant migration has 
already taken place \citep[e.g.,][]{offner2010,bate2012}. For example,
systems driving orthogonal outflows, but with close separations, like NGC 1333 IRAS2A \citep{tobin2015},
may have resulted from migration. Nevertheless, our sample of embedded multiples, especially the Class 0 
systems, should have a separation distribution that is closer to the initial separation
distribution than what would be obtained from more evolved 
sources. Thus, the VANDAM survey provides the best direct constraints on the origin of multiplicity thus far.

The most striking feature of the separation distribution for the full sample and Class 0 sources
in Figure \ref{comb-histo} is that the distribution appears bi-modal, with one peak near $\sim$75 AU 
and the other near $\sim$3000 AU. This feature is unlikely to be the result of any selection bias because
we have observed all the known protostars in the Perseus molecular clouds. Furthermore, our spatial resolution 
and sensitivity are sufficient to have detected multiples between 100 AU and 1000 AU if they were present.

An attractive interpretation of the bi-modal distribution is that the peaks are produced 
by two distinct mechanisms, namely disk and core fragmentation, respectively. Disk fragmentation
would naturally produce the multiples of $\la$ 300 AU scales and core fragmentation would then produce the
multiplicity on scales $>$ 1000 AU. Early studies of thermal (Jeans) fragmentation of 
dense cores concentrated on the effects of rotation and non-spherical 
shape \citep[e.g.,][]{bonnell1993,bb1993}. More recent calculations 
have focused on fragmentation induced by turbulence \citep{walch2009, offner2010, padoan2002}.
The complex structure and velocity fields often observed toward protostellar cores may provide some 
evidence for this picture \citep{tobin2011,pineda2011,pineda2015}. Furthermore,
wide multiples produced through turbulent fragmentation can tighten their 
separations through orbital migration on timescales as short as 10 kyr, 
potentially contributing to the close multiple population \citep{offner2010}.
However, if the close multiples are the result of migration, some mechanism must then cause them to 
accumulate at $\sim$ 75 AU rather than continuing to migrate inward.

The differences in the separation distributions for the Class 0 and Class I 
systems are suggestive of evolutionary effects. Class 0 systems have considerably
more wide multiples than Class I systems. 
The orbital period for a 4000 AU separation
binary system is $\sim$250 kyr (assuming 1 $M_{\sun}$), and if the system dissolves due
to internal dynamics, the timescale should be longer than an orbital period. This timescale is likely too long
for protostellar systems because the expected lifetime of a Class 0 system is only $\sim$160 kyr \citep{dunham2014}.
Therefore, we consider two additional interpretations related to the formation and evolution of these systems.

The first possibility is that wide multiples had formed initially, and as they evolved into the 
Class I phase the separations increased because the companions may have been unbound at the time of formation due
to initially large differential velocities 
as a result of turbulent fragmentation.
While it is true that systems are not binary/multiple if they
are not gravitationally bound, we are unable to assess whether 
or not all systems are bound. Therefore, we presently
consider all systems with projected separations less than 10000 AU as
a bound multiple system.
The boundedness of the widely separated systems is an active area 
of investigation \citep[e.g.,][ Lee et al. in prep.]{lee2015}, and
systems that are currently bound within their star forming cores 
may later become unbound as their envelope material is
dispersed by outflows \citep{arce2006,offner2014}.

The second possibility is that the wide multiples
dynamically evolved toward close separations, giving rise to the peak at $\sim$75 AU.
We regard the first possibility as more likely because, many of the wide Class 0
multiples are separated by more than 1000 AU, making it 
possible that some
of these systems would be unbound. In addition, the fraction of multiples at 
$<$ 300 AU scales
is comparable for both Class 0 and Class I sources. The similarity at scales $<$ 
300 AU
can be explained by either wide multiples not frequently migrating to $<$ 300 AU scales
or by the currently observed Class 0 multiples at $<$ 
300 AU 
migrating  to scales $<$ 15 AU (i.e., are now unresolved). The
$>$1000 AU companions would then need to migrate and fill-in the distribution at 
$<$ 300 AU
scales.

Turbulent fragmentation and disk fragmentation are expected to produce multiple systems that appear nearly coeval. 
On the other hand, the Class 0 sources with widely separated Class I or Class II
companions may also be evidence that significant, rapid orbital evolution does not happen
in all cases or that an additional process is at work.
A promising route to explain these systems is a dynamical 
ejection scenario \citep{reipurth2001,reipurth2010,reipurth2012}. In this scenario, a close triple system would 
have formed initially and dynamical interactions cause one member to be ejected into a very wide orbit. Even though
the ejected companion would be as young as the remaining compact binary, it might appear more
evolved because it would no longer be so deeply embedded and perhaps directly visible at near-infrared wavelengths. Thus,
the widely separated systems with different evolutionary states could be very young stars that were ejected from their cores.

\subsection{Multiplicity Evolution}

A principle conclusion of \citet{chen2013} was that multiplicity is decreasing with evolution,
decreasing from the Class 0 phase to the Class I phase within the separation range of 50 AU to 5000 AU. However, a limitation
of that survey was the inhomogeneous resolution (median resolution of 600 AU). In comparison, 
the VANDAM survey consists of a large, homogeneous sample at  $\sim$15 AU resolution.
With this large dataset, we can 
examine the multiplicity frequency of Class 0 and Class I systems separately.

We also showed the apparent differences between the Class 0 and Class I 
multiplicity distributions at separations $>$ 1000 AU in Figure \ref{comb-histo} (see Section 4.4), and
that there is marginal evidence for a statistical difference in the separations between the two populations. 
We can also compare the Class 0 and Class I populations in terms of their MF and CSF. 
Note that we count those Class 0 systems with a wide Class I or Class II companion in the MF
and CSF for the Class 0 sources only. Our main results are unchanged if these sources were also included in
the Class I statistics.

Across the full range of separations, from 15 AU to 10000 AU, we find 
that multiplicity is decreasing from the Class 0 to the Class I phase, in agreement with
\citet{chen2013} and in contrast with \citet{maury2010}. For example, we the 
MF = 0.57$\pm$0.09 for Class 0s and MF = 0.23$\pm$0.08 for Class Is.
If we then examine the separation range from 15 AU to 5000 AU \citep[the same outer limit as][]{chen2013}, we still find 
decreasing multiplicity from Class 0 to Class I (MFs of 0.55$\pm$0.09 and 0.24$\pm$0.08, respectively). The
same is true if we examine the separation range from 50 AU to 5000 AU \citep[the same range as][]{chen2013}, 
though we find that the MF for Class 0 sources is 0.45$\pm$0.09 and 0.24$\pm$0.08 for Class I sources. 
We note, however, that our value of Class I multiplicity
is consistent within the uncertainties with both the \citet{connelley2008} 
value of 0.35$\pm$0.03 and the value for field solar-type
stars from \citet{dm1991} for the separation 
range between 50 AU and 5000 AU as calculated by \citet{chen2013}. Thus, while we confirm a multiplicity decrease on these
scales from Class 0 to Class I, we do not confirm a further decrease from Class I to field stars from our data alone.

In contrast to the larger separations, the MF and CSF between 15 AU and 2000 AU of the Class 0 
and Class I subsamples are consistent within the uncertainties.
Thus, we conclude that on scales less than 2000 AU,
there does not appear to be multiplicity evolution taking place between the Class 0 and Class I 
phase. \citet{maury2010} had suggested 
that multiplicity increased from the Class 0
to the Class I phase
on these scales,
but this suggestion is not supported by our larger sample. Furthermore,
\citet{maury2010} suggested that there was no evidence for multiplicity
between 150 AU and 550 AU (400 AU). While multiples are clearly found within this range of separations
in our study and that of \citet{chen2013}, there is a 
deficit in multiples in this range of
separations relative to smaller and larger scales. Suffice it to say that there is, however, evidence
for
slightly lower multiplicity for both Class 0 and Class I systems
between 150 AU and 1000 AU.

\subsection{Evidence for Disk Fragmentation}

Three remarkable systems (IRAS 03292+3039/Per-emb-2, IRAS 03282+3035/Per-emb-5, and Per-emb-18) 
show multiplicity on scales $<$ 30 AU; see Figures \ref{IRAS03292}, \ref{IRAS03282}, and \ref{Per18}.
In each of these cases, the sources are surrounded by an extended structure and only become resolved into discrete
sources when imaged at higher resolution. IRAS 03292+3039/Per-emb-2 and Per-emb-18 have the largest continuum
structures detected in our survey, about 1\farcs5 and 1\arcsec\ in diameter, respectively. The A-configuration
data resolve-out the extended emission and reveal additional brightness peaks separated by $\sim$19 AU in both
cases. It is peculiar that the extended dust emission is only on the eastern 
side of Per-emb-18, having the appearance of a companion itself
when viewed at lower resolution. The dusty structure surrounding IRAS 03282+3035
is only $\sim$0\farcs5 in diameter.

Gravitational instability in a disk is the most likely mechanism for the production of
any substructures detected on scales $<$30 AU. This scale, however, this scale is near the inner
limit of where the disk is expected to cool quickly enough for gravitational
instability to make a bound object \citep{rafi2005,matzner2005}. 
Thus, these companions may have migrated to their current locations from initially larger
radii or the disks 
were cold enough to allow fragmentation on these scales due to the source luminosities being
low; L$_{bol}$ = 0.9, 1.3, and 2.8 for Per-emb-2, Per-emb-5, and Per-emb-18, respectively.

The masses 
associated with the extended structures
 on 0\farcs5 to 1\farcs5
 scales are estimated to be $>$ 0.1 $M_{\sun}$ from 1.3 mm
dust emission \citep{tobin2015b}. However, a missing piece of evidence
 is the dense gas kinematics,  which is necessary to determine whether or not these clumps
are the result of a fragmenting, rotationally supported disk. In the
 case of IRAS 03292+3039, there is evidence of
inner envelope rotation \citep{schnee2012,yen2015}, suggesting that
 a rotationally supported disk is possible for this source.
There have also been molecular line data for IRAS 03282+3035 \citep{arce2006, yen2015}, but a rotation signature
 is unclear toward this source and Per-emb-18 does not yet have existing observations.

The clumpy structure observed toward IRAS 03292+3039 on $>$ 0\farcs5 
scales appears real, sub-peaks within this structure
have close coincidence with peaks observed at 1.3 mm \citep{tobin2015b}. However, the 1.3 mm data
have a much smoother appearance, a possible indication that the dust emission is optically 
thick at 1.3 mm, but optically thin at 8 mm and 1 cm. It is unclear
 if the clumpy structures surrounding the source have formed or are 
likely to form protostellar objects. The peaks observed north and south of the
main protostar(s) are also present at 1.3 mm and when the 8 mm data are imaged at higher resolution 
(with lower S/N).

While we are confident that the structures observed on $<$30 AU scales are real, it is uncertain if they
were formed in their current locations, given that fragmentation via gravitational instability is difficult
at this scale. Furthermore, the ultimate fate of these structures is uncertain. For instance, gravitationally
unstable disk models often show clumps that have yet to collapse into stellar objects
migrating inward \citep{vorobyov2006,vorobyov2010}. Some clumps can be tidally disrupted if they have
not formed a bound object, or they may be accreted on to the protostar \citep{zhu2012}.
The accretion of these clumps results
in an increased luminosity and could be an explanation for the large spread 
observed in the luminosity distribution of young stellar objects \citep{dunham2014}. 
If each of the observed structures is associated with a stellar object, 
then it is unlikely for them to merge together.
Thus, these structures could be transient or they might reflect the 
formation of close companions. 

Another way to produce substructures in the dust emission is the Rossby Wave Instability 
(RWI) \citep{barge1995, klahr1997}. \citet{bae2015} showed
that RWI can be triggered in protostellar disks by the velocity
shear of the material falling onto the disk. This process could possibly explain some of the features we observe, e.g.,
the asymmetric dust clump around Per-emb-18. However, the RWI only concentrates the dust and not gas, and 
the largest dust grains are more highly concentrated than the smaller grains. Thus, in this scenario,
the detection of clumps would not necessarily be related to multiple star formation \citep[e.g.,][]{vdmarel2015}.
Observations of molecular line kinematics will help elucidate the nature of the small-scale substructures
and these sources are close enough that orbital motion can possibly be observed
in just a few years time.

\subsection{Orientation of Multiple Systems}

Figure \ref{posangles} shows the distribution of relative position angles 
between the close companions (separations $<$ 500 AU)
and the outflow axis of the protostars; the list of position angles is given in Table 7. 
The disk around the protostar is assumed to be oriented normal to 
the outflow direction (at least the portion driving the jet); therefore, if close companions have 
formed in the rotational plane
as a result of disk fragmentation or fragmentation of the rotating envelope this should be reflected in the 
distribution of relative position angles. For comparison, we also draw the distributions
for a uniform distribution of angles and the 
distribution of relative position angles for a random distribution of
binary orbital phases and inclinations.

Without performing any statistical tests, 
it is apparent that
the observations have a small 
excess of sources with small relative position angles over what would 
expected for randomly oriented circular orbits
(dotted line in Figure \ref{posangles}).
This is a random distribution of 
companion orbital phase and viewed with a random inclination, consistent
with companions being located in the plane of the
disk, normal to the outflow direction. The bottom panel of Figure \ref{posangles}
shows a scatter plot of companion separation versus position angle and there are
no apparent trends. The average relative 
position angle is 50\degr\ in the observations, while the average angle for 
randomly oriented circular orbits is $\sim$70\degr. Elliptical orbits in the 
disk plane would not help resolve the inconsistency because the companion
would spend more time at apastron and more sources would be expected to have 
relative PAs closer to 90\degr. Close companions formed via turbulent fragmentation
are not expected to follow a preferred orbital configuration and could be partly 
responsible for the disagreement. 
Note, however, that the multiple system NGC 1333 IRAS2A (Per-emb-27) has two orthogonal outflows and
we only list the dominant north-south outflow in the table, resulting 
in a small relative position angle. Including
the east-west outflow as an independent point or instead of the north-south outflow would reduce the excess.
Thus, the number of close companions with measured outflow 
position angles is currently too small to currently draw definitive conclusions. 
However, the distribution of relative outflow position angles for both the close and wide multiple systems is
being investigated further with new outflow data for the entire sample of Perseus protostars (Lee et al. in prep.).

Follow-up of these sources will enable proper motions to be measured and determine
if companions are co-moving or rapidly moving away. If the companion sources are 
found to be moving away rapidly, then they are most likely to 
be blobs of free-free emission in the jet and not true companion sources; an event like
this has been observed in the source IRAS 16293-2422A \citep{pech2010}. However, the spectral indices
of the companions are positive and not consistent with being optically thin free-free knots. Furthermore, the
distribution of spectral indices for companions is indistinguishable from
that of the single protostars, see Section 5.6.

\subsection{Comparison to T Tauri Multiples in Taurus}

The best characterized group of young stellar multiples is in Taurus, where \citet{kraus2011}
combined new observations down to 3 AU scales with previous multiplicity 
searches. This survey of visual multiples is sensitive to separations
between 3 AU and 5000 AU for primary star masses ranging between 0.25 $M_{\sun}$ and 2.5 $M_{\sun}$. The large number
of stars enabled the sample to be divided into low-mass and high-mass sub-samples. The high-mass
sub-sample has a roughly uniform distribution of 
companions (in log(separation)) out to 5000 AU, while the low-mass sub-sample
has very few companions at scales $>$ 200 AU. 

We compare our results to their sample in the cumulative distribution 
shown in Figure \ref{cumulat-taurus} for separations ranging between 
15 AU and 5000 AU. Both the full sample and low-mass sub-sample ($M_*$ $<$ 0.7 $M_{\sun}$) from \citet{kraus2011}
appear inconsistent with the distribution of multiples 
found in our sample, having AD probabilities of 0.025 and 0.00015,
respectively. The high-mass subsample (2.5 $M_{\sun}$ $\ge$ $M_*$ $>$ 0.7 $M_{\sun}$), 
on the other hand, appears consistent with our distribution 
having an AD probability of 0.80. The primary difference between the low-mass sample and the high-mass sample
is a lack of wide multiples in the low-mass sample. Considering the Class 0 and Class I systems separately,
the Class 0s agree best with the high-mass sample with an AD probability of 0.71 and the Class Is agree best
with the full sample with an AD probability of 0.21. For the other possible combinations with the Taurus sample, 
the Class 0 and Class I are likely to not have been drawn from the same 
distributions with AD probabilities less than 0.085.

The separation distribution agreement
with the high-mass subsample (2.5 $M_{\sun}$ $\ge$ $M_*$ $>$ 0.7 $M_{\sun}$) and the strong disagreement with the 
full/low-mass sample ($M_*$ $<$ 0.7 $M_{\sun}$) is quite striking and can be interpreted in several ways.
If we assume that companion separations do not significantly evolve between
 the Class 0 to Class II/III phases, then one could infer that the 
multiple protostar systems that we detect are going to be progenitors of high-mass systems.
While protostellar mass measurements are not available, the closest available proxies for stellar 
mass are either luminosity or core mass. 
\citet{mckee2010} and \citet{offner2011} examined both the protostellar mass function and 
protostellar luminosity functions. The observed protostellar luminosities can be reproduced with a
mass function closely following the Chabrier IMF; the two component turbulent core model (2CTC in their Figure 3). With 
this underlying mass function, the typical protostellar mass is $\sim$0.2 $M_{\sun}$ and only $\sim$14\% of sources should have
masses between 0.7 and 2.5 $M_{\sun}$, thus our sample should be comprised of mostly sources $<$ 0.7 $M_{\sun}$.
Furthermore, we observe no specific trend in multiplicity with respect 
to bolometric luminosity and there is no obvious trend with core mass.
However, core mass will change with evolution as the protostars accrete material and
outflows remove material from the envelope; most Class 0 systems have core
masses $>$ 0.5 $M_{\sun}$ \citep{enoch2009}.

If the protostars we observe in our sample are indeed characteristic of the low-mass Taurus systems, 
then they must
have undergone significant dynamical evolution since their formation and the binary orbits have
contracted. Therefore, the separation distribution that we observe toward the protostars could evolve toward what is observed 
for the low-mass Taurus systems. If this interpretation is true, then by inference the high-mass systems 
in Taurus may
have not undergone significant dynamical evolution.
We caution that these statements implicitly assume that the Perseus and Taurus multiples will follow 
the same evolutionary path.

The paucity of low-mass Taurus
systems with wide separations could imply that low-mass systems do not typically fragment on large-scales or that the
low-mass systems cannot hold onto wide companions. It is unknown if the wide multiple systems in 
Perseus are bound. If these wide systems drift apart over time, then the distribution of separations in Perseus
would become less consistent with the high-mass Taurus sample and more consistent with the low-mass Taurus sample. It was also
argued in \citet{kraus2011} that the high-frequency of close companions in the low-mass Taurus sample could be indicative
of disk fragmentation occurring preferentially on a 50 AU - 100 AU scale. This finding is consistent with our
distribution of close separations which peaks at $\sim$ 75 AU.

Furthermore, \citet{kraus2011} argued that the mass ratio of close companions being consistent with a log-flat
distribution is suggestive of disk fragmentation taking place after the primary has accumulated most of its mass,
accounting for $\ga$ 1/2 the entire core mass. If disk fragmentation occurred early in protostellar evolution,
\citet{kraus2011} argued that the mass ratio would be skewed toward unity rather than log-flat. It is unclear if such
a signature is present in our sample given that the Class 0 and Class I systems have similar numbers of companions
separated by $<$ 300 AU.

Finally, it is also possible that the agreement and/or disagreement between the Perseus multiples
and Taurus multiples is completely coincidental and reflects the different properties of the
two clouds and their YSO populations. The two regions have significantly different clustering
properties, gas densities, temperatures, and ratios of protostars to pre-main sequence stars. 
Thus, the differences
could simply result from the different initial conditions.

\subsection{Nature of 8 mm and 1 cm Emission}

Continuum emission at 8 mm and 1 cm likely include significant contributions from both thermal dust emission and 
thermal free-free (Bremsstrahlung) are likely. The dust emission is likely tracing the 
protostellar disk and/or inner envelope and the free-free emission is thought 
to be produced from ionized gas resulting from shocks at the base of the protostellar
jet on scales $\la$ 10 AU \citep[e.g.,][]{curiel1989,anglada1998}. 
Free-free emission typically becomes dominant, with respect to dust emission
at $\lambda$ $>$ 1 cm. 
Discrete sources detected at centimeter wavelengths toward protostellar
cores, have also been 
associated with multiple star formation \citep{rodriguez2000,anglada2004,reipurth2002,reipurth2004}.
This is because the presence of free-free
emission enhances the detectability of protostellar sources, where dust emission
may be faint depending on the properties of the source. However, there are some
examples of spatially-extended free-free emission with multiple 
clumps along the outflow \citep{rodriguez1989,rodriguez1990,curiel1993}, but
these sources have luminosities $>$ 50 L$_{\sun}$ and they 
have observed proper motion in the outflow direction
\citep{rodriguez2000,rodriguez1989}. Furthermore, 
clumps of emission created along free-free jets should be preferentially aligned with outflows, 
which is not observed survey (see Figure \ref{posangles}).

Figure \ref{cumulat-spindex} shows the distributions of spectral indices 
from the Ka-band observations. The histograms and
cumulative distributions of spectral indices calculated from the integrated 
flux densities and peak flux densities are quite comparable.
The spectral index of optically thick emission (dust or free-free)
will be $\sim$2. Optically thin free-free emission will have a spectral index of $\sim$ -0.1 and
optically thin dust emission will have a spectral index of $\sim$2 + $\beta$; 
$\beta$ in dense dusty disks or inner protostellar envelopes is typically $\la$ 1 \citep[e.g.,][]{testi2014,kwon2009},
in contrast to interstellar medium dust which has $\beta$ $\sim$ 2 \citep[e.g.,][]{draine1984}.
Most spectral indices are less
than 2, indicative of an at least partially optically thin free-free contribution to the Ka-band flux density ,
causing spectral indices that are flatter than pure thermal dust emission.

The detection of compact free-free emission is strong evidence for the presence of
a protostar, given the requirement of a jet-driving source. However, a lack
of detected free-free emission is not evidence for the absence of a protostellar
source. Nearly all the candidate companions exhibit a combination of free-free
and dust emission at 8 mm and 1 cm, with varying levels of strength. The spectral index
of the free-free emission is indicative of the physical conditions, extended jet emission
toward higher-luminosity sources is generally optically thin (spectral index $\sim$-0.1), while
protostellar sources typically have spectral indices between 0 and 2. The source NGC 1333 IRAS2A
VLA2 described in \citet{tobin2015} has a spectral index of 1.7 and models of free-free emission \citep{ghavamian1998}
indicate that shock velocities in excess of 150 \kms\ and/or densities $\sim$ 10$^9$ cm$^{-3}$ are 
required to have such optically thick free-free emission. Such extreme conditions favor generation of emission
on small-scales
close to a protostellar source rather than from a jet impacting the surrounding medium. A detailed
analysis of the dust and free-free contributions using the 4 cm and 6 cm for the full sample
and newly detected multiples will be analyzed in a future paper (Tychoniec et al. in prep.).

Each component of the multiple systems identified in this paper
represents spatially compact emission on scales less than 30 AU. 
The densities required to produce detectable dust emission and the necessity
of a jet driving source to produce free-free emission makes it highly probable that the 
detected sources are indeed protostellar in origin.

\subsubsection{VLA-detected Companions at Millimeter Wavelengths}

Many of the multiple systems uncovered by our VLA data do not currently
have data available with comparable resolution at millimeter wavelengths
to verify pure dust emission associated at shorter wavelengths.
Sensitive, high-resolution data (0\farcs3; 5x coarser than the VLA 
A-configuration data) do exist for a few sources. L1448 IRS2 (Per-emb-22),
L1448 IRS3 (A,B, and C), and NGC 1333 IRAS2A (Per-emb-27) were observed at 0\farcs3 resolution at
1.3 mm by \citet{tobin2015,tobin2015b}. NGC 1333 IRAS4A was observed by \citet{looney2000} and \citet{jorgensen2007}
at 2.7 mm and 1.3 mm, respectively.

For L1448 IRS2, the companion separated by 0\farcs75 that was detected in the VANDAM data is not detected
at 1.3 mm. The 1.3 mm emission is, however, extended toward the
 companion position. L1448 IRS3B was found to be a triple system in the VANDAM data, 
with a close pair separated by 0\farcs26 and separated from the main source
by 0\farcs9 \citep{tobin2015b}. At 1.3mm, the main source is detected and the close
pair is well-resolved from the main source; the close pair themselves are marginally resolved. 
L1448 IRS3C, which is identified in the VANDAM data as a close binary (0\farcs25)
is marginally resolved at 1.3 mm, having a deconvolved
position angle consistent with the orientation of the multiple system.
NGC 1333 IRAS2A (Per-emb-27) has a companion separated by 0\farcs62 and there
is a marginal (5$\sigma$) detection at 1.3 mm at the companion location. Lastly, the
more widely separated companion toward NGC 1333 IRAS4A (1\farcs8) 
is detected at 2.7 mm, 1.3 mm, and at $\sim$850 \micron\ 
\citep{lay1995}.

While these are only a few examples, as a whole we can suggest that multiplicity
at millimeter wavelengths is often also reflected at 8~mm and 1~cm, but not in all cases.
The companions toward L1448 IRS2 and NGC 1333 IRAS2A (Per-emb-22-B and Per-emb-27-B) 
were undetected or marginally detected at 1.3~mm, 
indicating that there is less than $\sim$0.001 $M_{\sun}$, of compact mass traced by dust emission
surrounding both of these companions (assuming \citet{ossenkopf1994} 
dust opacities, a dust to gas mass ratio of 1:100, and a dust
temperature of 20 K). The estimated flux densities from dust 
emission at 8~mm from these two sources would be 22 \microjy\ and 
8 \microjy\ for Per-emb-22-B and Per-emb-27-B, respectively. 
This would be below our sensitivity, thus the detected emission
at 8 mm must have a significant contribution from free-free emission. 
At longer wavelengths,  
IRAS2A VLA2 (Per-emb-27-B) is detected at 4~cm, but not 
6.4~cm \citep{tobin2015}.

The detection of free-free emission is strong evidence for the presence of
a protostellar source driving at jet. Therefore, the most likely explanation
for their lack of strong millimeter emission is a lack of concentrated circumstellar 
dust or compact, optically thick circumstellar dust. This may imply that any circumstellar
disk around these companions is quite low mass.
Without significant circumstellar dust, it is unlikely that ALMA will be able to detect
such companion sources out of the extended circumstellar dust surrounding
the primary protostars. Therefore, to gain a complete picture of multiplicity,
observations at longer wavelengths where other emission mechanisms contribute to the 
emergent flux are crucial to fully characterizing protostellar multiplicity.

\subsubsection{Potential Bias in Millimeter/Centimeter Characterizations of Multiplicity}

The methods of characterizing multiplicity toward Class 0 protostars are fundamentally different
from those typically employed to characterize multiplicity of Class I and more-evolved systems. Direct-detection
of stellar or inner-disk radiation can be employed in most cases to characterize the multiplicity 
toward Class I and more evolved systems. Moreover,
adaptive optics and other high-contrast imaging techniques can be used to identify close multiples, in addition to 
radial velocity monitoring. These techniques cannot be employed for Class 0 and early Class I protostellar
systems due to the high extinction through the protostellar envelopes and the large amount of scattered light emission
typically associated with young stars.

Characterizing multiplicity in protostellar systems relies on the 
indirect methods of detecting circumstellar dust emission and/or free-free 
emission, with the implicit assumption that multiple sources in the emission maps reflect
discrete protostellar objects. Dust emission is expected to reflect dense
concentrations of dust in the form of a circumstellar disk or 
an inner envelope near the individual protostars \citep[e.g.,][]{rodriguez1998, terebey1997,looney2000,chen2013}.
However, a clump of material in a protostellar disk or envelope does not
necessarily harbor a protostellar source and it is unknown whether such clumps will form a protostellar source.
For free-free emission, it is necessary to assume that the
emission is originating from shocks near the base of the protostellar jet on $<$ 10 AU scales \citep{curiel1989,anglada1998}.

Multiplicity characterized by dust emission assumes that the protostellar sources either have increasing 
density toward them or compact, emission from a circumstellar disk. The fundamental problem with using dust emission
to characterize multiplicity lies in the assumption that peaks in the millimeter dust emission harbor protostellar
sources. 
We fail to detect some previously reported multiplicity in some Class 0 systems (see Appendix C.13),
but our results are consistent with the millimeter detections in most cases, 
especially for the wider multiples detected by \citet{looney2000} and
\citet{chen2013}. We emphasize that caution will be needed when interpreting continuum data in the ALMA era given
that sensitivity is typically $>$10$\times$ better than was possible with PdBI/NOEMA, CARMA, and the SMA. ALMA data are already
enabling many more low-surface brightness features to be detected, which may or may not harbor or go on to form protostars
\citep[e.g., L1521F; ][]{tokuda2014}. 

The dust emission detected in our VLA survey is largely immune from the detection 
of low-surface brightness, extended sources due to our observations
at high resolution, requiring that dust emission be quite concentrated to be detected, let alone not be resolved-out. However,
note the case of Per-emb-18 where we detected an extended dust structure that does not appear to harbor a companion, but its appearance is
quite distinct from that of point-like emission. Per-emb-2 also has a very extended and apparently clumpy structure and most of this extended
emission is resolved-out at higher resolution. Moreover, we are typically detecting a 
combination of dust and free-free emission at Ka-band. This is demonstrated by the spectral index being less
than 2 (Figure \ref{cumulat-spindex}). The addition of free-free emission is advantageous because
it enhances the detectability of the protostars.

We can conclude that there are not likely many false detections in our sample from 
free-free emission associated with outflow shocks. The 
strong outflows toward SVS13, IRAS2A, L1448C, IRAS4A, and HH211 do not 
yield detections of Ka-band emission associated with shock knots in the
outflows. Furthermore, the close companion to SVS13 (SVS13A2) does not exhibit 
significant proper motion away from SVS13A \citep{carrasco2008b}, nor is it associated
with a shock feature in high-resolution near-infrared imaging \citep{hodapp2014}. 
We have also examined the distribution
of spectral indices in the Ka-band of the close multiples ($<$ 500 AU separations) 
relative to the single sources, and the histograms
and cumulative distributions are quite comparable. We performed an AD test on 
the samples to see how likely the are to be drawn from the same distribution.
The resulting probabilities are 0.7 and 0.35 for the integrated and peak spectral indices, respectively, indicating
that the spectral indices for the singles and multiples are likely drawn from the same distribution. Thus, we argue that
the Ka-band data are unlikely to have significant numbers of false companions.

The characterization of multiplicity from both dust emission and free-free emission has potential
pitfalls. Both methods can lead to detections of sources that are not truly protostellar and both can also 
yield non-detections toward genuine protostellar sources. Therefore, the biases associated 
with the characterization of multiplicity are difficult to quantify
and correct for.

\subsubsection{Completeness Limits}

The discussion in Section 5.6 illustrates the difficulties in quantifying the incompleteness of
our multiplicity detections given that dust and free-free emission process are not directly
connected to physical properties of the source, i.e., protostellar mass. The best, albeit poor, proxy
for mass is L$_{bol}$, and this value can only be determined for the system as a whole, not individual
components of multiple systems separated by $<$~1000~AU. Figure \ref{sample-flux8} shows the histograms of L$_{bol}$ and plots of 8~mm flux
density versus L$_{bol}$. These figures show that the few non-detections in our Class 0 and Class I protostellar
samples are typically the lower luminosity sources. In the case of the Class 0 sources, many of the non-detections are
candidate FHSCs. However, we cannot simply assign a minimum luminosity that we can detect
because some of the lowest luminosity sources are well-detected. Furthermore, the single sources similarly show 
a broad scatter in terms of 8 mm flux density at a given luminosity. This makes us unable to assign a lower-limit to the
luminosity of companions that we are able to detect from single sources, nor can we provide a lower mass limit.

Therefore, it is possible that there may be non-detections of some 
companions due to a lack of concentrated dust emission and/or
free-free emission toward some companions. Given that we are examining
multiplicity from indirect methods that do not directly correlate with source
properties, the level of incompleteness cannot be quantified with 
any degree of accuracy, nor can we give sensitivity limits as a 
function of mass ratio and separation. Thus, the multiplicity 
statistics derived from millimeter/centimeter studies 
should be further considered as lower limits. 
Moreover, we cannot examine multiplicity at scale 
smaller than 15 AU, making our MF and CSFs lower limits.

While we cannot reliably quantify our level of incompleteness, we do 
detect all the currently known infrared companions \citep[i.e., EDJ2009-183 and L1448 IRS1][]{connelley2008}
and most millimeter companions (except for VeLLOs/candidate FHSCs). Furthermore, 
our observed CSF of Class I protostars is consistent with the near-infrared studies. Thus,
incompleteness may not be a serious issue given the agreement of our results with 
those obtained from independent techniques.

\section{Conclusions}
We have conducted a multi-wavelength VLA survey (8~mm, 1~cm, 4~cm, and 6.4~cm) of all known protostellar systems in
the Perseus molecular cloud and presented our results on the
multiplicity of the protostellar systems based on our 8 mm and 1 cm data.
Our survey observed an unprecedented number of systems
with uniform sensitivity and resolution in a single star forming region. The high-resolution
data taken in A and B configurations have enabled us to carry out a 
relatively unbiased characterization of
protostellar multiplicity down to 15 AU scales for all protostars in the Perseus molecular cloud. 
We note, however,
that the MF, CSF, and companion frequencies with separation 
given in this paper are most likely lower limits. 
This is because of
the inherent bias associated with detecting multiplicity from the presence of dust 
or free-free emission toward companions and
there are likely unresolved systems at separations smaller than our resolution limit.

1. We detect 18 multiple systems with 
separations between 15 AU and 500 AU, 
of which 16 are new detections by the VANDAM survey. This increases 
the number of known Class 0 systems with companion
separations between 15 AU and 500 AU 
by more than a factor of two.

2. The distribution of protostellar companion separations in Perseus appears
bi-modal or double-peaked, with one peak at $\sim$75 AU and the second peak
at $\sim$3000 AU. We argue that the double-peaked distribution is suggestive of two formation mechanisms for the 
wide and close multiple systems, disk fragmentation for scales $\la$ 300 AU and core/turbulent fragmentation 
for scales larger than 1000 AU.

3. The MF and CSF for separations from 15 AU to 5000 AU (and 10000 AU)  are larger for Class 0 sources than Class I 
sources and field solar-type stars, confirming the results \citet{chen2013}. However, for separations
 $\le$ 2000~AU, the MF and CSF of Class 0 sources are consistent with Class I sources 
and field solar-type stars.

4. The distribution of separations for the Class 0 and Class I sources appear different.
There is a clear deficit of wide companions toward Class I sources relative to
Class 0 sources, which we interpret as evidence for evolution of companion 
separations between the Class 0 and Class I phases.
Systems could either form or become unbound as the star forming gas is dispersed. Alternatively,
wide companions could migrate inward.
However, the MF and CSF for Class 0 and I protostars are consistent with each other
on scales between 15 AU and 2000 AU. 
Therefore, a significant fraction of multiples may not be 
migrating inward from $>$1000 AU separations. 

5. We detect companions embedded within extended dust continuum structures toward 3 systems 
(Per-emb-2/IRAS 03292+3039, Per-emb-5/IRAS 03282+3035, and Per-emb-18); the companions
are separated by $\le$30 AU. We interpret this structure as evidence 
on-going disk fragmentation in these systems, given that the
companion sources are found within a larger surrounding structure. 
Though the surrounding structures are not confirmed to be rotationally-supported, this is the first observed
evidence for such small-scale substructure toward young protostars.

6. WE compared our distribution of separations to the Taurus pre-main sequence samples from \citet{kraus2011}.
The separation distribution for the Perseus sample is more consistent with the high-mass Taurus sub-sample 
(2.5 $M_{\sun}$ $\ge$ $M_*$ $>$ 0.7 $M_{\sun}$) than their low-mass 
sub-sample ($M_*$ $<$ 0.7 $M_{\sun}$). The primary difference is the number of wide companions.
If the wide systems in Perseus are not bound, then the separation distribution may evolve to be more consistent
with the Taurus low-mass sub-sample. 
We caution, however, that the comparisons of Perseus
and Taurus may not be valid given the differences in the star formation conditions.

7. While millimeter observations of the newly discovered multiple systems are not complete,
both L1448 IRS2 and NGC 1333 IRAS2A have existing 1.3 mm observations with sufficient resolution
to resolve the VLA-detected companions. However, the companions are not convincingly detected
at 1.3 mm. This result carries the implication that ALMA may not be able to completely 
characterize protostellar multiplicity because not all protostellar companions
will be detectable. Moreover, optically thick dust emission on $<$100 AU scales 
may inhibit the detection of embedded companions separated by $<$50 AU at 
millimeter/submillimeter wavelengths.

8. We demonstrate that close companions are likely to be real sources and not knots of free-free emission in the outflows. 
First there is a lack of correlation between 
companion separation and relative position angle between the outflow and companions, and secondly the distribution of spectral indices
for companion sources is consistent with having the same distribution as single protostellar sources.
While we cannot absolutely rule-out the possibility that some companions are features of the protostellar jet interacting
with the surrounding cloud, proper motion studies can be carried out on timescales of a few years to determine
whether or not the sources move in the jet direction or if the companion sources are co-moving. Orbital motion will likely require longer
time baselines of order 10-20 years to determine.

We thank the anonymous referee for constructive suggestions that improved the
quality of the manuscript. The authors wish to thank K. I. Lee, A. Stutz, B. Reipurth, J. Jorgensen, A. Kraus, and L. Tychoniec
for useful discussions regarding this work. J.J.T. is 
currently supported by grant 639.041.439 from the Netherlands
Organisation for Scientific Research (NWO).
J.J.T. acknowledges past support provided by NASA through Hubble Fellowship 
grant \#HST-HF-51300.01-A awarded by the Space Telescope Science Institute, which is 
operated by the Association of Universities for Research in Astronomy, 
Inc., for NASA, under contract NAS 5-26555. 
L.W.L. acknowledges support from the Laboratory for Astronomical 
Imaging at the University of Illinois and the NSF under grant AST-07-09206.
Z.Y.L. is supported in part by NSF1313083 and NASA NNX14AB38G.
C.M. acknowledges financial support from the U.S. National Science Foundation through award AST-1313428.
M.M.D. acknowledges support from the Submillimeter Array through an SMA postdoctoral fellowship.
This research made use of Astropy, a community-developed core Python package 
for Astronomy (Astropy Collaboration, 2013, http://www.astropy.org).
This research made use of APLpy, an open-source plotting package for 
Python hosted at http://aplpy.github.com. This research has made use 
of NASA's Astrophysics Data System. The National Radio Astronomy 
Observatory is a facility of the National Science Foundation 
operated under cooperative agreement by Associated Universities, Inc.

{\it Facilities:}  \facility{VLA}

\appendix

\section{Notes on Multiple Sources with $<$ 30 AU Separations}

\subsection{Per-emb-2/IRAS 03292+3039}
Figure \ref{IRAS03292} shows the source Per-emb-2 (IRAS 03292+3039) at 9 mm (top) and the 8 mm - 1 cm spectral
index maps (bottom), generated using the \textit{nterms=2} option in the CASA \textit{clean} task. The A+B configuration
images, tapered at 1000 k$\lambda$ are shown in the left panel, emphasizing the structured 
extended emission; the extent of the resolved structure agrees with the 1.3 mm observation
presented by \citet{tobin2015b}. When zooming in on the inner region in the middle panel 
with the A-configuration-only image, the extended structure is resolved-out. 
The main source appears extended and another source clearly detected with a separation of 18.5 AU
when imaged with Briggs weighting, shown in the right panel of Figure \ref{IRAS03292}. 
The spectral index is $\sim$2 at the peak intensity, 
but the extensions north and east have steeper spectral slopes indicative of dust emission. In the higher resolution
images, the eastern source has a spectral index of $\sim$3, while the western source has a spectral index of $\sim$1.5.
The shallower spectral index is an indication that both dust and free-free emission are contributing to the
source flux.

\subsection{Per-emb-5/IRAS 03282+3035}
Per-emb-5 (IRAS 03282+3035) is shown in Figure \ref{IRAS03282}; it is resolved
at lower resolutions in the left panels of Figure \ref{IRAS03282},
but the emission is not as extended as Per-emb-2. The middle panels of
Figure \ref{IRAS03282} begin to show double-peaked structure with a separation
of 22.3 AU. The highest resolution data in the upper right panel also
shows that the eastern peak is resolved. The eastern source also has a steeper spectral index ($\sim$3),
indicative of dust emission being the dominant emission component. The spectral index of the western
source is shallower ($\sim$1), suggesting that it has a larger contribution from free-free emission than the eastern source. 

\subsection{Per-emb-18}
Per-emb-18 (NGC 1333 IRAS 7), shown in Figure \ref{Per18}, 
shows an apparent secondary source in the low-resolution 
image in the left panels of Figure \ref{Per18} with a separation of $\sim$84 AU. When viewed 
at higher resolution in the middle panels, the apparent secondary source appears
resolved-out and is most likely an extended dust structure; thus, we do not consider it as a companion source.
However, the main source then shows evidence of 
resolved structure.  The upper right panel of Figure \ref{Per18} appears 
double peaked with a separation of 19.6 AU. Both sources have shallow
spectral indices indicating that free-free emission is dominating at the source location
and there is evidence for a steeper spectral index between the sources, suggesting dust emission between the two sources.

\section{Possible Close Multiples}

In addition to the clear multiples that are presented in the main text, there are several sources for which
resolved structure is apparent, but the significance of the detections are below the 5$\sigma$ criteria or only
detected in one band. We show these sources in Figure \ref{poss-multiples}. These sources are only revealed at 
the highest resolutions or only with the increased sensitivity of the combined A and B configuration data. These may be
resolved disk structures or possibly bonafide companions and their nature may be further revealed by higher sensitivity imaging.

\section{Notes on Multiple Sources with $>$ 500 AU Separations and Specific Multiple Systems}

\subsection{IC348 MMS}
IC348 MMS/Per-emb-11 was previously identified as a multiple system with 15\arcsec\ separation by \citet{chen2013};
the VLA 9 mm image is shown in Figure \ref{IC348MMS-wide}. 
\citet{rodriguez2014} also identified another source separated by $\sim$3\arcsec\ southwest 
at 2.1 cm and 3.3 cm (JVLA3a); the brightest source in Figure \ref{IC348MMS-wide} is associated with Per-emb-11-A
at the center of the image and denoted JVLA3b by \citet{rodriguez2014}. 
JVLA3a/Per-emb-11-B appears coincident with the MIPS 24 \micron\ source, and it is directly 
between outflow cavities evident at shorter wavelengths \citep{pech2012}. JVLA3b/Per-emb-11-A seems to be
located to the side of the outflow cavity and \citet{pech2012} 
showed CO outflow possibly misaligned with knots observed at IRAC wavelengths. Finally,
\citet{tobin2015b} presented 1.3 mm imaging toward IC348 MMS with 0\farcs3 resolution and 1 mJy sensitivity;
the 1.3 mm emission toward JVLA3b is well-detected, but JVLA3a is not. 
A reexamination of the image finds that
there is indeed a 3$\sigma$ peak at the position of JVLA3a at 1.3~mm. 
The spectral index of JVLA3a/Per-emb-11-B
is positive in Ka-band as well as from 3.3~cm to 2.1~cm, but overall shallow. Thus, JVLA3a may be protostellar in
nature but with faint continuum emission at 1.3 mm. Finally, JVLA 3c from \citet{rodriguez2014} 
also coincides with the wide companion
Per-emb-11-C, a possible proto-brown dwarf suggested by \citet{palau2014}.

\subsection{NGC 1333 IRAS4B}
The wide multiple system of IRAS 4B and IRAS 4B$^{\prime}$ is well detected 
by the VLA at 9mm (Figure \ref{IRAS4B-wide}), with a separation of 2450 AU. 
Both sources have resolved structure at this wavelength. Compact outflows originating from the two
sources were identified by \citet{hull2014}, and they are in nearly orthogonal directions.

\subsection{Per-emb-16 and Per-emb-28}
Per-emb-16 and Per-emb-28 comprise a wide multiple system in the IC348 region. Per-emb-16 is a Class 0 object and Per-emb-28 is classified
as a Class 0/I source. Per-emb-28 is notable because it shows a high degree of periodic variability, possibly
from pulsed accretion \citep{muzerolle2013}. Both sources are rather faint at 9 mm, see Figure \ref{Per16-Per28-wide}.

\subsection{NGC 1333 IRAS7}
Within the system collectively known as NGC 1333 IRAS7, Per-emb-18, Per-emb-21, 
and Per-emb-49 are found to comprise a quintuple system, see Figure \ref{IRAS7-wide}.
Both Per-emb-18 and Per-emb-49 have companions separated by less than 100 AU and Per-emb-21 is single.
Per-emb-49 appears to be a Class I source, while Per-emb-18 and Per-emb-21 are Class 0 objects. Per-emb-18
appears to drive a long system of HH objects, while Per-emb-21 has a relatively compact outflow \citep{davis2008}.

\subsection{L1448C}
We detected both components of the L1448C system, Per-emb-26 (L1448C-N) 
and Per-emb-42 (L1448C-S), see Figure \ref{L1448C-wide}. These
sources were previously resolved by \textit{Spitzer} observations \citep{jorgensen2006,tobin2007}
and at submillimeter and millimeter wavelengths \citep{hirano2010,maury2010}. 
Per-emb-42/L1448C-S is located in the direction of the outflow
from Per-emb-26/L1448C-N; however, an independent outflow is found to 
originate from Per-emb-42/L1448C-S \citep{hirano2010}. Per-emb-42 may be more evolved than 
L1448C, given that it is classified as a Class I by \citet{enoch2009}.

\subsection{L1448N/IRS3}
The L1448 IRS3 (also known as L1448-N) system comprises Per-emb-33 (L1448 IRS3B), L1448 IRS3A, and L1448NW, the
three components are within a radius of 5000 AU, see Figure \ref{L1448N-wide}. 
A companion in the Class 0 system L1448 IRS3B (Per-emb-33) had been reported 
by \citet{tobin2015b}, but the VLA observations have resolved the secondary into two distinct sources making
Per-emb-33 a triple.
L1448NW is also found to be a binary in our study, the previous CARMA observations of \citet{tobin2015b} 
had noted that this source was extended. Thus, the system as a whole is a sextuple.
L1448NW is the most widely
separated system from the rest, 4946 AU from Per-emb-33 and 3749 AU from L1448 IRS3A. Per-emb-33 and L1448NW 
are both Class 0 systems and L1448 IRS3A is likely Class I.
Then Per-emb-26 (L1448C) is $\sim$18500 AU south of Per-emb-33 and Per-emb-22 is $\sim$41000 AU (0.2 pc) to the west.
The kinematics and outflows of this system are analyzed in detail by \citet{lee2015}.

\subsection{SVS13}
The SVS13 (HH 7-11 region)is comprised of
three main sources: SVS13A, B and C, as denoted by \citet{looney2000}, see Figure \ref{SVS13-wide}.
Per-emb-44 (SVS13A) is the driving source of HH 7-11 and the companion SVS13B was first
tentatively identified by \citet{grossman1987}; subsequent maps of the region by \citet{chini1997} 
confirmed the detection of SVS13B and identified the source that is now known as SVS13C as MMS3. \citet{looney2000}
confirmed all of these detections with interferometry at 2.7 mm. SVS13B is $\sim$3400 AU 
from SVS13A and SVS13C is $\sim$ 4500 AU from SVS13B. 

Per-emb-44 (SVS13A) itself is a close binary \citep[first discovered by ][]{rodriguez1999}
and it has another companion 1222 AU away which we refer to as SVS13A2 \citep[also known as VLA3 ][]{anglada2004}.
\citet{carrasco2008} examined the proper motions of sources in this region and found that SVS13A2 is indeed co-moving
with the other sources and not likely to be an outflow ejection.
Per-emb-44 is classified as Class I, but its outflow power
is comparable to Class 0 sources \citep{plunkett2013}. SVS13B and SVS13C 
also appear to be Class 0 sources \citep{sadavoy2014}, but their nature
is more uncertain given their close proximity Per-emb-44, a 
bright source from the near to far-infrared, making photometry difficult.
Finally, another source is apparent northeast of Per-emb-44 in 
Figure \ref{SVS13-wide}, denoted VLA20 by \citet{rodriguez1999}. This
source had been classified as a YSO by those authors given its 
rising spectral index, but no counterpart is detected in the
infrared \citep[e.g.,][]{jorgensen2006} or the millimeter \citep{looney2000}.
 We find that this source has a flat spectral index in Ka-band; this and 
the lack of counterparts at millimeter and infrared wavelengths indicate
that this source is most likely extragalactic in nature and we do not 
consider it in the multiplicity statistics.

\subsection{NGC 1333 IRAS2B}

NGC 1333 IRAS2B has an apparent companion separated by 3\farcs8 ($\sim$870 AU), in addition to the 72.5 AU companion
source, see Figure \ref{IRAS2B-wide}. However, the more widely separated source does not appear to be physically associated
with the IRAS2B system.
Its position is coincident with an optically visible star (BD +30 547)
that appears to be illuminating the near side of the dark cloud and
not embedded like IRAS 2B, suggesting that this is a line-of-sight
alignment \citep{rodriguez1999}. Therefore, we do not consider the more 
widely separated source in the multiplicity statistics. However, BD +30 547 could be a Class III
source given that it appears to be located physically close to the molecular cloud. BD +30 547 also has
a negative spectral index, possibly indicative of gyrosynchrotron emission from an active corona \citep{dulk1985}.

\subsection{ Per-emb-8 and Per-emb-55}
Per-emb-8 and Per-emb-55 form another Class 0 - Class I wide binary system separated by 
$\sim$2200 AU. 
Per-emb-55 itself is a close binary,separated by 142 AU,
see Figure \ref{Per8-wide}. Per-emb-55 is
quite bright in the IRAC bands, while Per-emb-8 is faint and shows some diffuse emission.

\subsection{Barnard 1-b Region}
The B1-b region comprises the three sources B1-bN, B1-bS, and 
Per-emb-41, see Figure \ref{B1-wide}. Both B1-bN and B1-bS have been suggested
to be candidate first hydrostatic core objects and are faint even at 70 \micron; B1-bS is also
quite faint at 100 \micron\ \citep{pezzuto2012}. Per-emb-41 appears to be more 
evolved and is classified as Class I, having bright emission at IRAC wavelengths.

\subsection{Per-emb-37}
Per-emb-37 was mentioned in Section 2 because it is a Class 0 source that had been incorrectly associated with a brighter
IRAC source and subsequently classified as Class I. \citet{sadavoy2014} also detected this source in their search for
Class 0 sources in Perseus. We do not detect all the infrared-associated sources in the 8 mm image, see Figure \ref{Per37-wide}, 
but two sources are detected at 8~mm with separations of 2428 AU and 7752 AU. 
Thus, Per-emb-37 is the only new wide multiple
reported in this study.

\subsection{Per-emb-35}
Per-emb-35, also known as NGC 1333 IRAS1, is located on the western outskirts of NGC 1333. It is particularly
interesting that this source is found to be a $\sim$440~AU binary (Figure 1), 
given that it is also found to have a apparent S-shaped outflow
in the \textit{Spitzer} image of the region \citep{gutermuth2008}. 
Thus, the companion could pay a role in shaping the outflow morphology.

\subsection{Non-Detections Toward Previously Reported Multiples}
There are a few cases where multiplicity has been previously reported toward sources
in Perseus, but we do not confirm the presence of companion sources. One notable case is NGC 1333 IRAS2A (Per-emb-27) where
a candidate companion was reported by \citet{codella2014} and \citet{maury2014}, along
with a non-protostellar continuum source. As discussed extensively in \citet{tobin2015},
we did not detect these sources in the VLA data nor at 1.3 mm and 850 \micron, despite having 
sufficient sensitivity at all wavelengths given their measurement of the spectral index. We did, however,
discover a new companion with our VLA data separated by 142 AU. Thus, it is likely that the
new VLA-detected companion is the driving source of the second outflow from this system
\citep{sandell1994} and that the other reported sources are spurious; see
\citet{tobin2015} for more details.

HH211-mms (Per-emb-1) was reported to have a candidate companion separated by 0\farcs3 ($\sim$69~AU)
in \citet{lee2009} and \citet{chen2013} with 870 \micron\ data. We failed to detect a companion toward this 
source in our VLA data. Moreover, the companion is not detected in CARMA 3 mm A-array
data with 0\farcs3 resolution (H.-F. Chiang, private communication). The lack of detection at multiple wavelengths
casts doubt on the robustness of the claimed companion. It was only detected in an
image generated with super-uniform weighting and could be spurious.
 The flux density of the companion
at 870~\micron\ is 25~mJy; assuming optically thin emission and a dust opacity spectral index of 3, the
estimated peak flux density at 8.1~mm is 31 \microjy\ at about our 3$\sigma$ sensitivity limit. Therefore, 
the source detected by \citet{lee2009} could be a dust clump that is most apparent at 870 \micron.

Finally, \citet{chen2013} reported a companion toward IRAS 03282+3035 (Per-emb-5) separated by 1\farcs5 (345 AU) 
at 870 \micron. We did not detect a companion at this
location in our VLA data, nor did \citet{tobin2015b} detect this source at 1.3 mm. Thus, the lack of
detection at multiple wavelengths suggests that this companion is likely spurious.

\begin{small}
\bibliographystyle{apj}
\bibliography{ms}
\end{small}

\clearpage
\begin{figure}[!ht]
\figurenum{1b}
\begin{center}
\includegraphics[scale=0.33]{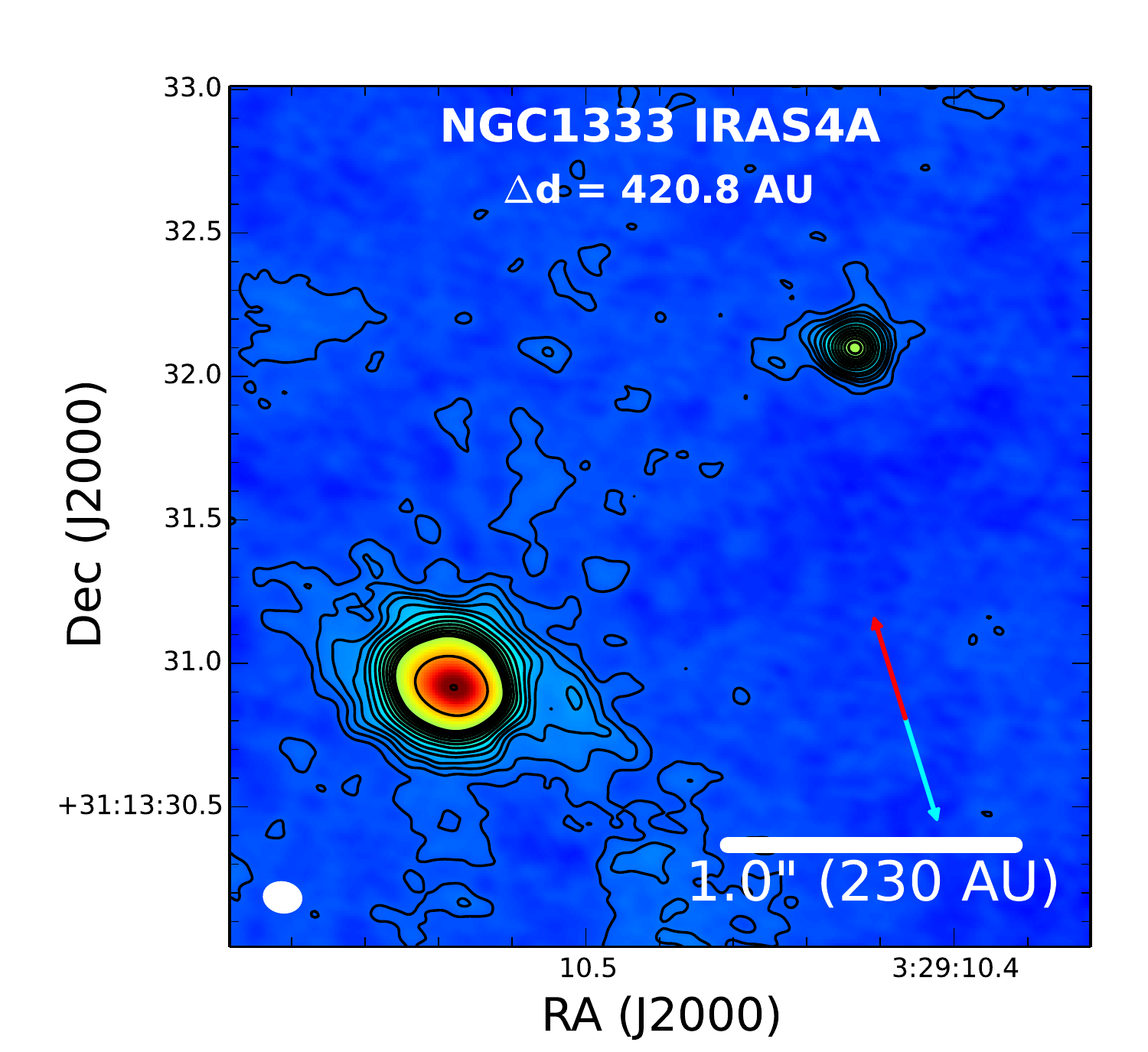}
\includegraphics[scale=0.33]{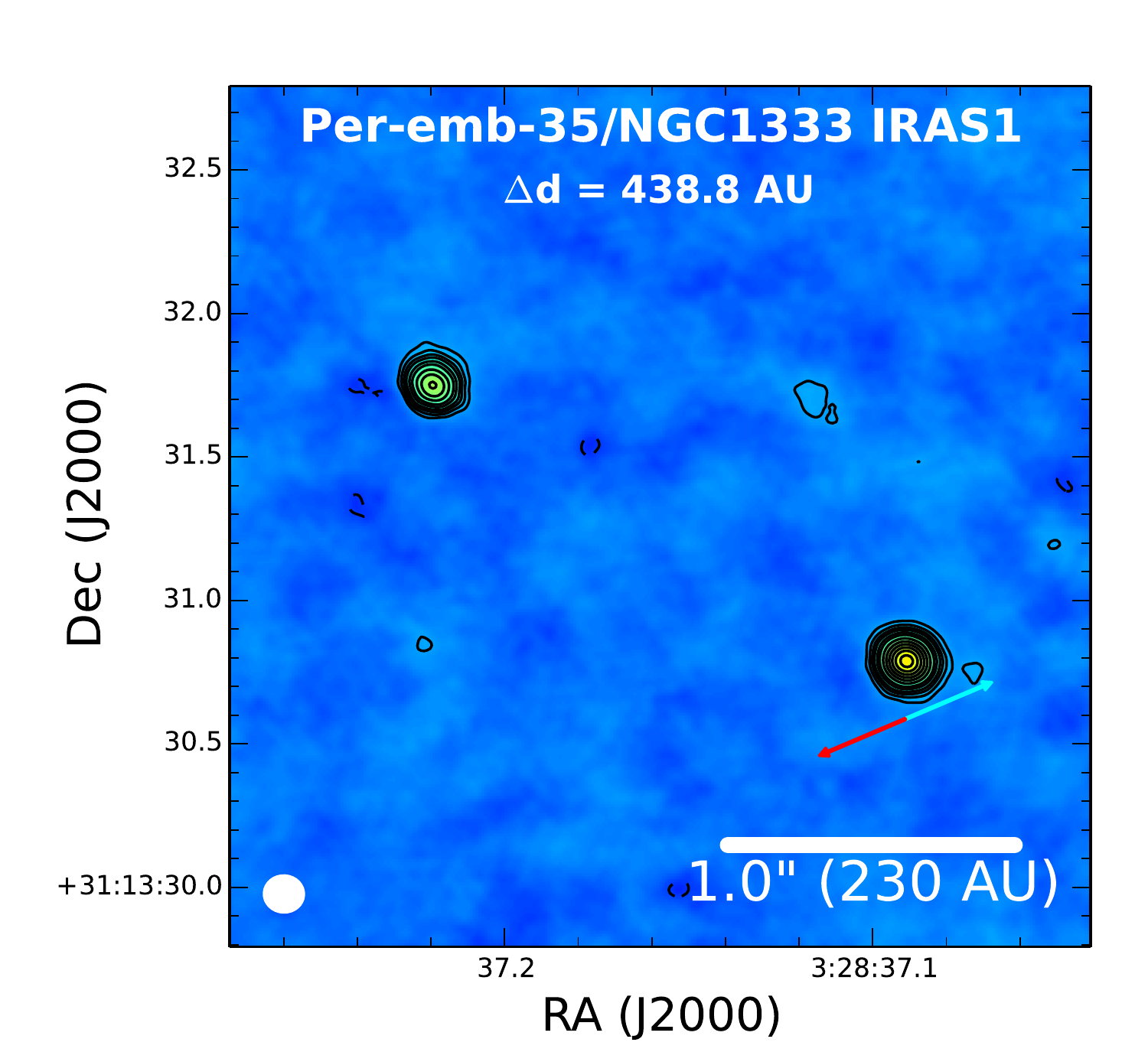}
\includegraphics[scale=0.33]{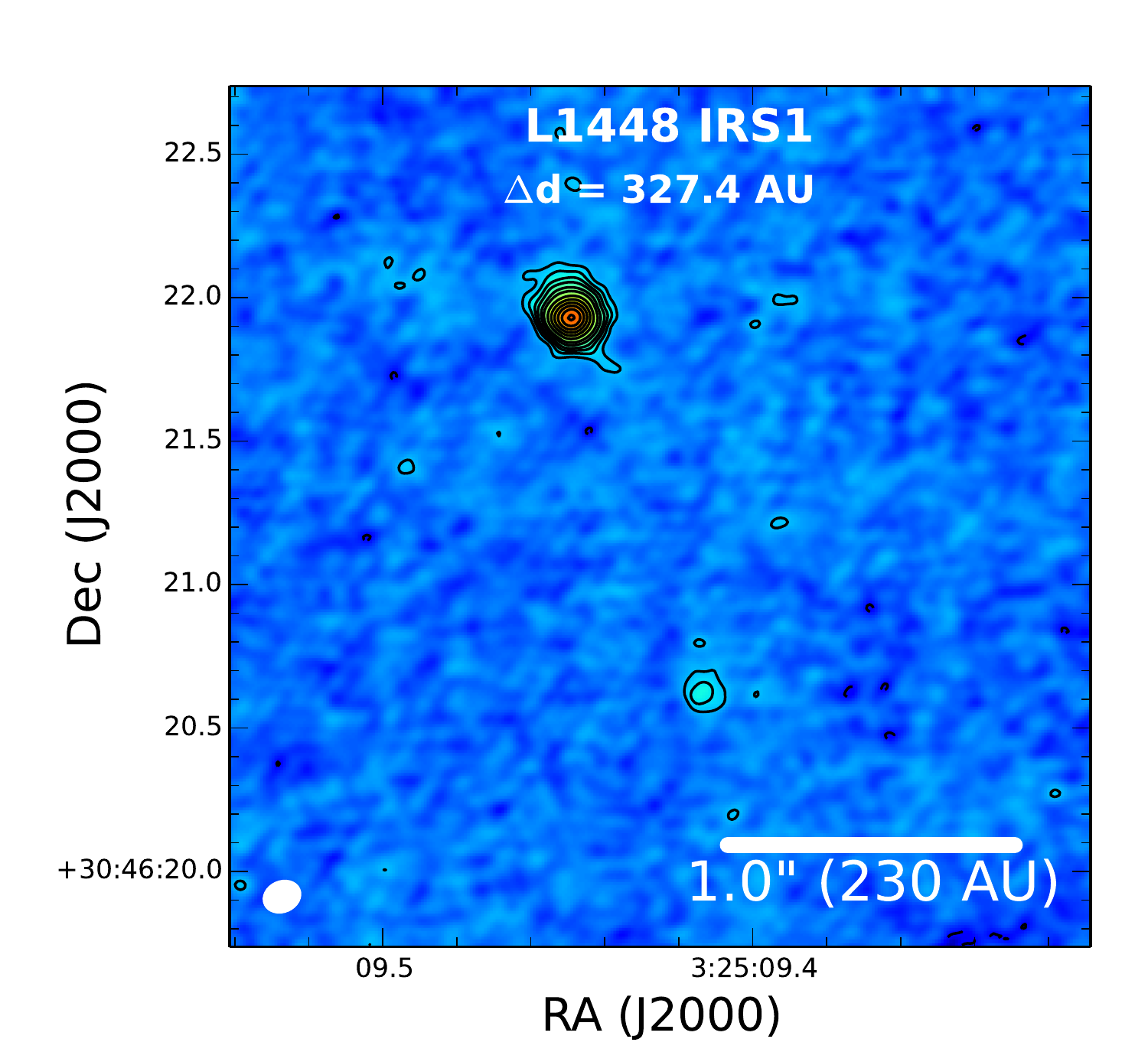}
\includegraphics[scale=0.33]{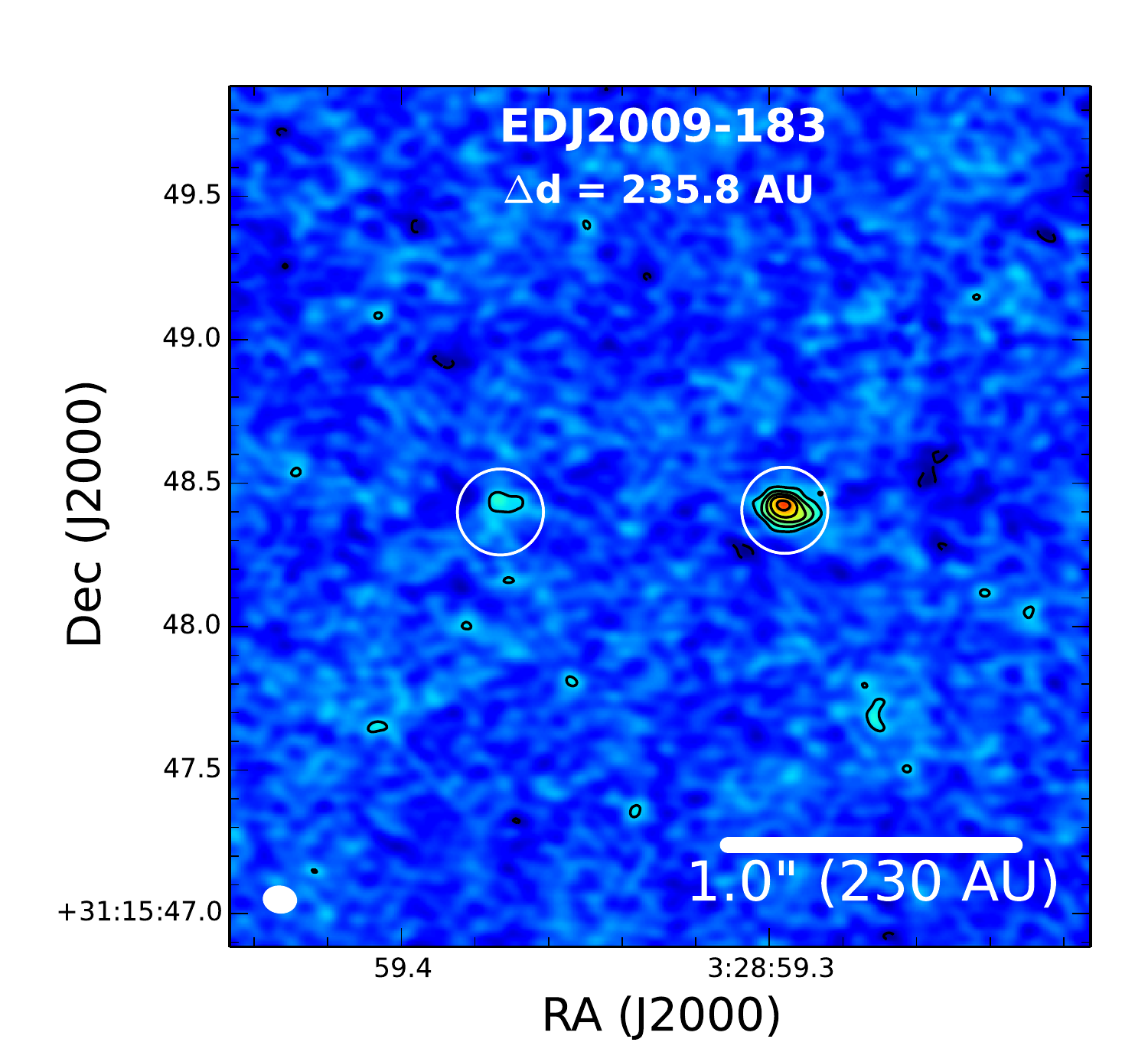}
\includegraphics[scale=0.33]{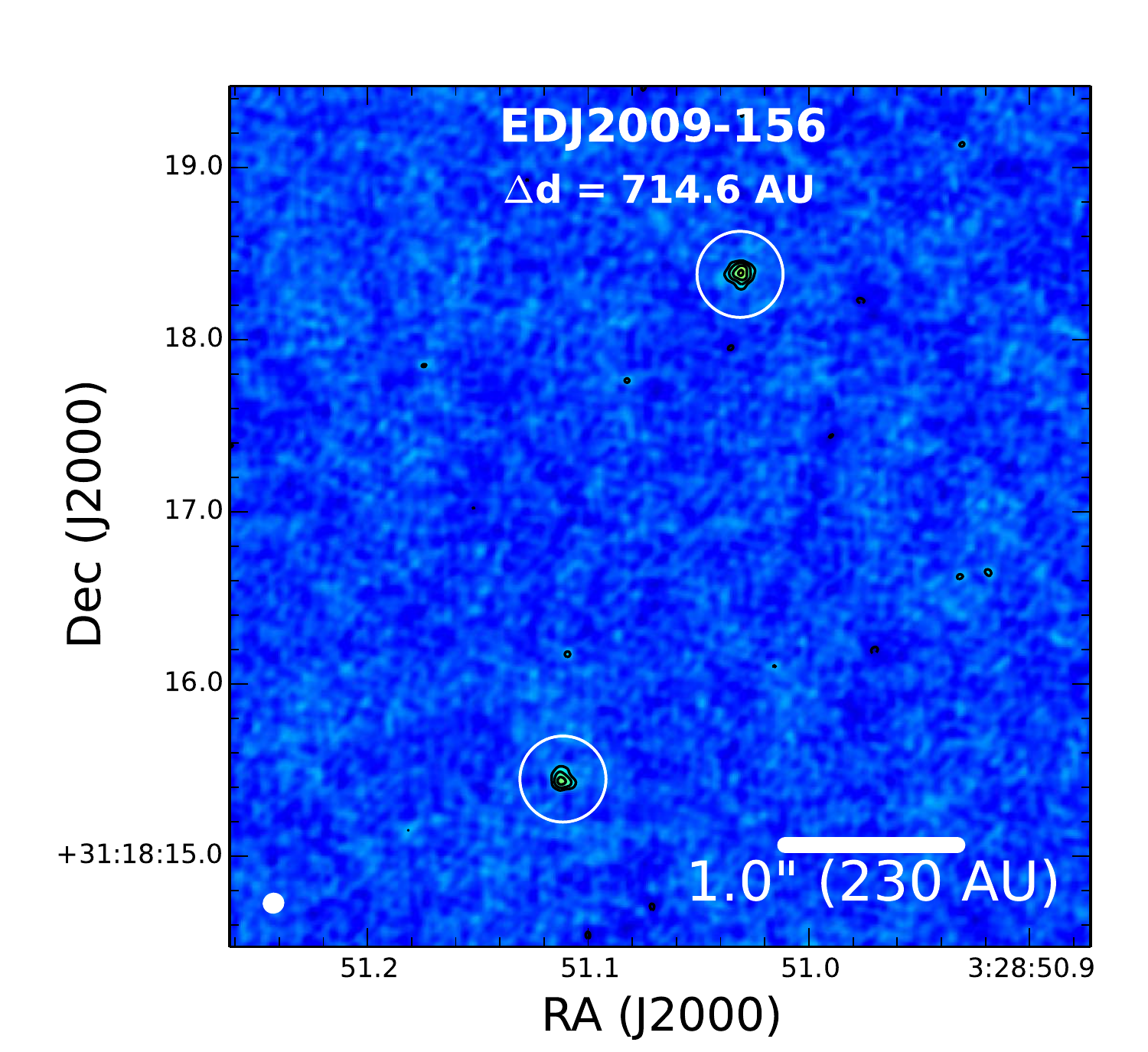}
\end{center}
\caption{}
\end{figure}

\begin{figure}[!ht]
\begin{center}
\includegraphics[scale=0.25]{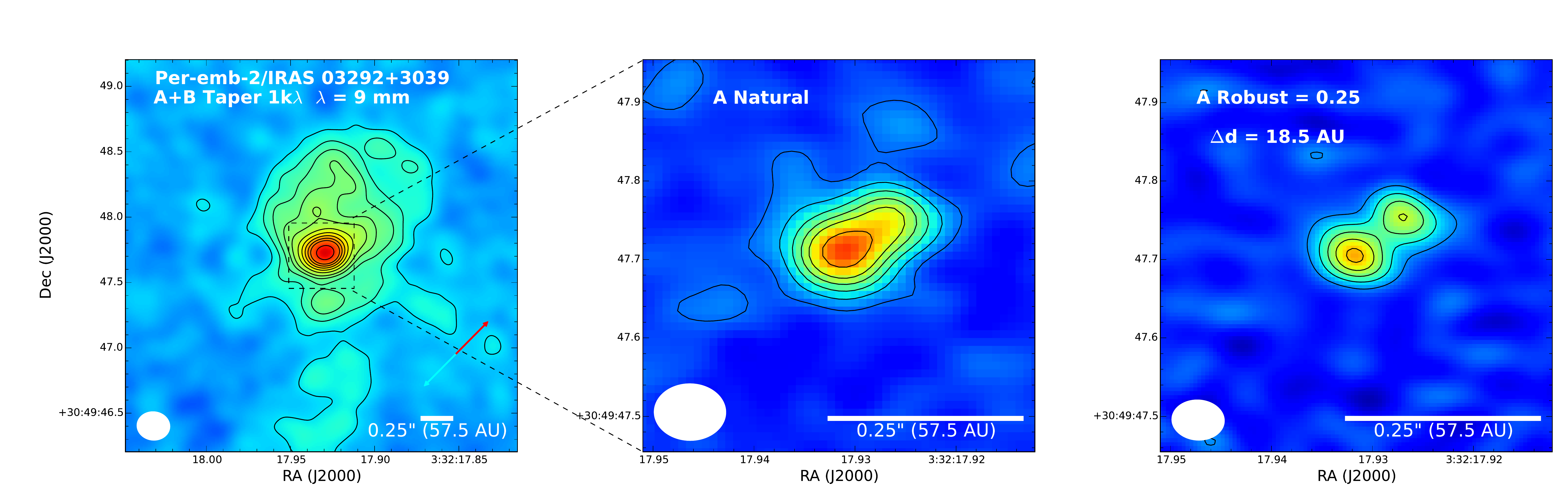}
\includegraphics[scale=0.25]{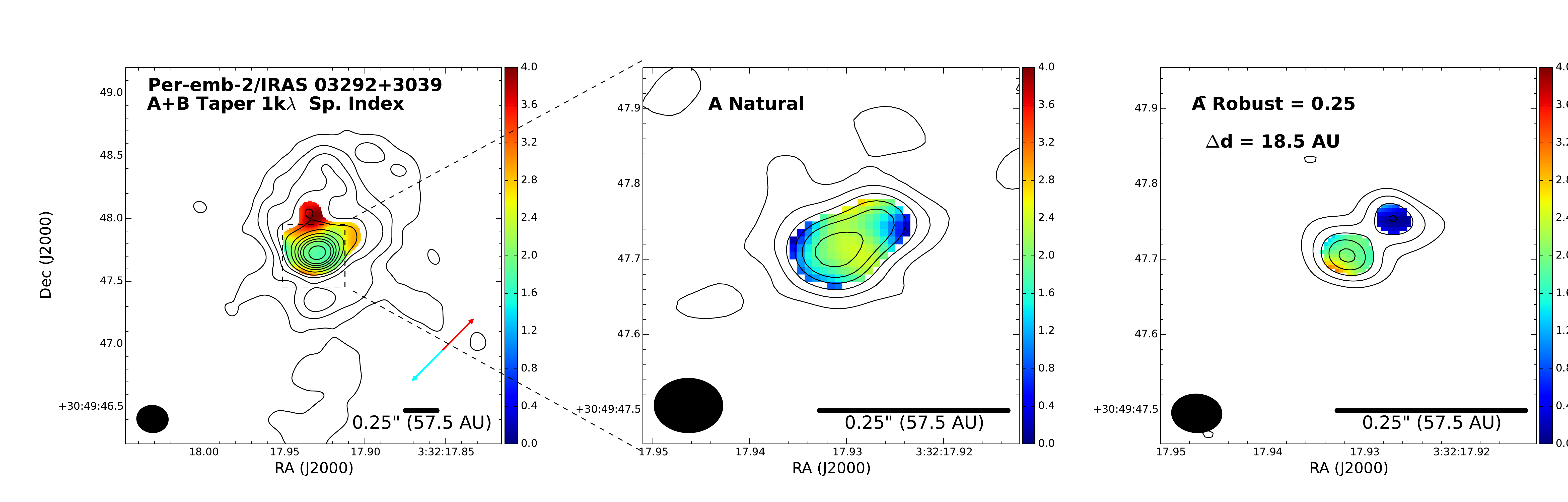}
\end{center}
\caption{Images of Per-emb-2 (IRAS 03292+3039) at 9 mm (top) and the 8 mm to 1 cm spectral index map (bottom) 
at increasing resolution from left
to right. The left panels with the lowest resolution and most sensitivity 
to extended structure show significant, structured emission surrounding
a bright source that we interpret as the position of the main protostar; 
the middle and right panels zoom-in on the region outlined
with a dashed box. The middle panels with higher resolution have 
resolved-out the extended structure and only detect the bright peak at the position 
of the protostar; however, the source appears extended at this resolution.
The highest resolution images in the right panels show that the
source is resolved into two sources separated by 18.5 AU.
The contours in each panel are [-6, -3, 3, 6, 9, 12, 15, 20, 25, 
30, 35, 40, 50, 60, 70, 80, 90, 100, 150] $\times$ $\sigma$, where
$\sigma$ = 7.3 \microjy, 9.6 \microjy, 11.9 \microjy\ from left 
to right at 9 mm. The spectral index maps are only drawn in regions where the S/N $>$ 10.}
\label{IRAS03292}
\end{figure}

\clearpage
\begin{figure}[!ht]
\begin{center}
\includegraphics[scale=0.25]{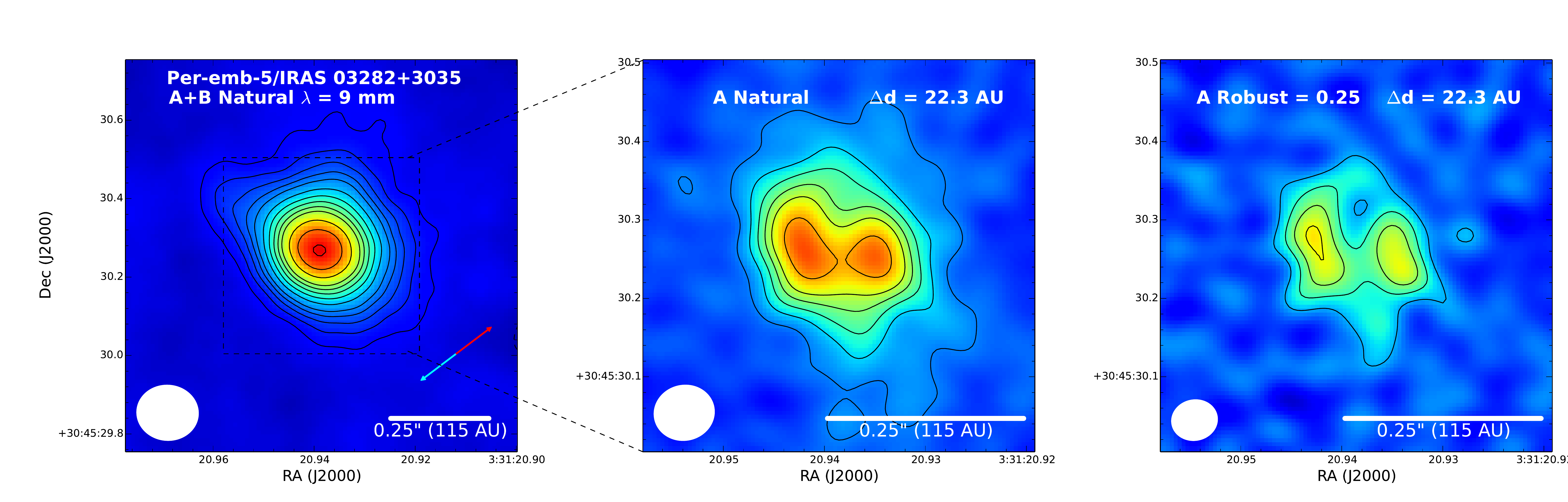}
\includegraphics[scale=0.25]{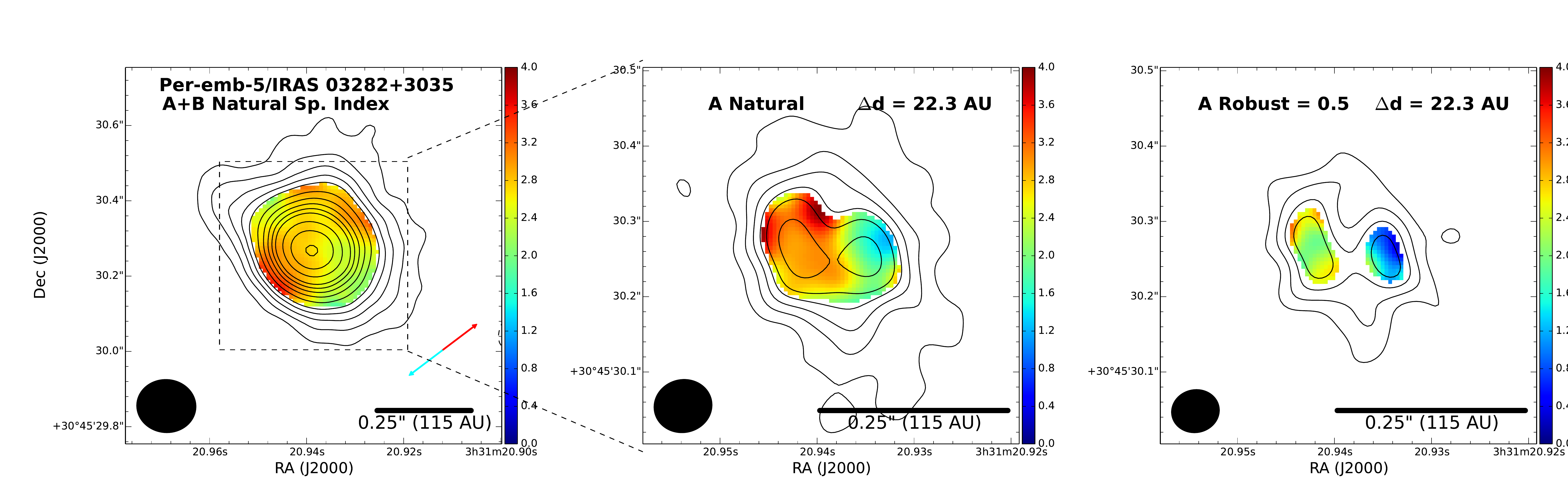}
\end{center}
\caption{Images of Per-emb-5 (IRAS 03282+3035)  at 9 mm (top) and the 8 mm to 1 cm spectral index map (bottom) 
 at increasing resolution from left
to right. The left panels with the lowest resolution show marginally-resolved 
emission, but the middle panels with higher resolution show
that this source breaks into a double-peaked structure at higher resolution.
The highest resolution images in the right panels show that the eastern peak 
is elongated in the north-south direction; 
the two sources are separated by 22.3 AU. The contours in each panel are [-6, -3, 3, 6, 9, 12, 15, 20, 25, 30,
35, 40, 50, 60, 70, 80, 90, 100, 150] $\times$ $\sigma$, where $\sigma$ = 6.35 \microjy, 8.4 \microjy, 9.3 \microjy\ 
from left to right at 9 mm. The spectral index maps are only drawn in regions where the S/N $>$ 10.}
\label{IRAS03282}
\end{figure}
\clearpage

\begin{figure}[!ht]
\begin{center}
\includegraphics[scale=0.25]{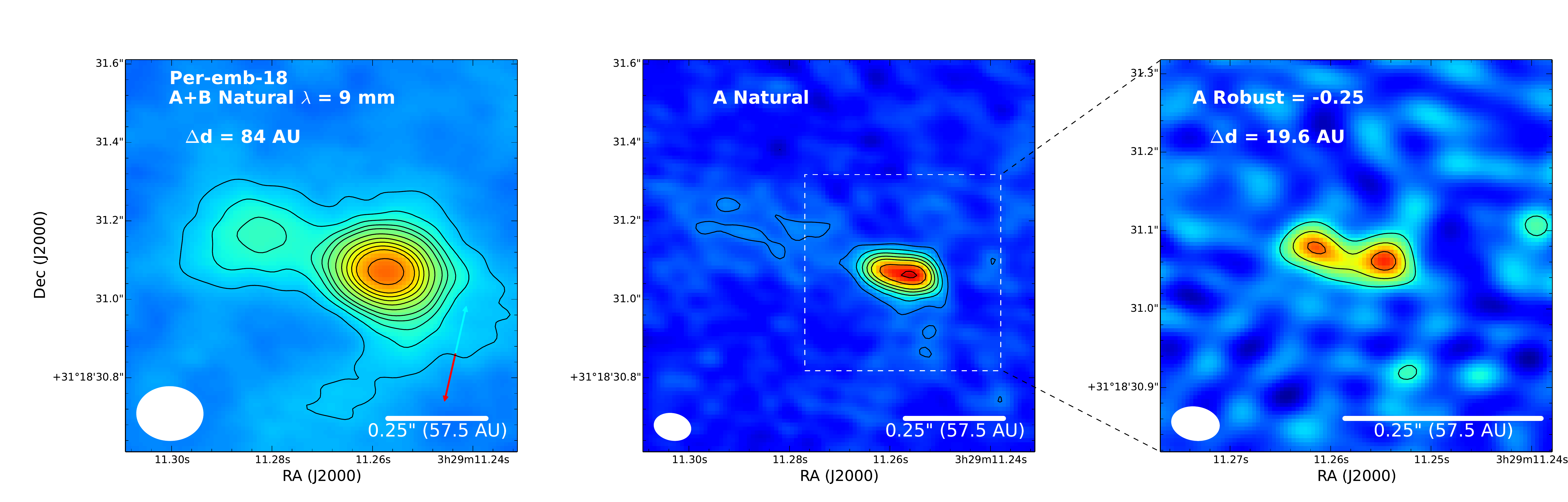}
\includegraphics[scale=0.25]{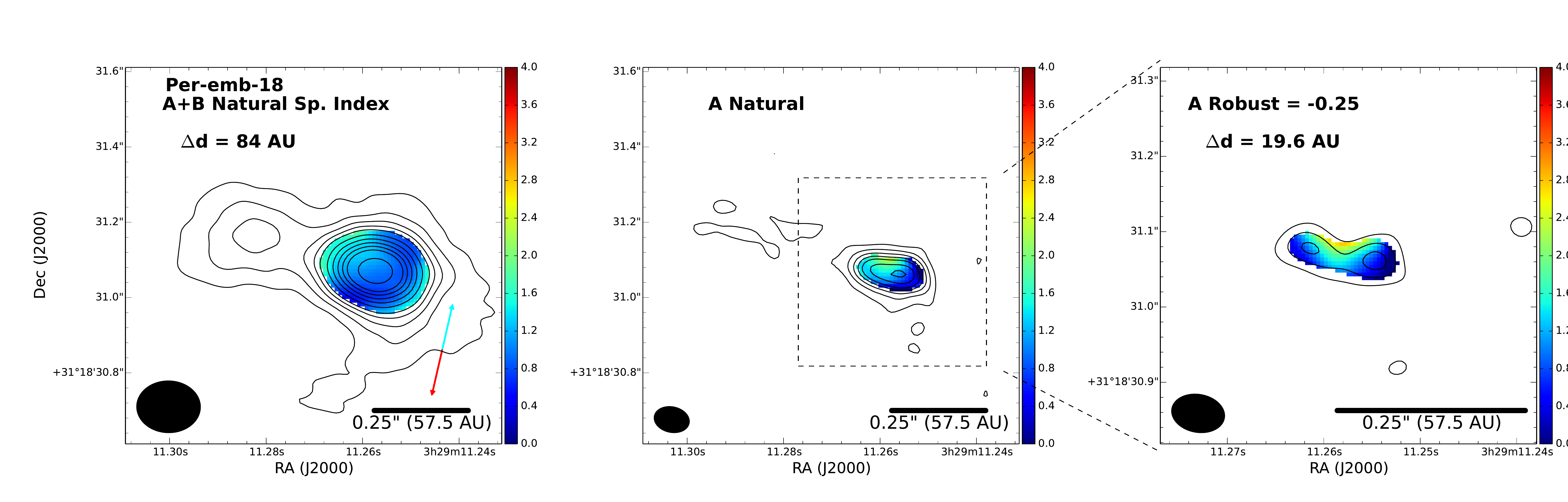}
\end{center}
\caption{Images of Per-emb-18 (NGC 1333 IRAS7) at 9 mm (top) and the 8 mm to 1 cm spectral index map (bottom) 
at increasing resolution from left
to right. The left panels with the lowest resolution show double-peaked 
emission with the eastern source being significantly fainter
than the western source. Higher resolution data are shown in the middle 
panels and the eastern source is now absent, indicating
that it has been resolved-out at higher resolution. However, 
the western source is resolved at this scale.
The highest-resolution view is shown in the right panels as a 
zoom-in on the dashed-box shown in the middle panel toward the western
source. At 9 mm the source is clearly double-peaked, separated 
by 19.6 AU. The contours in each panel are 
[-6, -3, 3, 6, 9, 12, 15, 20, 25, 30, 35, 40, 50, 60, 70, 80, 90, 100, 150] $\times$ $\sigma$, where
$\sigma$ = 6.56 \microjy, 8.8 \microjy, 9.8 \microjy\ from left to right at 9 mm. 
The spectral index maps are only drawn in regions where the S/N $>$ 10.}
\label{Per18}
\end{figure}
\clearpage

\begin{figure}[!ht]
\begin{center}
\includegraphics[scale=0.4]{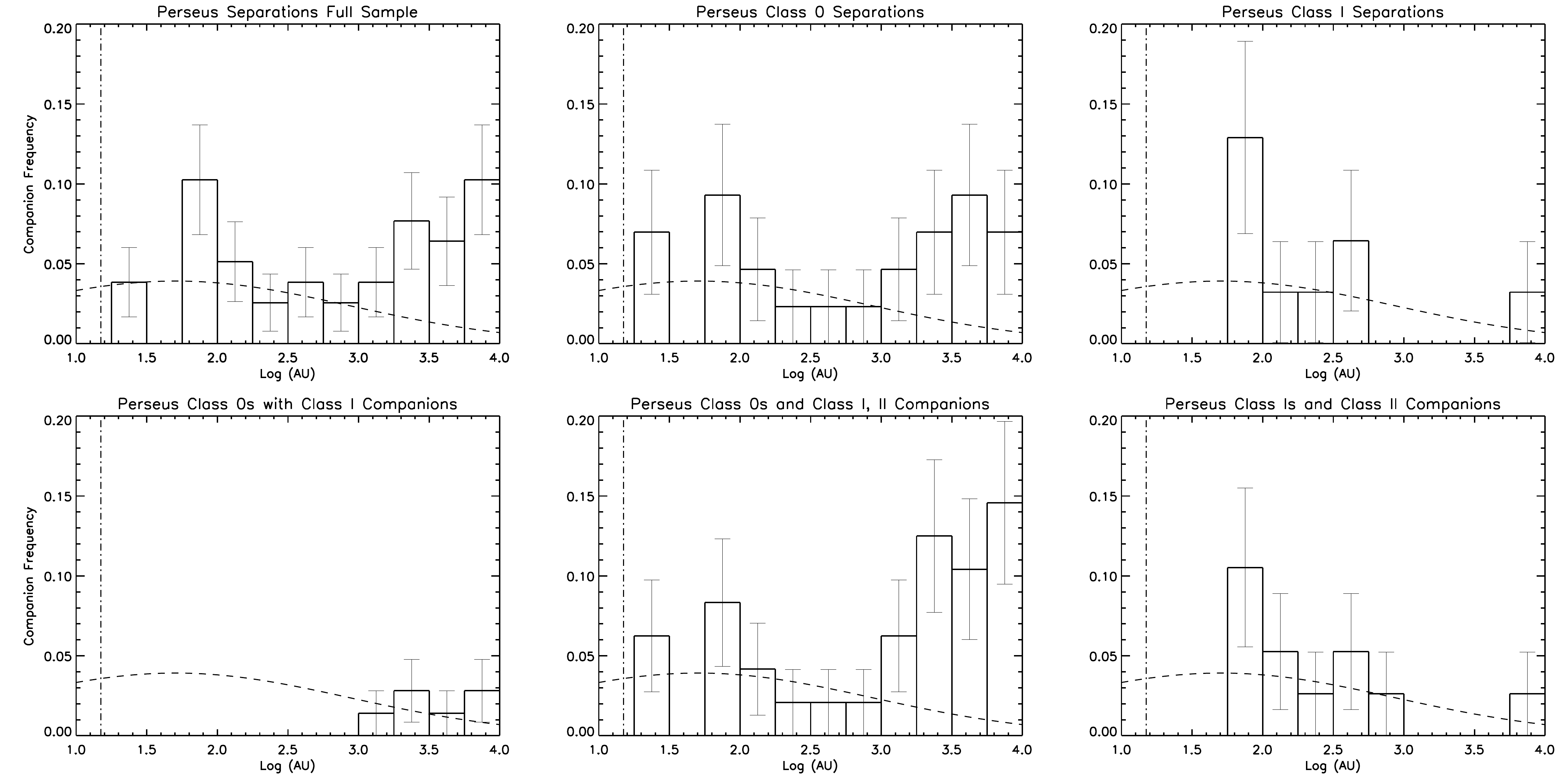}
\end{center}
\caption{Histograms of companion frequency versus separation for 
multiple sources in Perseus. The top left panel shows the distribution
for all sources in the sample; the top middle and top right panels break the distribution
into sources that are only comprised of Class 0 protostars and Class I protostars, respectively.
The bottom left panel shows only the multiple systems comprised of Class 0 and I sources, the
bottom middle shows the separation distribution of all systems with a Class 0 primary source and
the bottom left panel shows the same, but with a Class I primary. The systems comprised of
a Class 0 and Class I protostar are not included in the Class I plot in the bottom right.
 Note the apparent bi-modal distribution
for the full sample and Class 0 samples and the apparent deficit of wide companions for the Class I 
systems. In all plots, the dashed
curve is the Gaussian fit to the field star separation distribution 
from \citet{raghavan2010} and the vertical dot-dashed line
corresponds to the approximate resolution limit of 15 AU.}
\label{comb-histo}
\end{figure}

\begin{figure}[!ht]
\begin{center}
\includegraphics[scale=0.8]{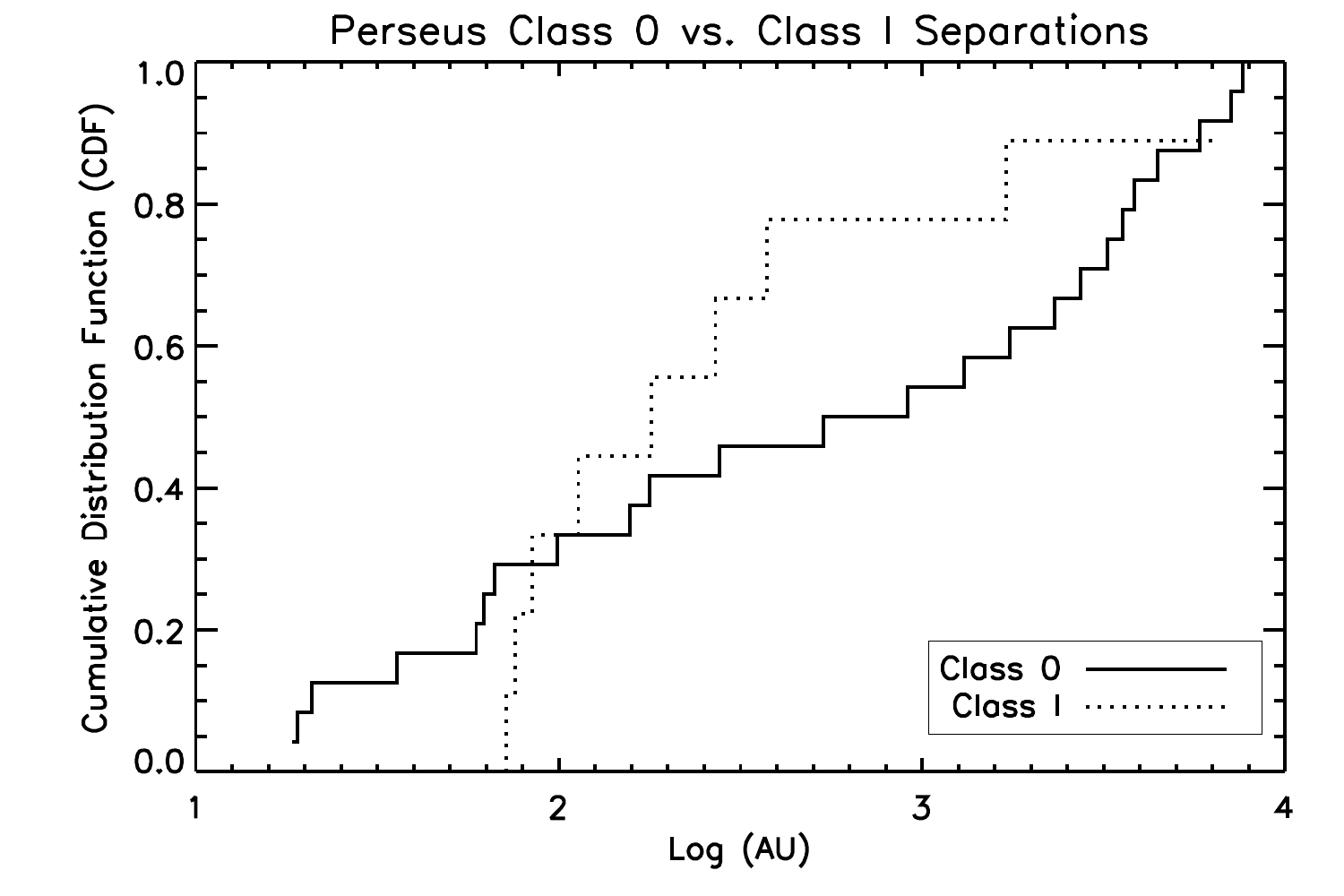}
\end{center}
\caption{Cumulative distribution function versus separation for the Class 0 
and Class I protostars. There is a large difference between
the two functions and the results from the Anderson-Darling (AD) test on the two samples indicates a probability
 of only 0.16 that they are drawn from the same
distribution. The Class I sources have substantially fewer wide companions 
relative to the Class 0s (also see Figure \ref{comb-histo}), this may be
 indicative of wide companions either migrating inward or moving apart 
as sources evolve to the Class I phase.
The Class 0 sources with wide Class I or Class II companions are not included in either of 
the cumulative distributions.
}
\label{cumulat-class}
\end{figure}

\clearpage

\begin{figure}[!ht]
\begin{center}
\includegraphics[scale=0.55]{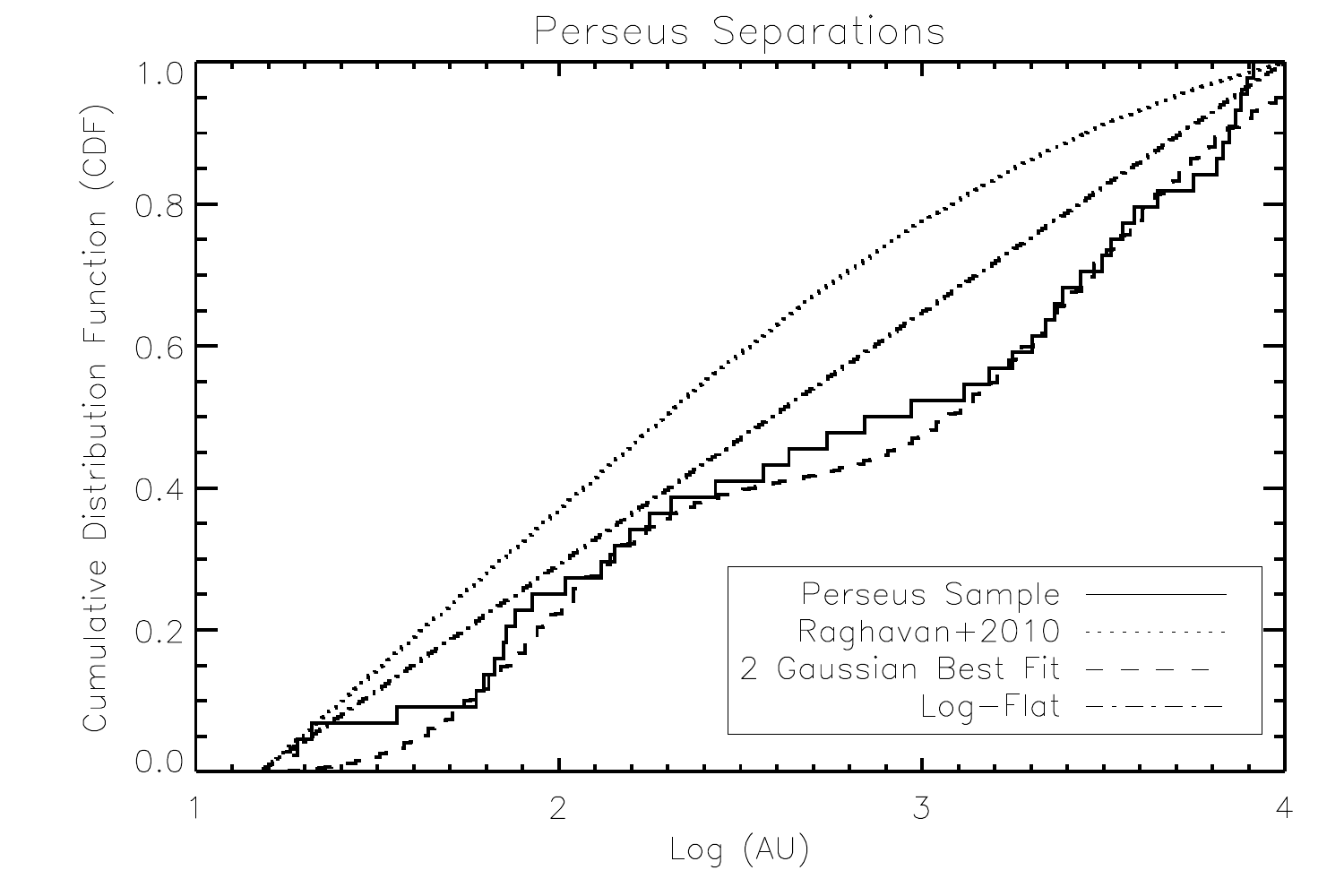}
\end{center}
\caption{Cumulative distribution function of the full sample of Perseus multiples
compared to different empirical and observed separation distributions.
The comparisons shown are for a log-flat distribution, the \citet{raghavan2010} 
distribution, and a distribution defined by 2 Gaussians. The
AD test probabilities for the log-flat distribution and the \citet{raghavan2010} 
distribution are 0.1 and 0.00015 respectively, meaning
that the Perseus separations are most likely not drawn from either of these distributions. 
Two Gaussians fit the data well, but
the parameters of the second Gaussian at large 
separations are poorly constrained.
}
\label{cumulat-full}
\end{figure}

\begin{figure}[!ht]
\begin{center}
\includegraphics[scale=0.7]{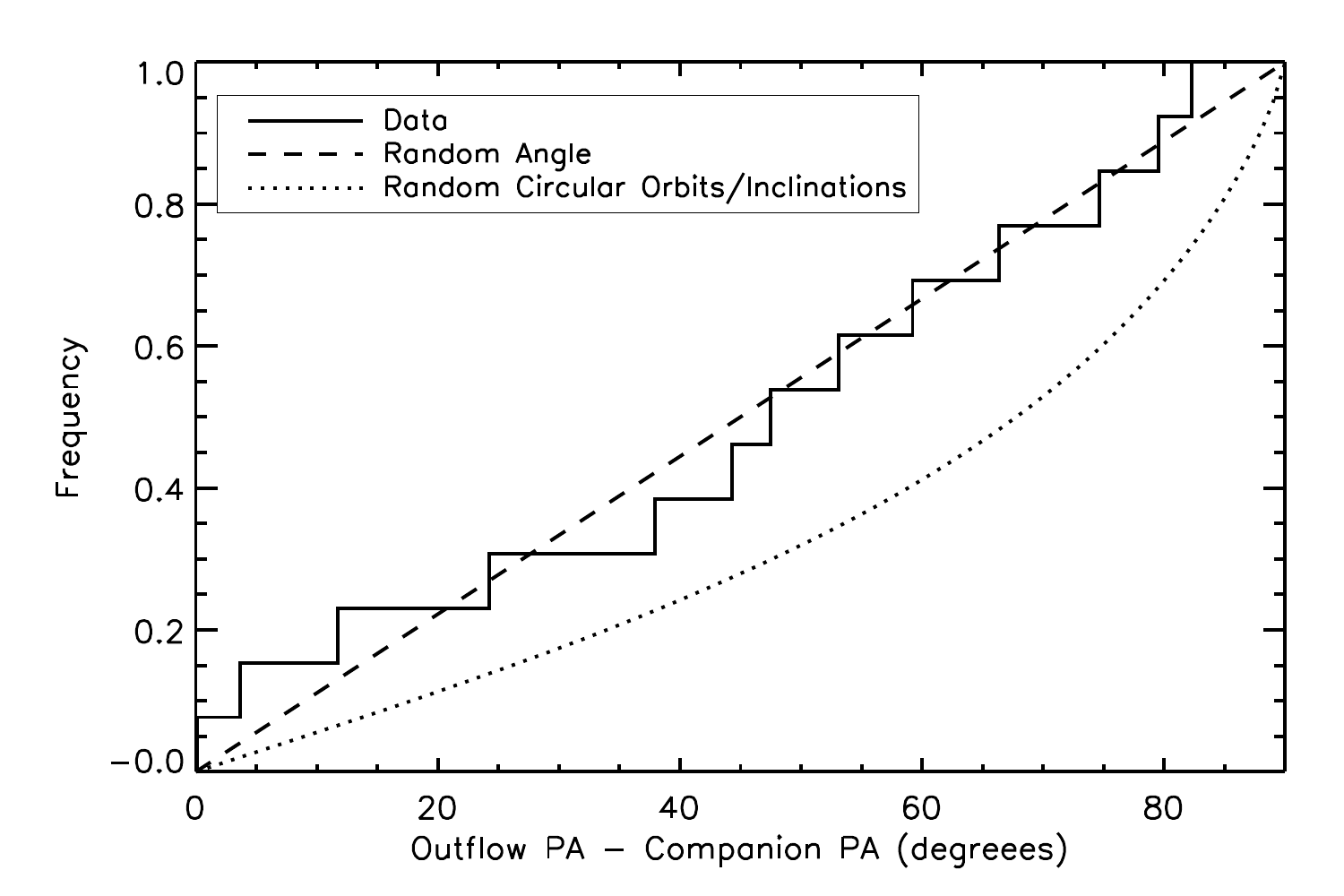}
\includegraphics[scale=0.55]{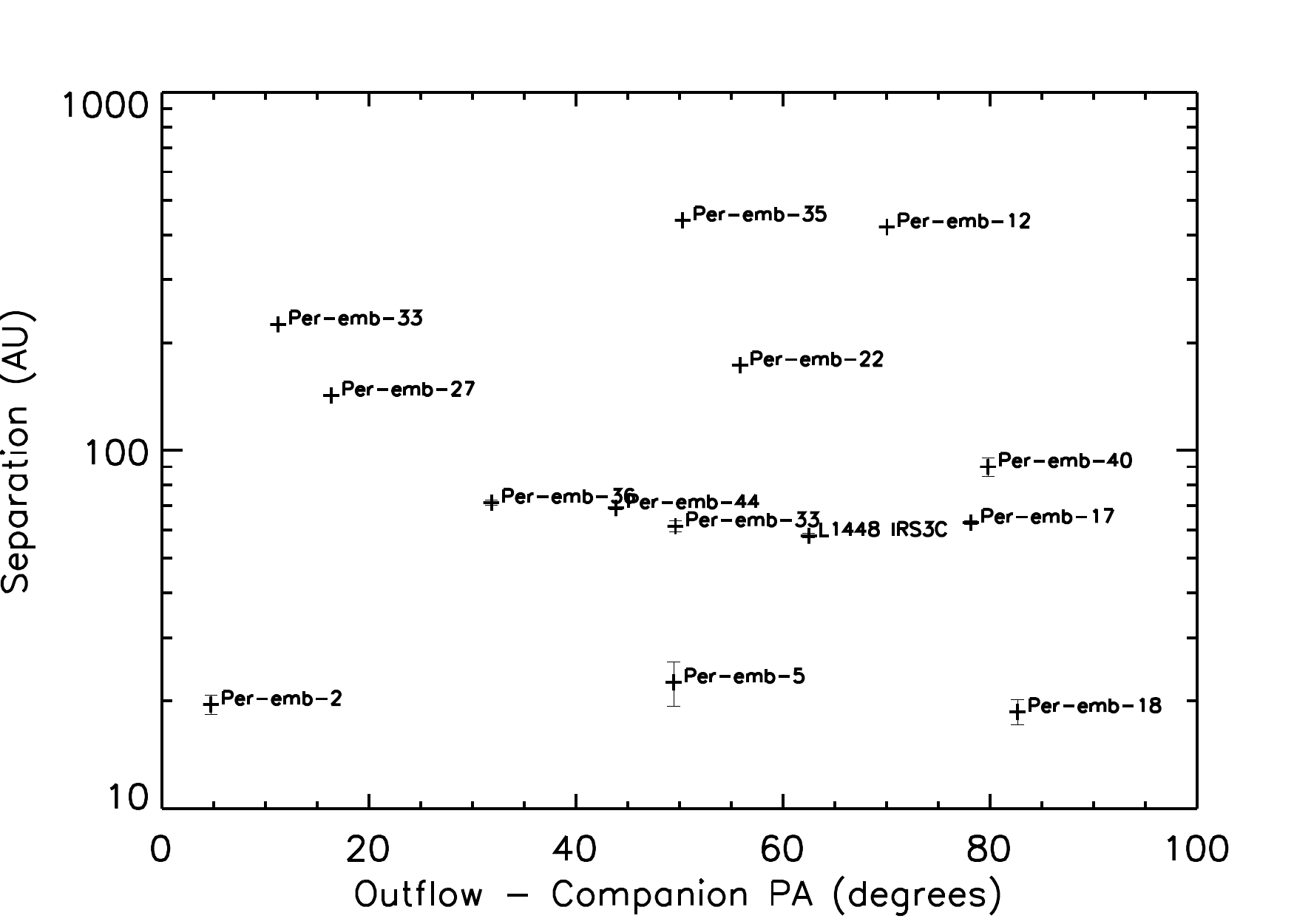}
\end{center}
\caption{Cumulative distribution of companion position angles relative to outflow 
position angles (top panel) for Class 0 and Class I sources
with separations $<$ 500 AU and known outflow position angles. The bottom panel
shows a plot of separation versus companion position angles relative to outflow position angles.
In the top panel, the solid line shows the data, the 
dashed line shows a random distribution of 
relative position angles, and the dotted line shows the distribution for 
position angles for companions at a random phase
in a circular orbit projected with a uniform distribution of inclinations.
The position angles, as measured on-sky, appear to be consistent with random.
The observations have a clear excess of companions at
position angles $<$ 40\degr\ relative to the expectation for random 
orientations and inclinations. However, the bottom plot shows that there is no
apparent correlation with relative position angle and separation; the apparent
excess may be due to small number statistics.
The outflow position angles are given in
Table 7.
}
\label{posangles}
\end{figure}

\begin{figure}[!ht]
\begin{center}
\includegraphics[scale=0.75]{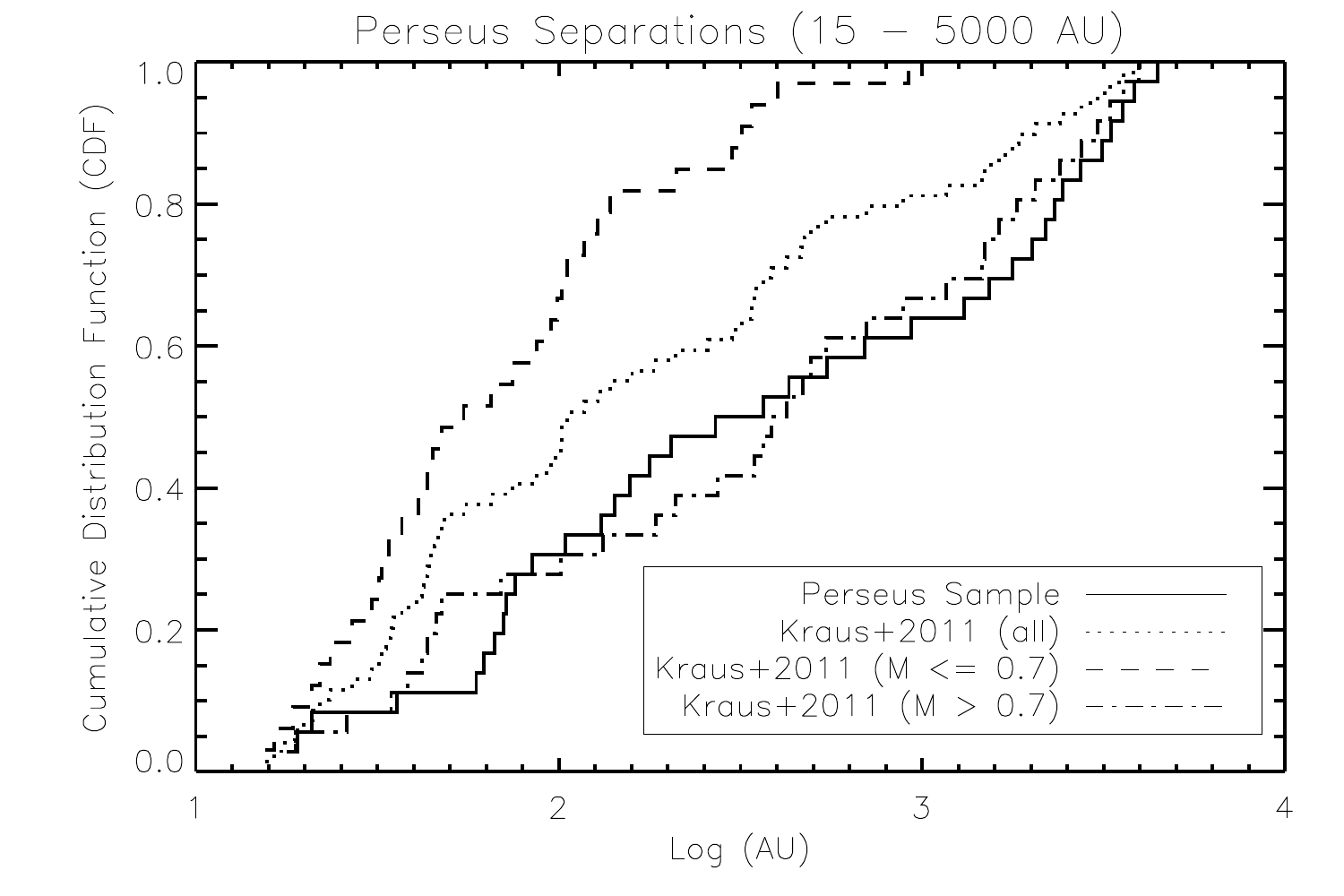}
\end{center}
\caption{Comparison of the Perseus multiples to the more-evolved multiple stars in Taurus from
\citet{kraus2011}. The whole Taurus sample and low-mass Taurus sub-sample are in disagreement with the
Perseus results, while the high-mass Taurus sub-sample is reasonably consistent with the Perseus sample.
The AD test results of the Taurus samples relative to Perseus indicate probabilities of 
being drawn from the same distribution of 0.024, 0.00015, and 0.80 for the full sample, low-mass, and high-mass
samples, respectively.
}
\label{cumulat-taurus}
\end{figure}

\begin{figure}[!ht]
\begin{center}
\includegraphics[scale=0.45]{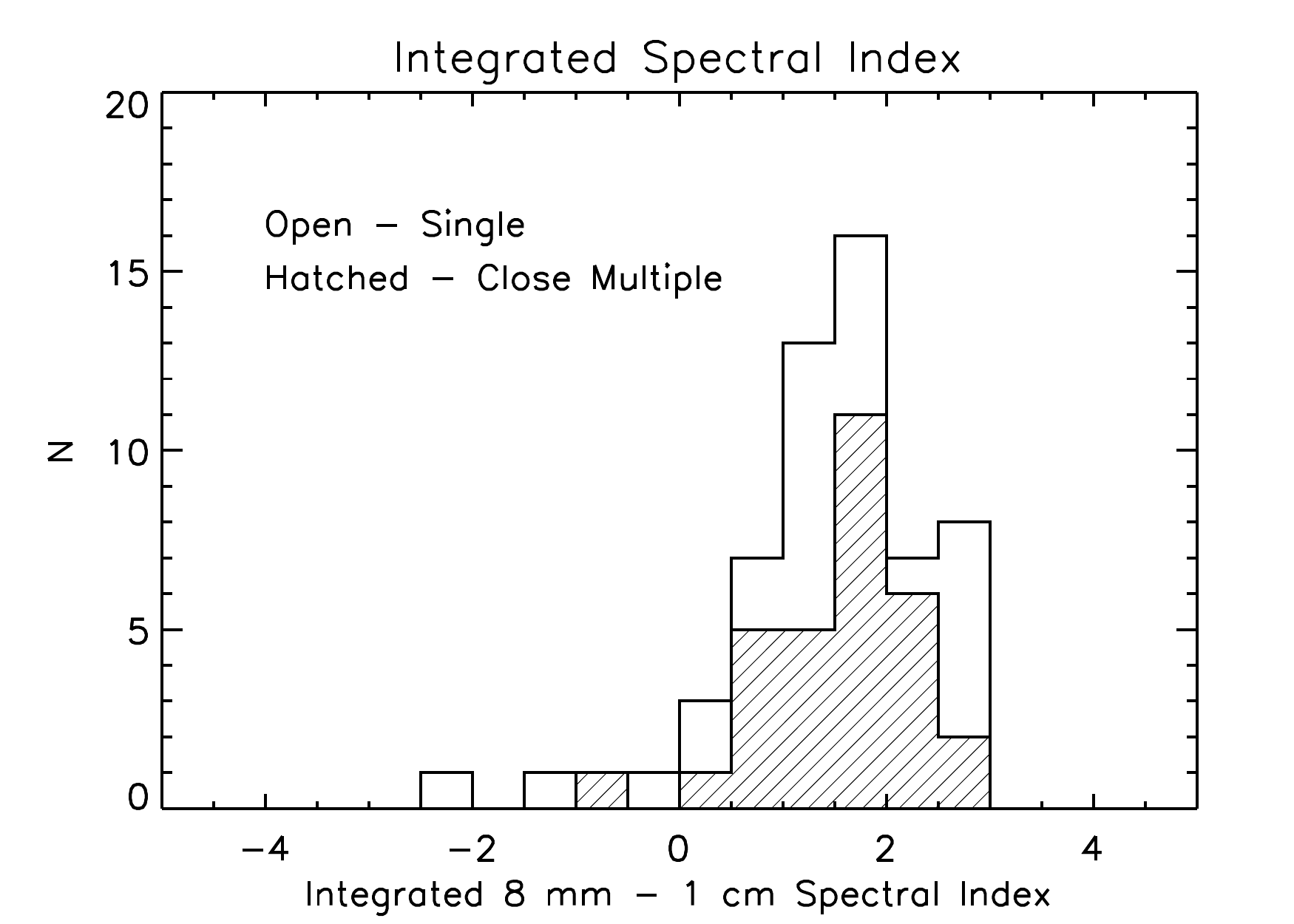}
\includegraphics[scale=0.45]{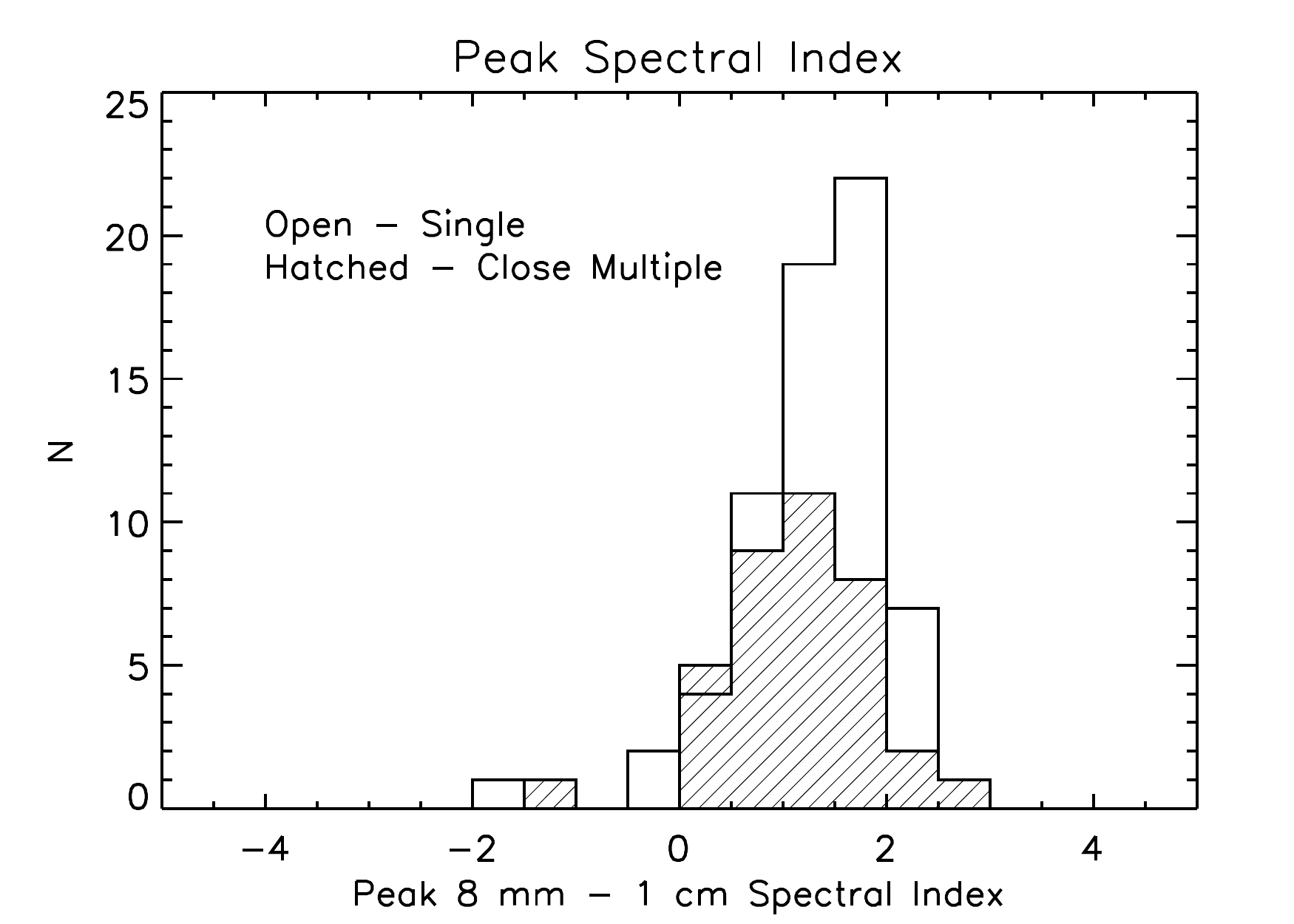}
\includegraphics[scale=0.45]{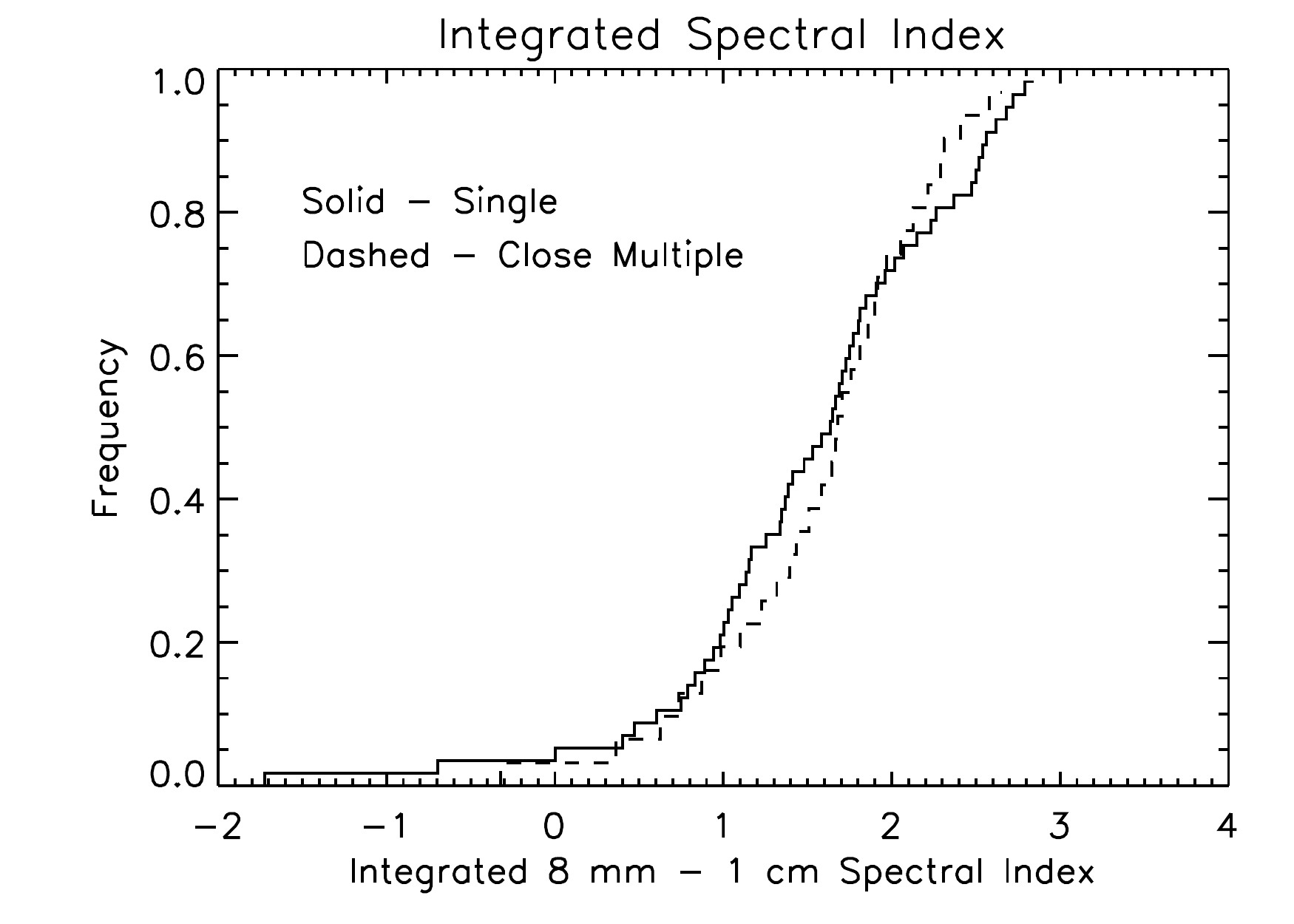}
\includegraphics[scale=0.45]{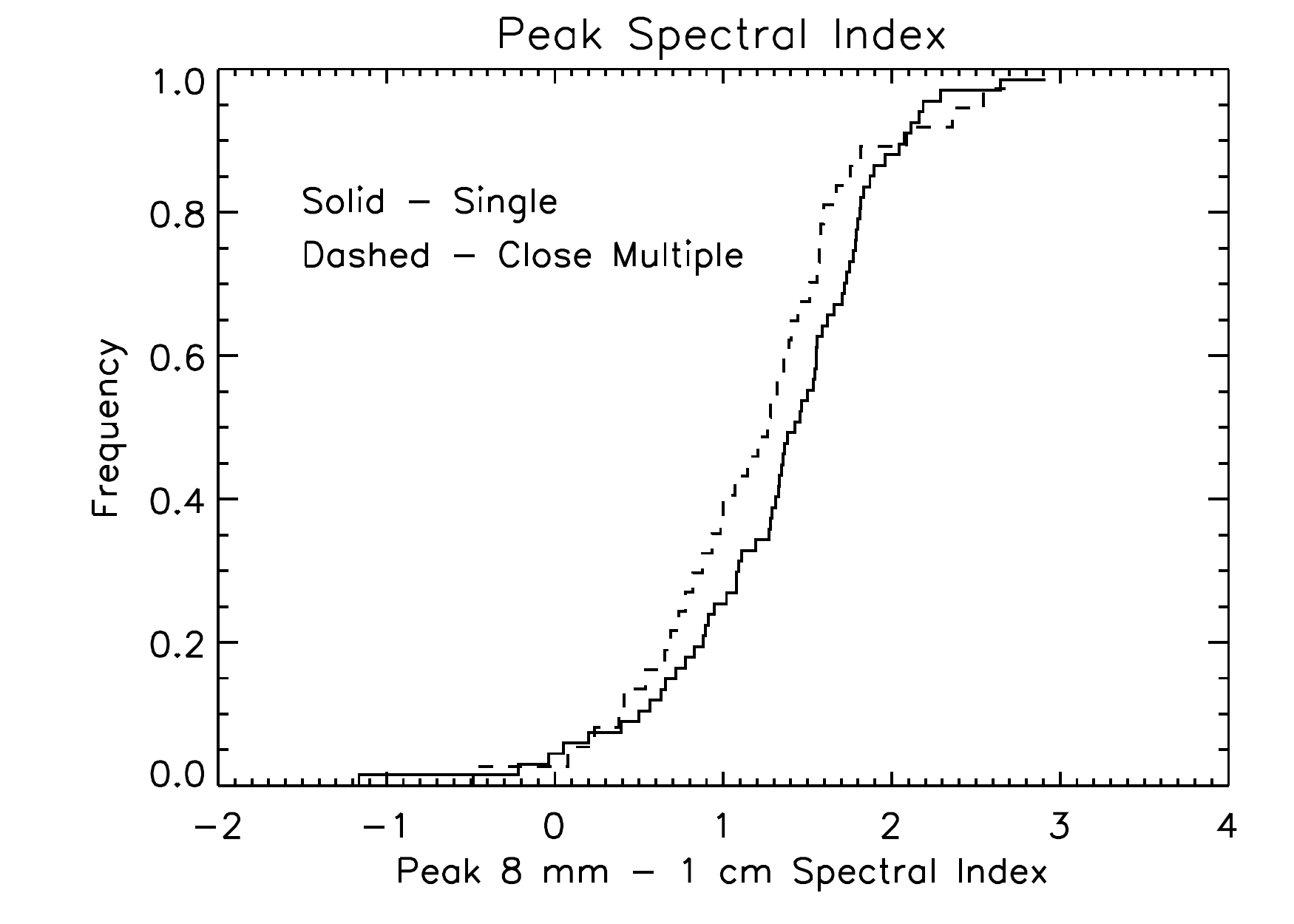}
\end{center}
\caption{Distributions of spectral indices for the single and multiple sources. The distributions
only include sources that have spectral index error less than 0.9. The histograms of the peak and
integrated spectral index are quite comparable for the single and multiple sources. The cumulative distributions
also show close correspondence of the two samples. Running the AD test on the distributions of 
integrated and peak spectral indices yield probabilities of 0.7 and 0.35, respectively, indicating
that the distributions for single and multiple sources are most likely drawn from the same sample.
Thus, the emission properties of single and multiple sources are statistically indistinguishable.
}
\label{cumulat-spindex}
\end{figure}

\begin{figure}[!ht]
\begin{center}
\includegraphics[scale=0.33]{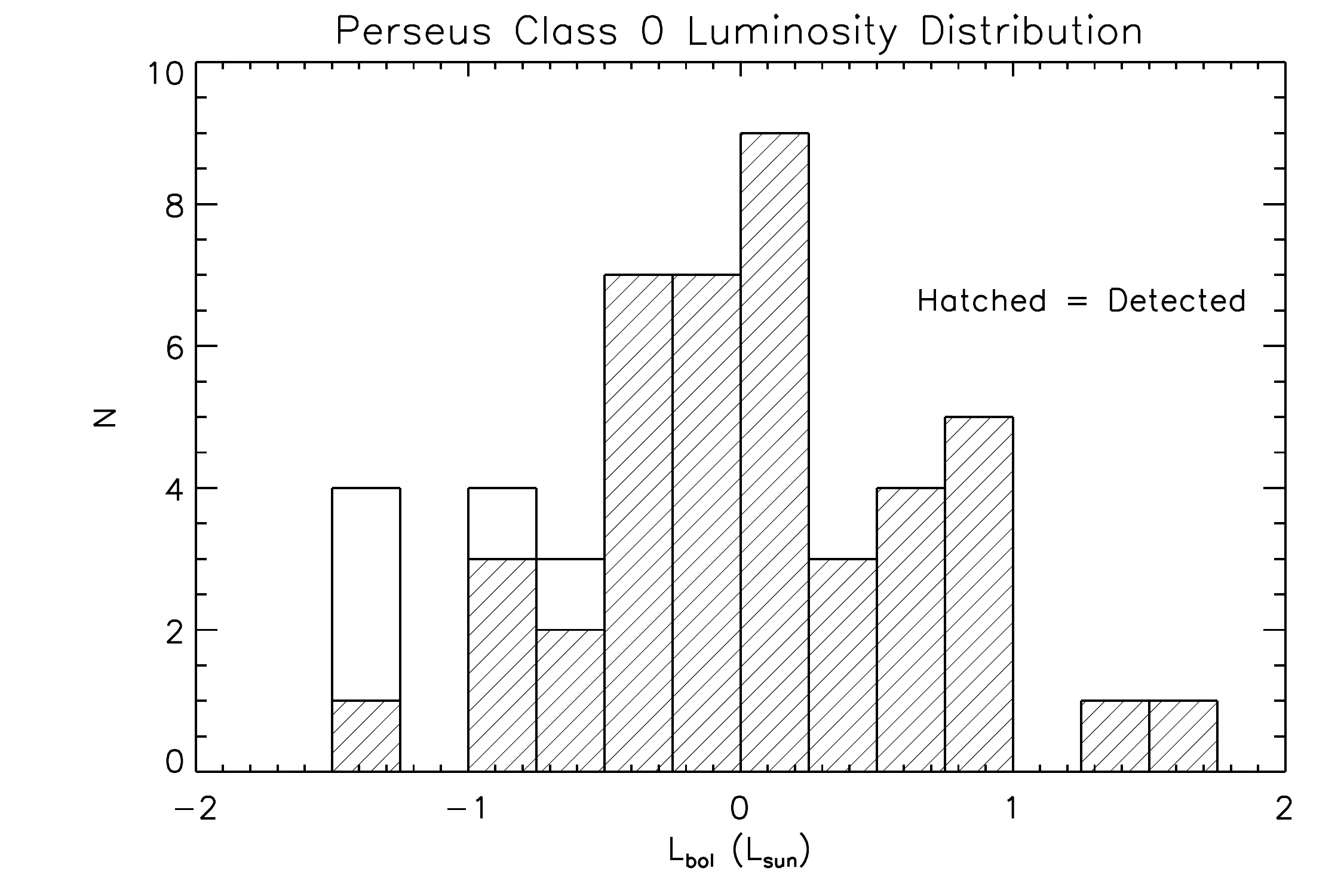}
\includegraphics[scale=0.33]{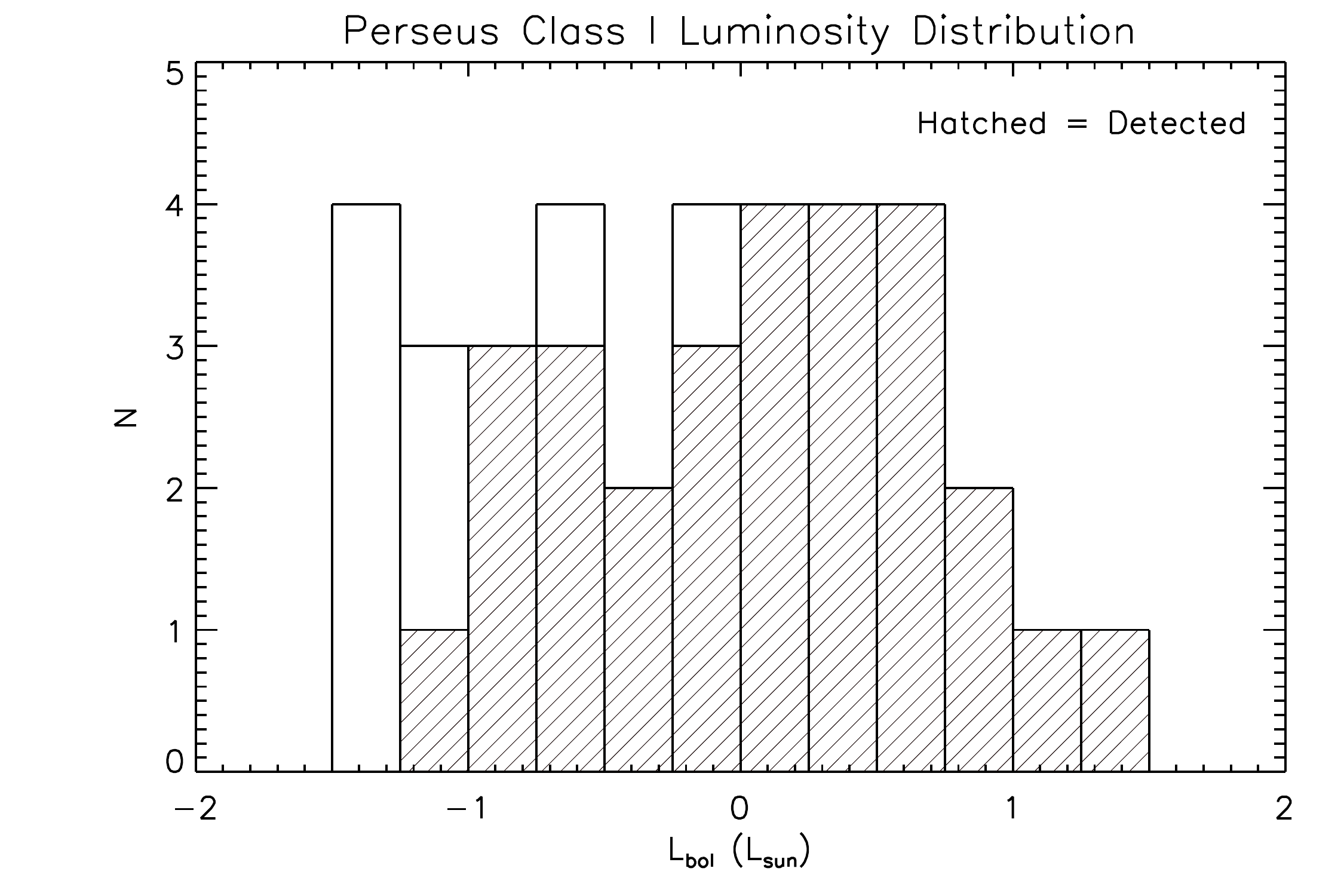}
\includegraphics[scale=0.33]{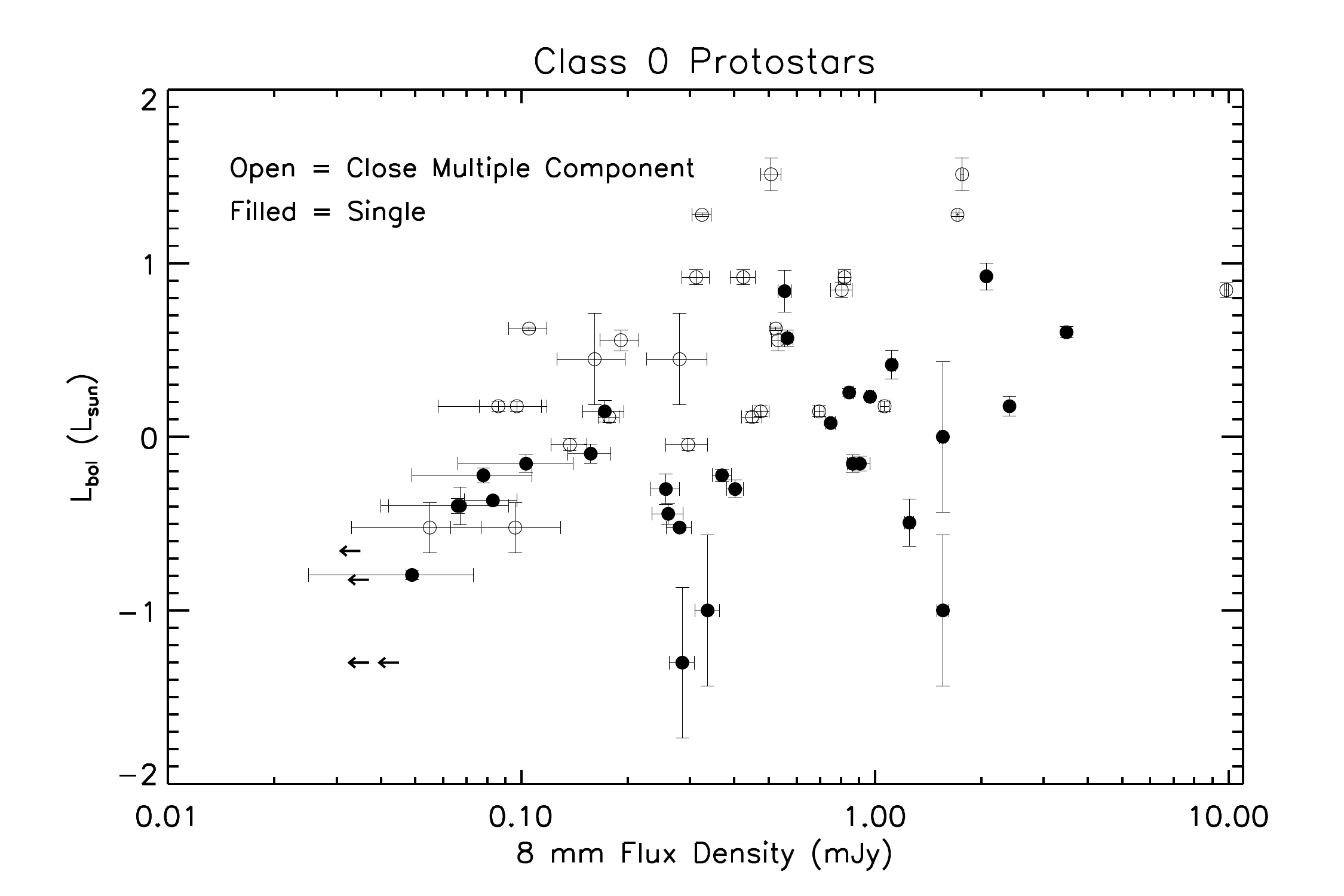}
\includegraphics[scale=0.33]{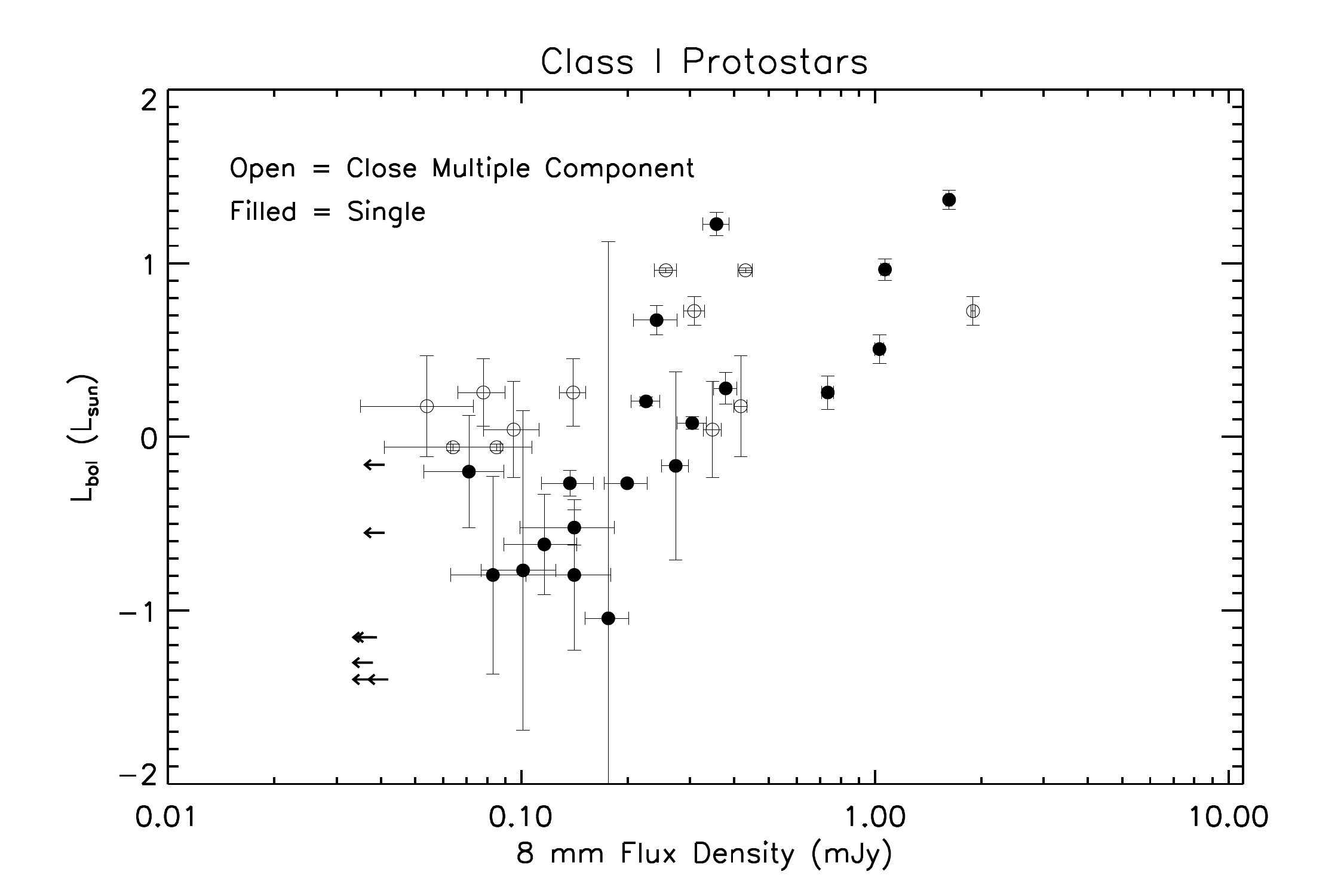}
\end{center}
\caption{The bolometric luminosity distributions of the Class 0 and Class I samples
are shown in the top left and top right panels, respectively. The hatched regions show the histogram
for sources that were detected in the VLA survey; we detect nearly all of the Class 0 and Class I samples
except for mainly a few low-luminosity sources. In the case of the Class 0 sample, most of these low-luminosity
sources are comprised of candidate first hydrostatic cores. The distribution of 8 mm flux densities
as a function of bolometric luminosity are then shown in the bottom left and bottom right
panels, respectively. The components of multiple systems are drawn as open circles and the
single sources are filled circles. There is an apparent weak correlation between 8 mm flux density and bolometric
luminosity, but there is significant scatter. 
The Class I sources show a more clear relationship since there are no 
low-luminosity sources with large 8~mm flux densities.
}
\label{sample-flux8}
\end{figure}

\begin{figure}[!ht]
\begin{center}
\includegraphics[scale=0.33]{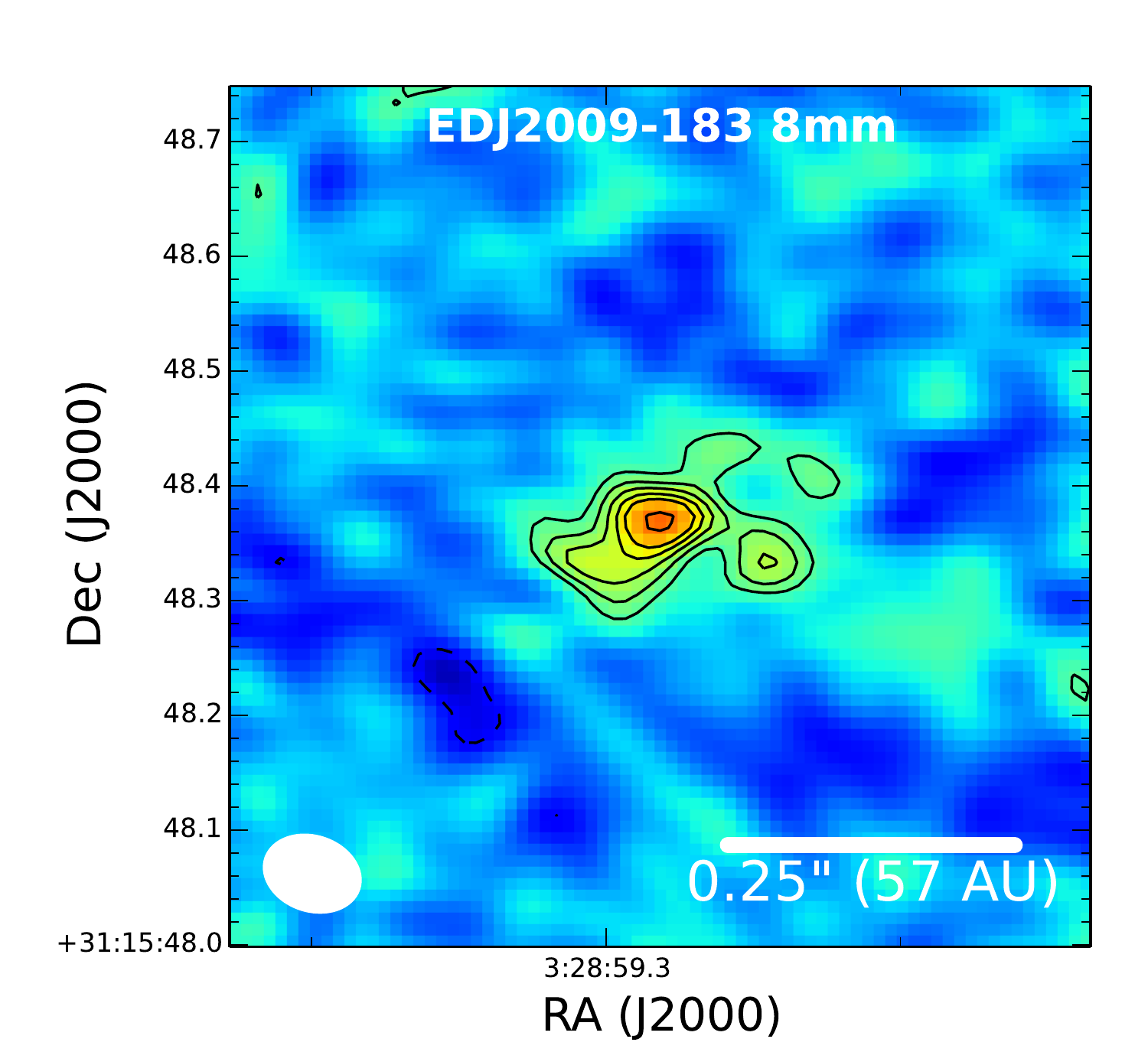}
\includegraphics[scale=0.33]{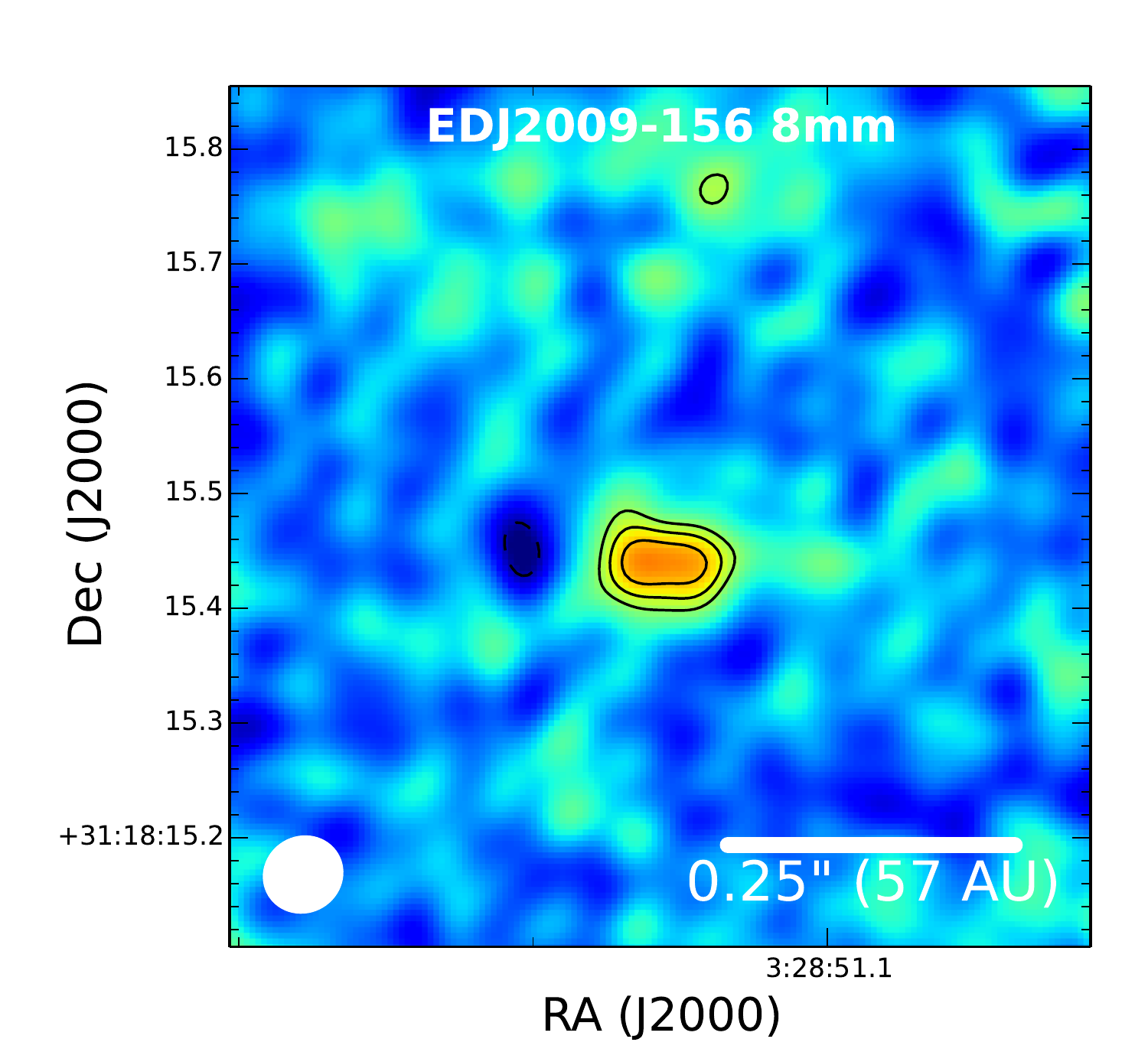}
\includegraphics[scale=0.33]{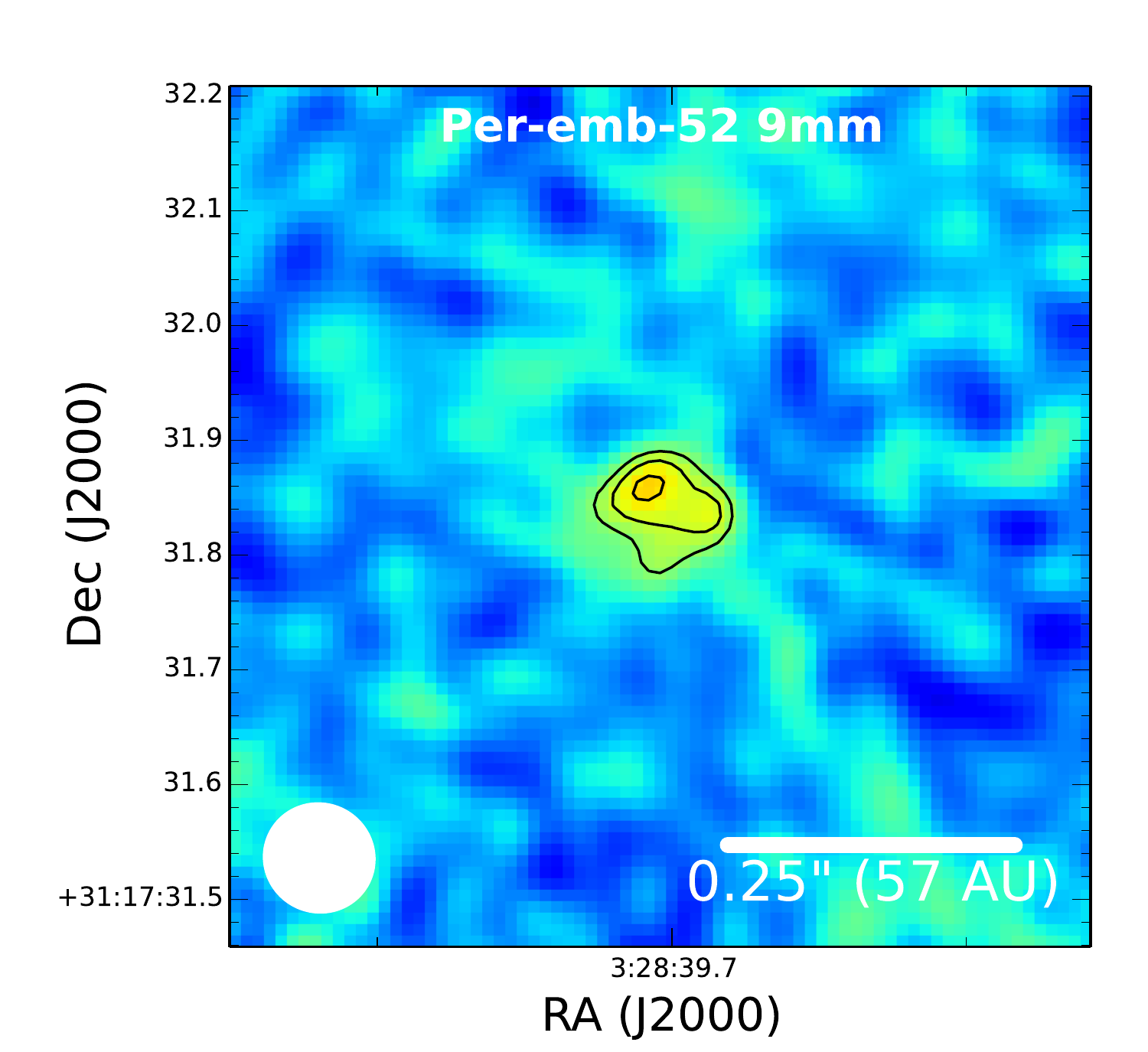}
\includegraphics[scale=0.33]{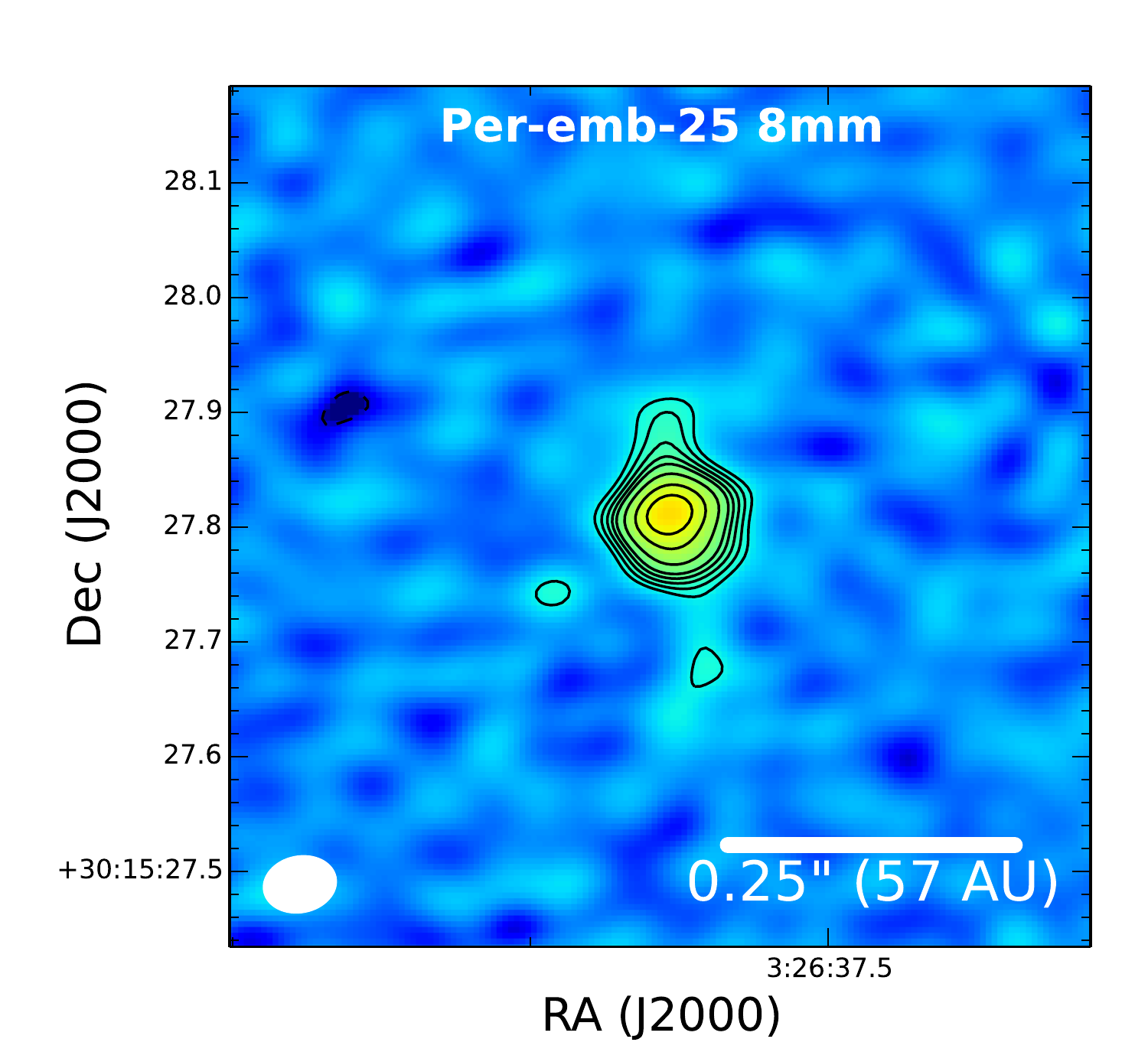}
\end{center}
\caption{Images of resolved structures that are not significant enough to be classified as companions. The contours in 
these images start at 3$\sigma$ and increase in 1$\sigma$ intervals; $\sigma$= 11.5 \microjy, 17.4\microjy, 11.0 \microjy,
 14.1 \microjy\ and 14.5 \microjy\ for EDJ2009-156, EDJ2009-183, Per-emb-52, and Per-emb-25, respectively. Images
are from combined A and B-configuration images, except for EDJ2009-156 and EDJ2009-183 where they are A-configuration data only.
}
\label{poss-multiples}
\end{figure}

\begin{figure}[!ht]
\begin{center}
\includegraphics[scale=0.8]{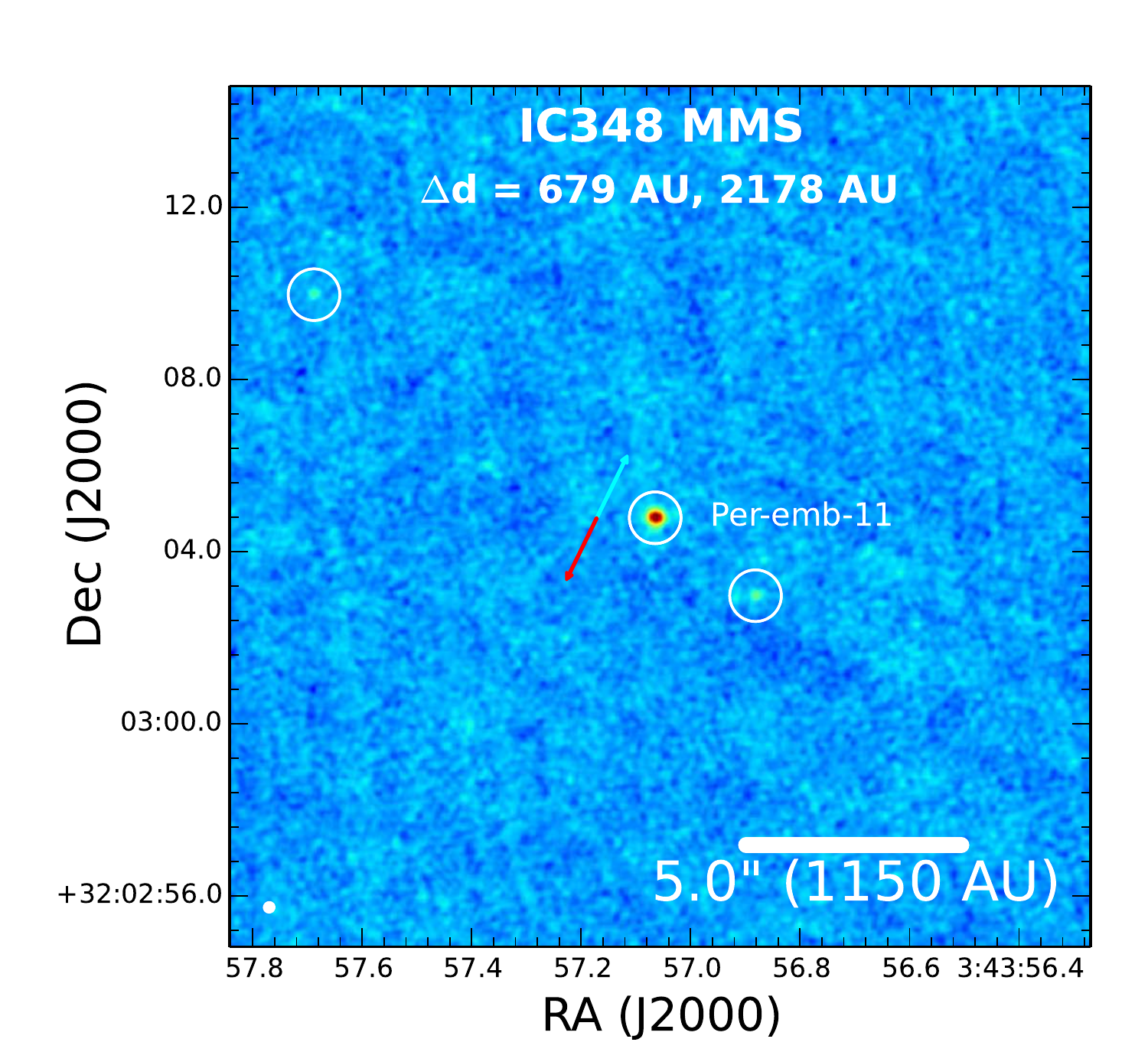}
\end{center}
\caption{Image of the wide multiple system toward IC348 MMS (Per-emb-11). The image is a B configuration image of the region and three
sources are detected. IC348 MMS1/Per-emb-11-A is the brightest source in the middle, JVLA3a from \citet{rodriguez2014} is 2\farcs95 
southwest and IC348 MMS2 (Per-emb-11-C) is separated by 9\farcs47. White circles are drawn around the companion sources.
Separations written inside the figure are relative to Per-emb-11 at the center. The blue and red arrows drawn near sources denote the
blue and red-shifted direction of the outflows when known.}
\label{IC348MMS-wide}
\end{figure}

\begin{figure}[!ht]
\begin{center}

\includegraphics[scale=0.8]{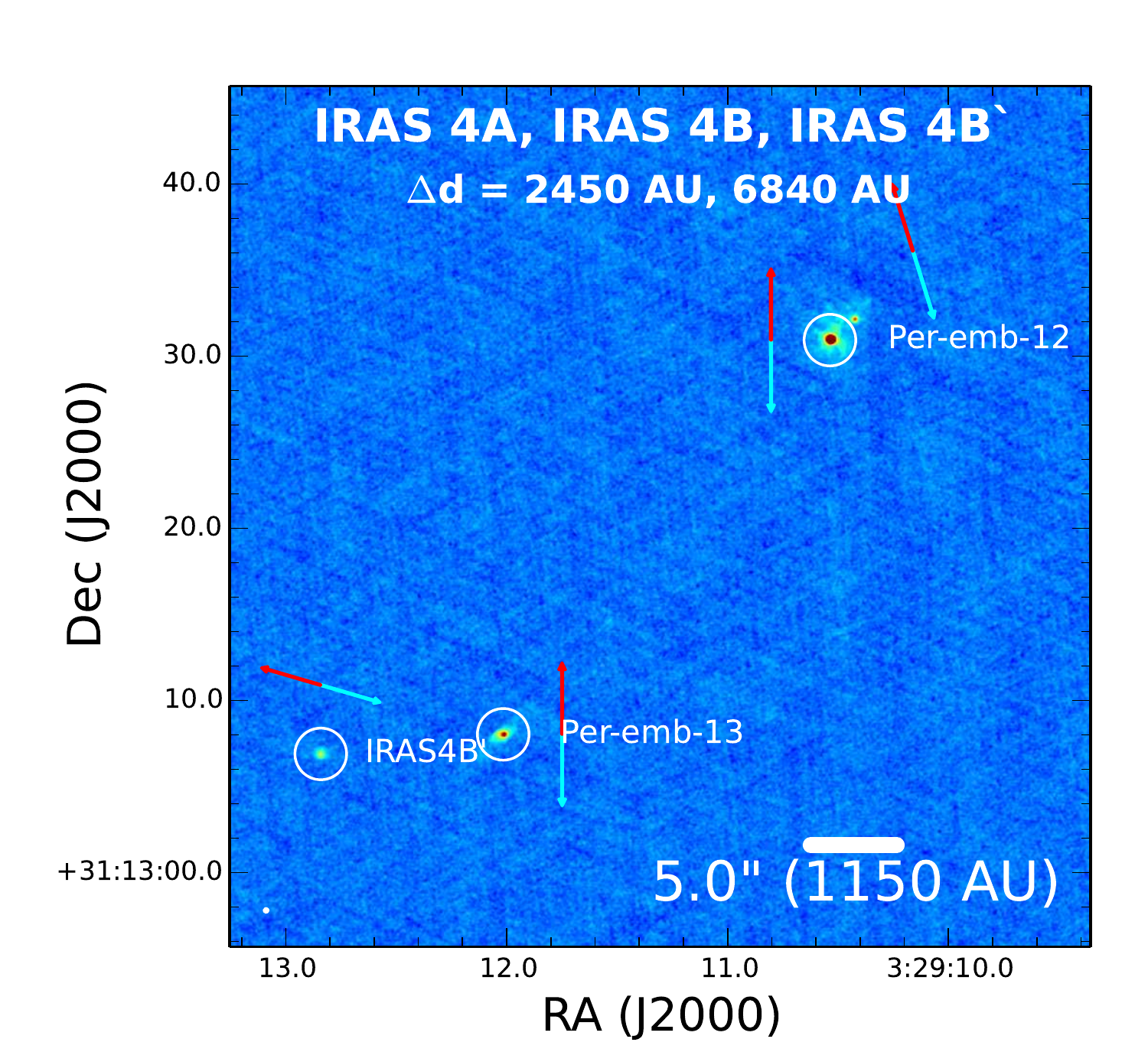}
\end{center}
\caption{B configuration image of NGC 1333 IRAS4A/Per-emb-12, NGC 1333 IRAS 4B/Per-emb-13, and NGC 1333 IRAS4B' 
\citep[IRAS4B' was also called IRAS4C by ][]{looney2000}.
Both sources appear resolved in the images, with IRAS4B being extended to the east. The blue and red arrows drawn near sources denote the
blue and red-shifted direction of the outflows when known.
\label{IRAS4B-wide}
}
\end{figure}

\begin{figure}[!ht]
\begin{center}
\includegraphics[scale=0.8]{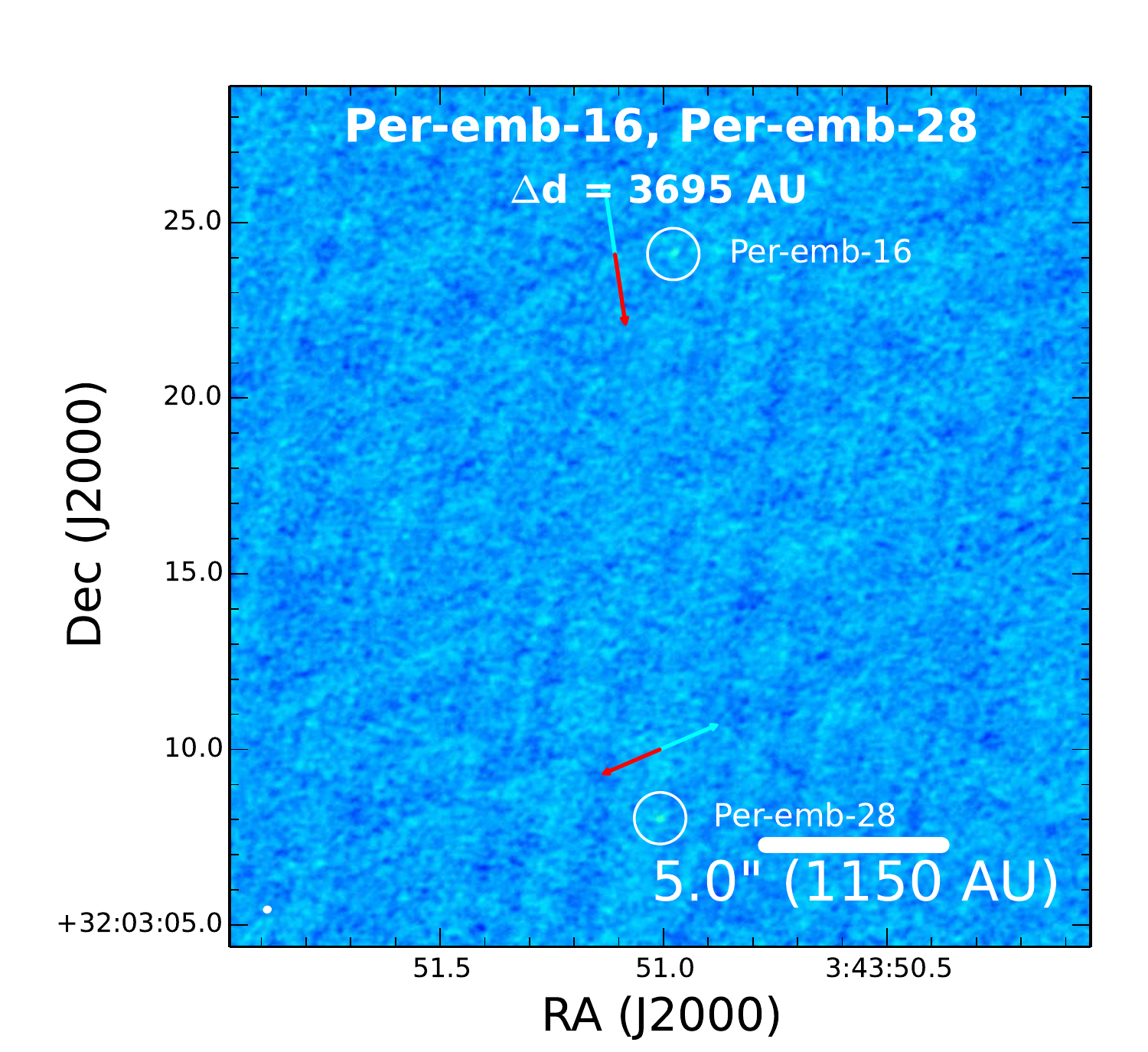}
\end{center}
\caption{A+B configuration image of Per-emb-16 and Per-emb-28, forming a wide multiple system. Per-emb-28 is also known as LRLL 54361
and has been shown to exhibit strong variability by \citet{muzerolle2013}. The blue and red arrows drawn near sources denote the
blue and red-shifted direction of the outflows. The position angle of the outflow
from Per-emb-16 is approximately in the north-south direction \citet{yen2015} and the outflow direction of Per-emb-28 is about 60\degr\ different
with a position angle of $\sim$300\degr\ \citep{muzerolle2013}. 
}
\label{Per16-Per28-wide}
\end{figure}

\begin{figure}[!ht]
\begin{center}
\includegraphics[scale=0.8]{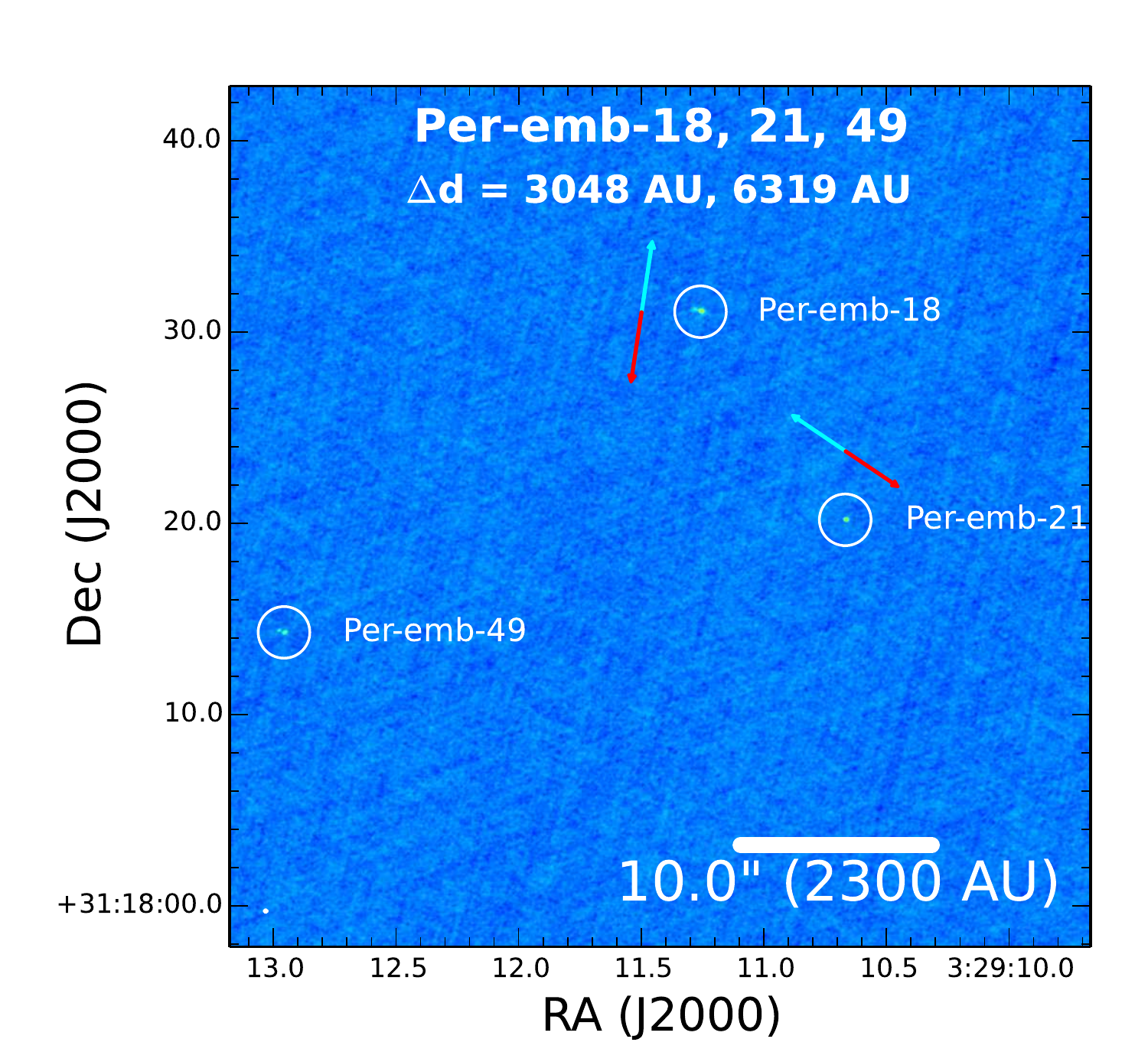}
\end{center}
\caption{Wide view of the NGC 1333 IRAS7 system in a 9 mm A+B configuration image. 
Both Per-emb-18 and Per-emb-49 are multiples with separations less than 100 AU and
Per-emb-21 is single to the limit of our resolution. The blue and red arrows drawn near sources denote the
blue and red-shifted direction of the outflows when known. The outflow direction of Per-emb-18 is approximately north-south, a position
angle of $\sim$159\degr\ is estimated from H$_2$ knots \citep{davis2008}. The outflow of Per-emb-21 is oriented
approximately orthogonal to that of Per-emb-18 with a position angle of 48\degr. Note that Per-emb-18 = YSO23 and Per-emb-21 = YSO24 from
\citet{davis2008}. The outflow position angle for Per-emb-49 is currently unknown.
Separations written inside the figure are relative to Per-emb-18.
}
\label{IRAS7-wide}
\end{figure}

\begin{figure}[!ht]
\begin{center}
\includegraphics[scale=0.8]{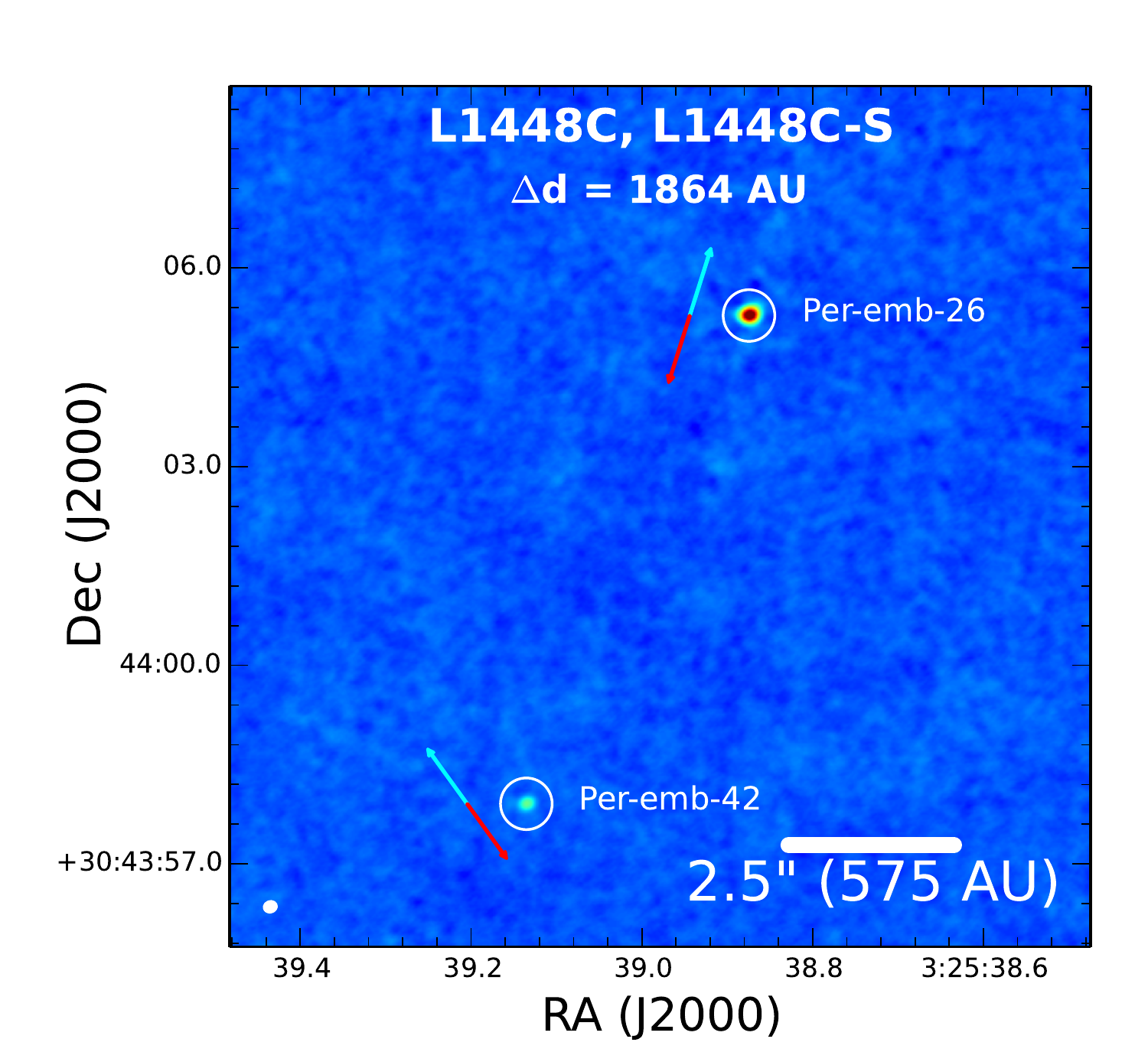}
\end{center}
\caption{Image of the region around L1448C/Per-emb-26, also known as L1448-mm. Per-emb-42 was referred to as L1448C-S in \citet{jorgensen2006}
where the source is detected by \spitzer\ with IRAC and MIPS \citep[see also][]{tobin2007}.  Both sources are found to drive
an outflow. The blue and red arrows drawn near sources denote the
blue and red-shifted direction of the outflows. The outflow from Per-emb-42 oriented with a position angle of 40\degr\ relative to the
$\sim$340\degr\ position angle of Per-emb-26 \citep{hirano2010}. }
\label{L1448C-wide}
\end{figure}

\begin{figure}[!ht]
\begin{center}
\includegraphics[scale=0.8]{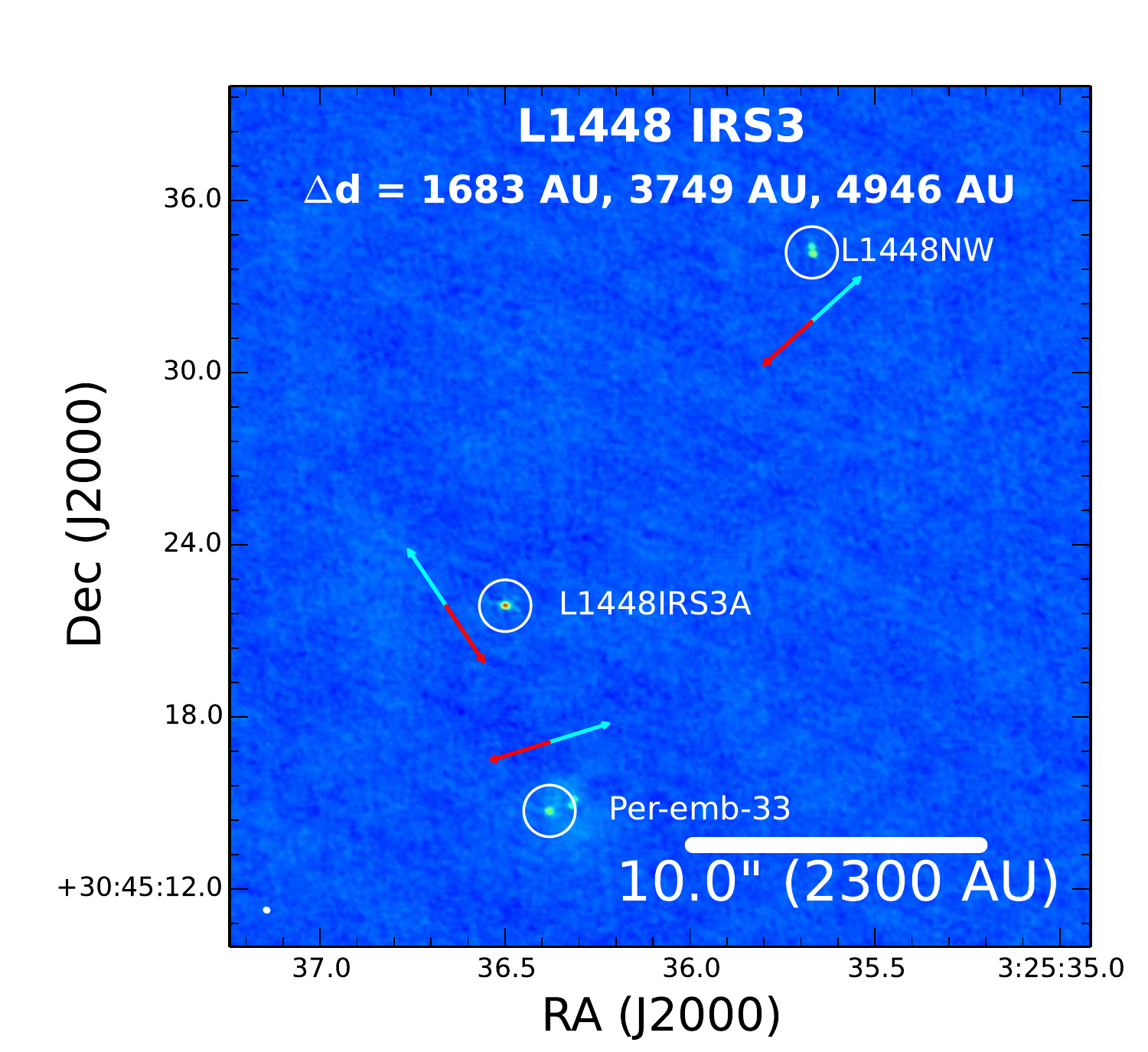}
\end{center}
\caption{Image of the L1448-N or IRS3 region; Per-emb-33 corresponds to L1448 IRS3B.
Both Per-emb-33 and L1448NW (L1448 IRS3C) are Class 0 sources and multiple systems with separations less than 200 AU; L1448 IRS3A is single at 
the limit of our resolution. The blue and red arrows drawn near sources given the
blue and red-shifted direction of the outflows when known.
The position angle of the outflow from Per-emb-33 is $\sim$275\degr\ and the outflow from L1448NW has a position angle of $\sim$330\degr. 
The outflow from L1448 IRS3A is uncertain \citep{tobin2015b}, but may have a position angle of $\sim$30\degr\ \citep{lee2015}.
Separations written inside the figure are relative to Per-emb-33.
}
\label{L1448N-wide}
\end{figure}

\begin{figure}[!ht]
\begin{center}
\includegraphics[scale=0.8]{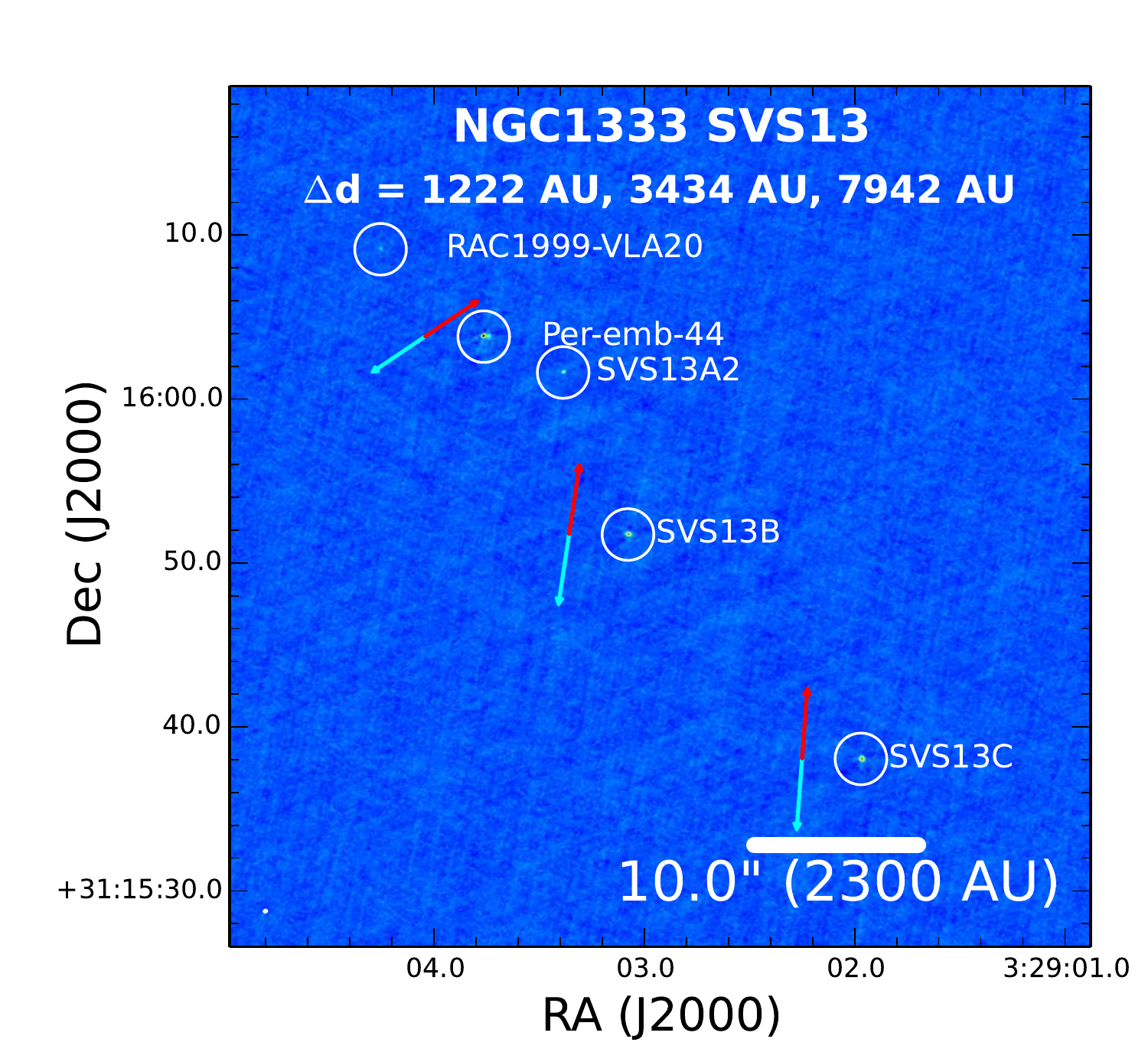}
\end{center}
\caption{NGC 1333 SVS13 region. Per-emb-44 corresponds to SVS13A and the close companion to
Per-emb-44 was first discovered by \citet{anglada2000} with the VLA at 
3.6 cm (VLA4) and later at 7 mm \citep{anglada2004}. 
SVS13A2 was discovered at 3.6 cm by \citet{rodriguez1997} 
and was referred to as VLA3. Thus far, outflows have only been
conclusively detected toward Per-emb-44 with a position 
angle of 120\degr\ \citep{plunkett2013} and SVS13B with a position angle
of $\sim$160\degr\ \citep{bachiller1998,bachiller2000}. SVS13C has a candidate outflow
position angle of $\sim$8\degr\ \citep{plunkett2013}. Separations 
written inside the figure are relative to Per-emb-44. The blue and red arrows drawn near sources given the
blue and red-shifted direction of the known outflows.
}
\label{SVS13-wide}
\end{figure}

\begin{figure}[!ht]
\begin{center}
\includegraphics[scale=0.8]{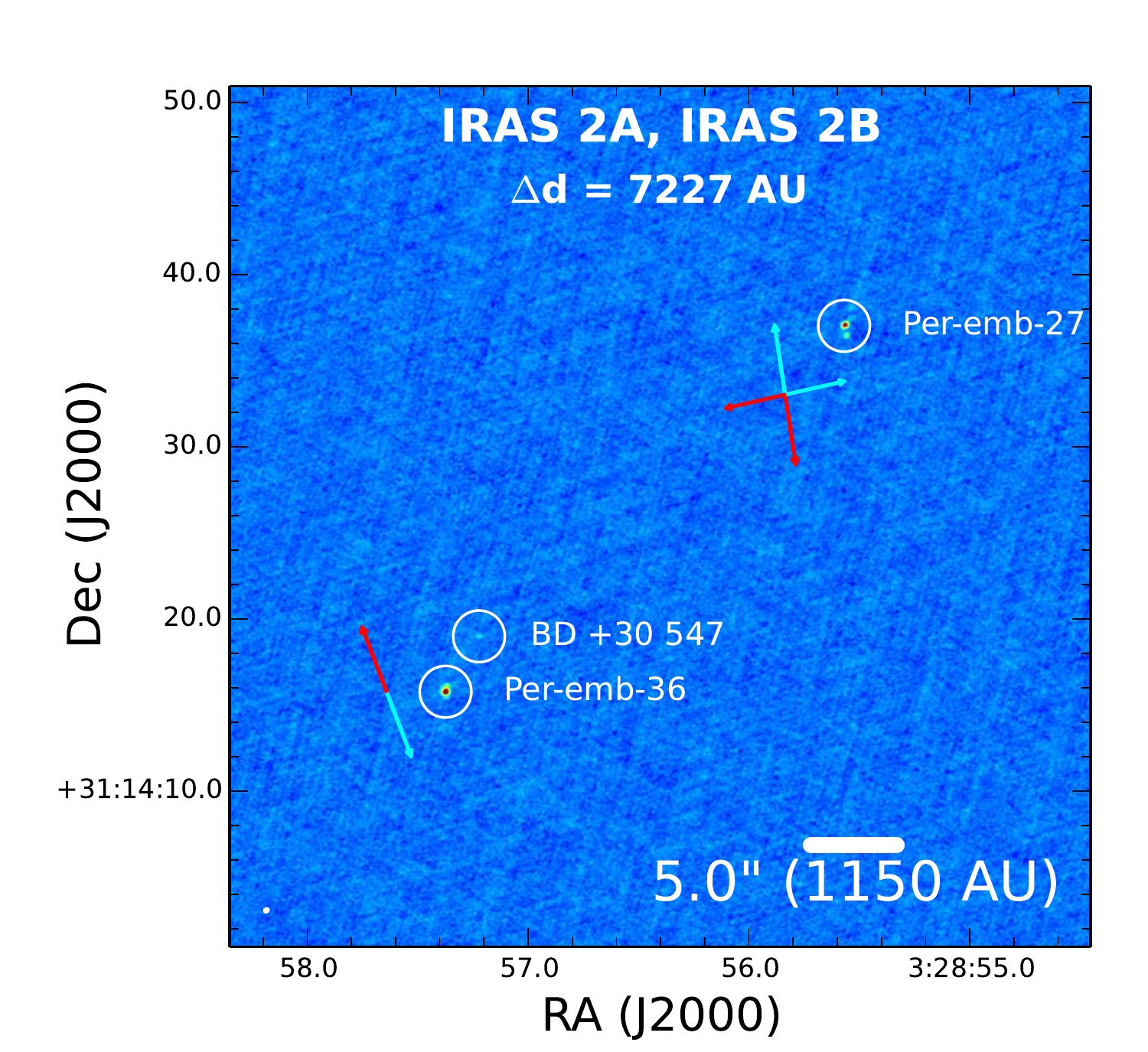}
\end{center}
\caption{NGC 1333 IRAS2 at 9 mm in A+B configuration. IRAS 2B is $\sim$7225 AU 
east of IRAS 2A and is most likely a
Class I protostar. The faint source northwest of IRAS2B is an optically-visible star (BD +30 547)
that appears to be illuminating the near side of the molecular cloud. We do not 
believe that it is physically associated with IRAS2B itself. The blue and red arrows drawn near sources given the
blue and red-shifted direction of the outflows identified by \citet{plunkett2013}. }
\label{IRAS2B-wide}
\end{figure}

\begin{figure}[!ht]
\begin{center}
\includegraphics[scale=0.8]{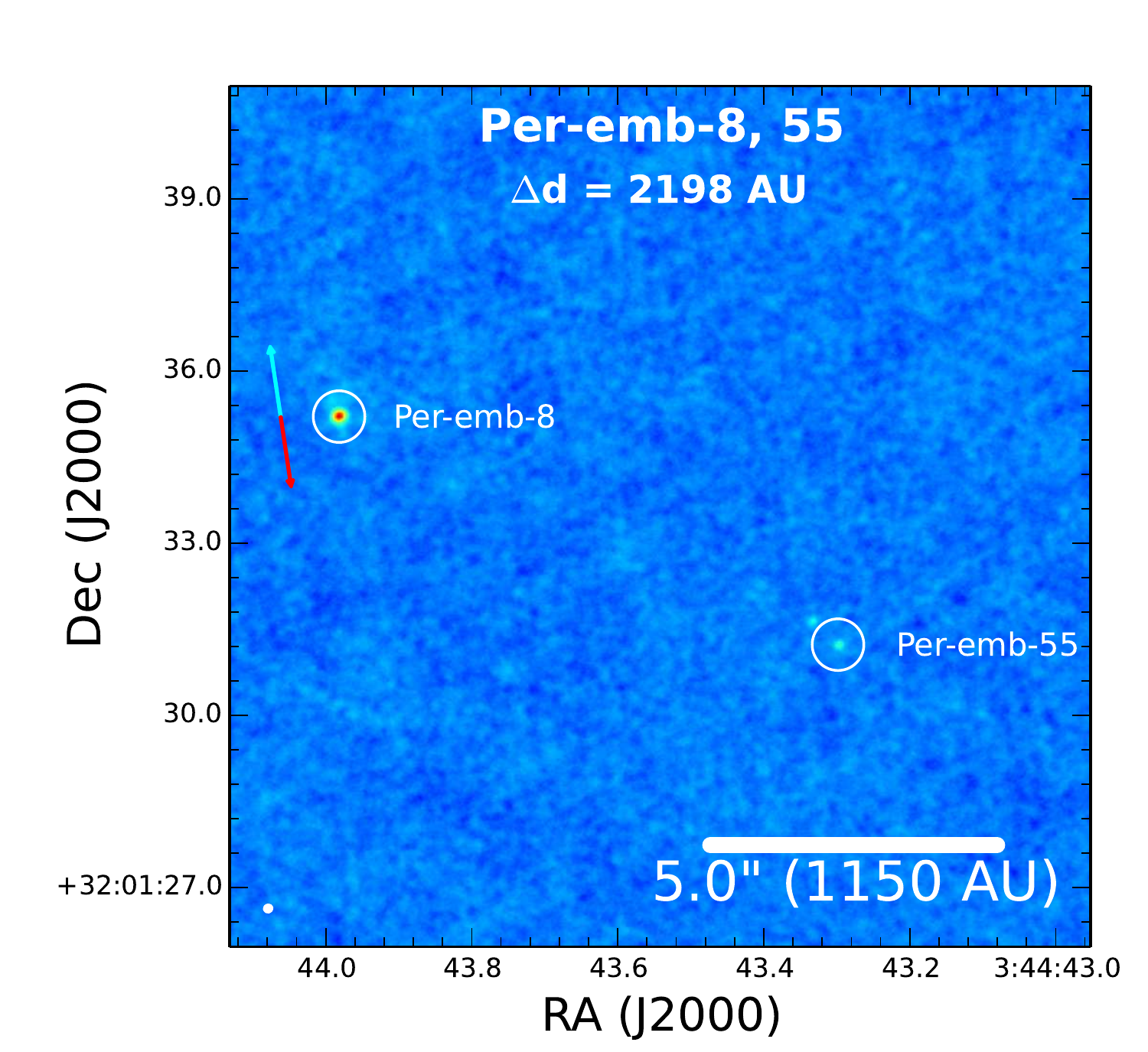}
\end{center}
\caption{Image of Per-emb-8 and Per-emb-55 at 9 mm in A+B configuration. 
Per-emb-8 is a Class 0 protostar, while Per-emb-55 is a Class I protostar with a close
companion (Figure 1) and much brighter at IRAC wavelengths \citep{jorgensen2006}. The relative 
outflow directions from these sources are unknown, but
we estimate that Per-emb-8 has a position angle of $\sim$30\degr\ from 
the IRAC scattered light morphology (denoted by red and blue arrows).}
\label{Per8-wide}
\end{figure}

\begin{figure}[!ht]
\begin{center}
\includegraphics[scale=0.8]{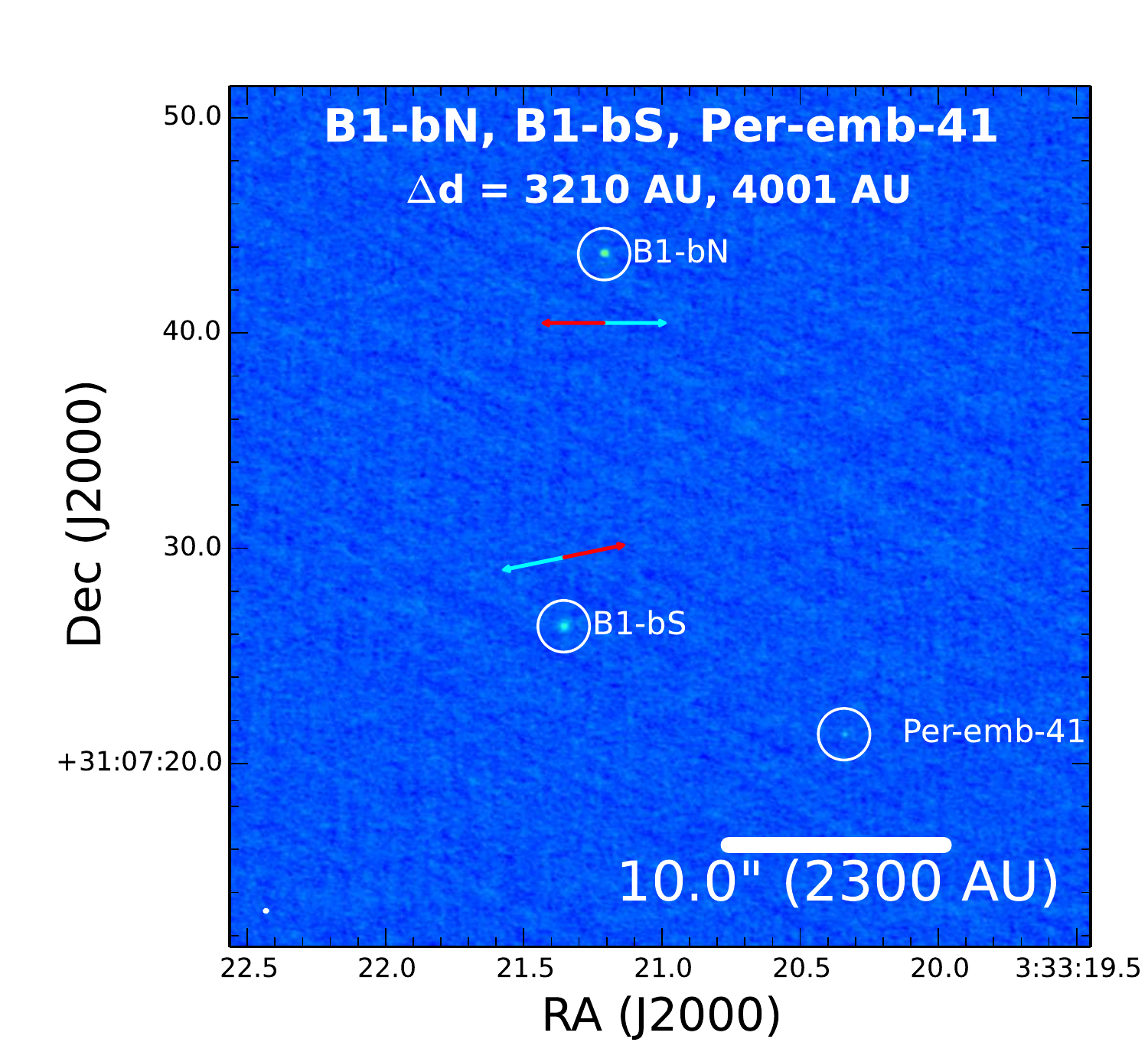}
\end{center}
\caption{B1-b region in A+B configuration at 9 mm. Both B1-bN and B1-bS are deeply embedded systems and candidate FHSC objects,
while Per-emb-41 is a Class I object. \citet{hirano2014} suggests that all three sources are driving misaligned outflows, but their
directions are difficult to discern. \citet{gerin2015} finds that the B1-bN and B1-bS are both driving outflows
in the east-west directions (blue and red arrows). Separations are with respect to B1-bS.}
\label{B1-wide}
\end{figure}

\begin{figure}[!ht]
\begin{center}
\includegraphics[scale=0.8]{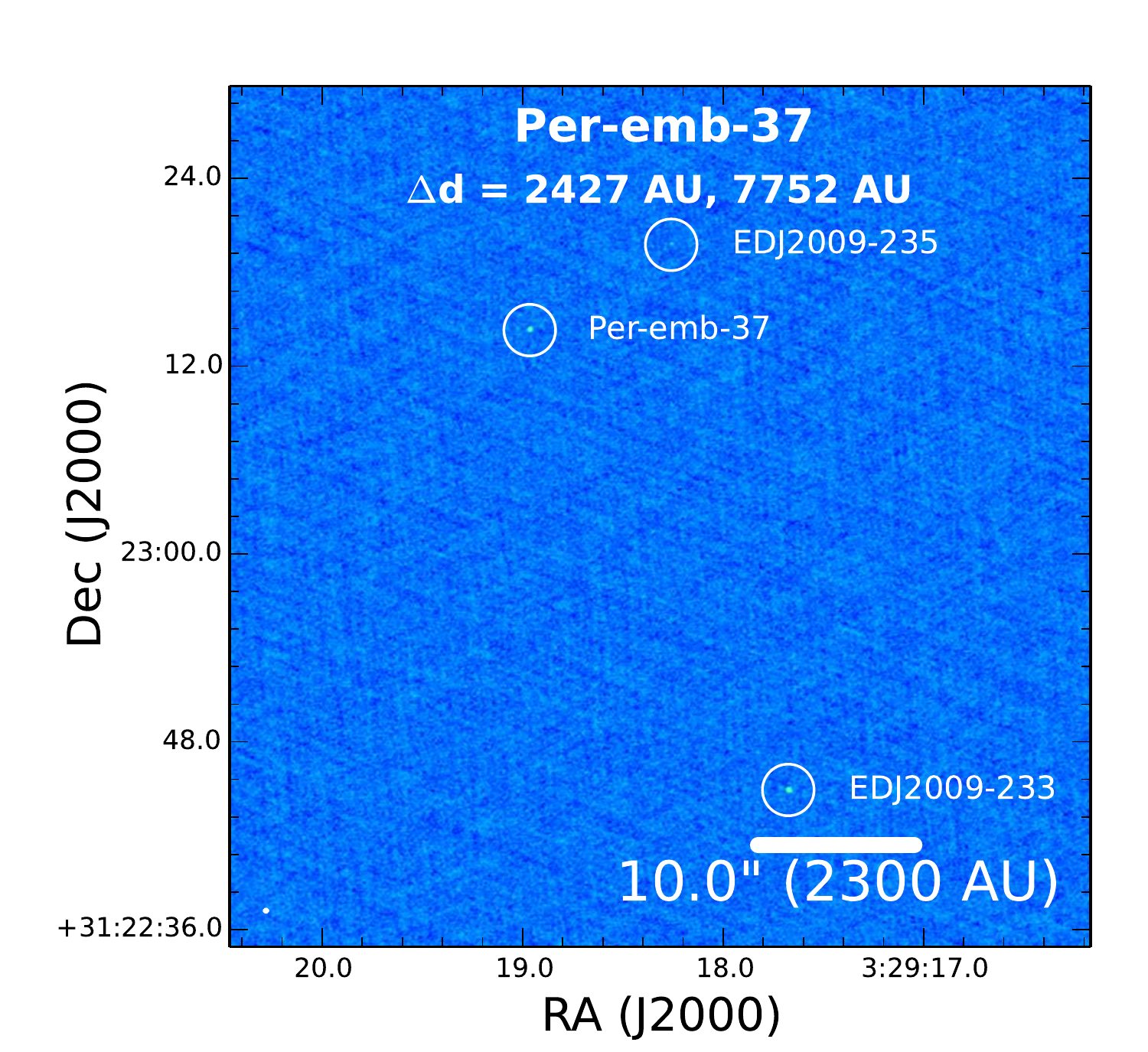}
\end{center}
\caption{Image of the Per-emb-37 region at 9 mm in A+B configuration. Per-emb-37 was associated with a brighter infrared source
located directly north by a few arcseconds in \citet{enoch2009}. However, \herschel\ imaging showed that the 70 \micron\
peak emission was not coincident with the IRAC position, therefore Per-emb-37 must be a deeply embedded Class 0 system, perhaps similar
to the PBRS in Orion \citep{stutz2013}. The source originally associated with Per-emb-37 was not detected in our 9 mm data; however,
we did detect two other sources associated with IRAC detections, but not \herschel\ 70 \micron\ or 100 \micron\ indicating that
they may be more-evolved Class II sources. No outflow has currently been detected toward Per-emb-37. Separations
are relative to Per-emb-37.}
\label{Per37-wide}
\end{figure}

\clearpage

\clearpage
\begin{deluxetable}{lllllllllll}
\tablewidth{0pt}
\rotate
\tabletypesize{\scriptsize}
\tablecaption{Source List}
\tablehead{
  \colhead{Source}  & \colhead{RA} & \colhead{Dec}      & \colhead{Other Names}    &  \colhead{Detected\tablenotemark{a}} & \colhead{Multiple\tablenotemark{b}}   & \colhead{Configuration\tablenotemark{c}}   & \colhead{Class\tablenotemark{d}} & \colhead{L$_{bol}$} & \colhead{T$_{bol}$}\\
                    & \colhead{(J2000)} &  \colhead{(J2000)}     &           &  \colhead{(Y/N)}     & \colhead{(Y/N)}     &  &  & \colhead{(L${\sun}$)} & \colhead{(K)} \\

}
\startdata
Per-emb-1 & 03:43:56.806 & +32:00:50.202 & HH211-MMS & Y & N & A,B & 0 & 1.80 $\pm$ 0.10 & 27.0 $\pm$ 1.0\\ 
Per-emb-2 & 03:32:17.928 & +30:49:47.825 & IRAS 03292+3039 & Y & Y & A,B & 0 & 0.90 $\pm$ 0.07 & 27.0 $\pm$ 1.0\\ 
Per-emb-3 & 03:29:00.575 & +31:12:00.204 &            & Y & N & A,B & 0 & 0.50 $\pm$ 0.06 & 32.0 $\pm$ 2.0\\ 
Per-emb-4 & 03:28:39.101 & +31:06:01.800 &            & N & N & B & 0 & 0.22 $\pm$ 0.03 & 31.0 $\pm$ 3.0\\ 
Per-emb-5 & 03:31:20.939 & +30:45:30.273 & IRAS 03282+3035 & Y & Y & A,B & 0 & 1.30 $\pm$ 0.10 & 32.0 $\pm$ 2.0\\ 
Per-emb-6 & 03:33:14.404 & +31:07:10.715 &            & Y & N & A,B & 0 & 0.30 $\pm$ 0.00 & 52.0 $\pm$ 3.0\\ 
Per-emb-7 & 03:30:32.681 & +30:26:26.480 &            & N & N & B & 0 & 0.15 $\pm$ 0.06 & 37.0 $\pm$ 4.0\\ 
Per-emb-8 & 03:44:43.982 & +32:01:35.210 &            & Y & Y & A,B & 0 & 2.60 $\pm$ 0.50 & 43.0 $\pm$ 6.0\\ 
Per-emb-9 & 03:29:51.832 & +31:39:05.905 & IRAS 03267+3128, Perseus5 & Y & N & A,B & 0 & 0.60 $\pm$ 0.06 & 36.0 $\pm$ 2.0\\ 
Per-emb-10 & 03:33:16.424 & +31:06:52.063 &            & Y & N & A,B & 0 & 0.60 $\pm$ 0.05 & 30.0 $\pm$ 2.0\\ 
Per-emb-11 & 03:43:57.065 & +32:03:04.788 & IC348MMS & Y & Y & A,B & 0 & 1.50 $\pm$ 0.10 & 30.0 $\pm$ 2.0\\ 
Per-emb-12 & 03:29:10.537 & +31:13:30.933 & NGC 1333 IRAS4A & Y & Y & A,B & 0 & 7.00 $\pm$ 0.70 & 29.0 $\pm$ 2.0\\ 
Per-emb-13 & 03:29:12.016 & +31:13:08.031 & NGC 1333 IRAS4B & Y & Y & A,B & 0 & 4.00 $\pm$ 0.30 & 28.0 $\pm$ 1.0\\ 
Per-emb-14 & 03:29:13.548 & +31:13:58.150 & NGC 1333 IRAS4C & Y & N & A,B & 0 & 0.70 $\pm$ 0.08 & 31.0 $\pm$ 2.0\\ 
Per-emb-15 & 03:29:04.055 & +31:14:46.237 & RNO15-FIR & Y & N & A,B & 0 & 0.40 $\pm$ 0.10 & 36.0 $\pm$ 4.0\\ 
Per-emb-16 & 03:43:50.978 & +32:03:24.101 &            & Y & Y & A,B & 0 & 0.40 $\pm$ 0.04 & 39.0 $\pm$ 2.0\\ 
Per-emb-17 & 03:27:39.105 & +30:13:03.068 &            & Y & Y & A,B & 0 & 4.20 $\pm$ 0.10 & 59.0 $\pm$ 11.0\\ 
Per-emb-18 & 03:29:11.258 & +31:18:31.073 & NGC 1333 IRAS7 & Y & Y & A,B & 0 & 2.80 $\pm$ 1.70 & 59.0 $\pm$ 12.0\\ 
Per-emb-19 & 03:29:23.498 & +31:33:29.173 &            & Y & N & A,B & 0/I & 0.36 $\pm$ 0.05 & 60.0 $\pm$ 3.0\\ 
Per-emb-20 & 03:27:43.276 & +30:12:28.781 & L1455-IRS4 & Y & N & A,B & 0/I & 1.40 $\pm$ 0.20 & 65.0 $\pm$ 3.0\\ 
Per-emb-21 & 03:29:10.668 & +31:18:20.191 &            & Y & Y & A,B & 0 & 6.90 $\pm$ 1.90 & 45.0 $\pm$ 12.0\\ 
Per-emb-22 & 03:25:22.409 & +30:45:13.258 & L1448-IRS2 & Y & Y & A,B & 0 & 3.60 $\pm$ 0.50 & 43.0 $\pm$ 2.0\\ 
Per-emb-23 & 03:29:17.211 & +31:27:46.302 & ASR 30 & Y & N & A,B & 0 & 0.80 $\pm$ 0.10 & 42.0 $\pm$ 2.0\\ 
Per-emb-24 & 03:28:45.297 & +31:05:41.693 &            & Y & N & A,B & 0/I & 0.43 $\pm$ 0.01 & 67.0 $\pm$ 10.0\\ 
Per-emb-25 & 03:26:37.511 & +30:15:27.813 &            & Y & N & A,B & 0/I & 1.20 $\pm$ 0.02 & 61.0 $\pm$ 12.0\\ 
Per-emb-26 & 03:25:38.875 & +30:44:05.283 & L1448C, L1448-mm & Y & Y & A,B & 0 & 8.40 $\pm$ 1.50 & 47.0 $\pm$ 7.0\\ 
Per-emb-27 & 03:28:55.569 & +31:14:37.025 & NGC 1333 IRAS2A & Y & Y & A,B & 0/I & 19.00 $\pm$ 0.40 & 69.0 $\pm$ 1.0\\ 
Per-emb-28 & 03:43:51.008 & +32:03:08.042 &            & Y & Y & A,B & 0 & 0.70 $\pm$ 0.08 & 45.0 $\pm$ 2.0\\ 
Per-emb-29 & 03:33:17.877 & +31:09:31.817 & B1-c & Y & N & A,B & 0 & 3.70 $\pm$ 0.40 & 48.0 $\pm$ 1.0\\ 
Per-emb-30 & 03:33:27.303 & +31:07:10.160 &            & Y & N & A,B & 0/I & 1.70 $\pm$ 0.01 & 78.0 $\pm$ 6.0\\ 
Per-emb-31 & 03:28:32.547 & +31:11:05.151 &            & Y & N & A,B & 0/I & 0.16 $\pm$ 0.01 & 80.0 $\pm$ 13.0\\ 
Per-emb-32 & 03:44:02.403 & +32:02:04.734 &            & Y & Y & A,B & 0 & 0.30 $\pm$ 0.10 & 57.0 $\pm$ 10.0\\ 
Per-emb-33 & 03:25:36.379 & +30:45:14.728 & L1448IRS3B, L1448N & Y & Y & A,B & 0 & 8.30 $\pm$ 0.80 & 57.0 $\pm$ 3.0\\ 
Per-emb-34 & 03:30:15.163 & +30:23:49.233 & IRAS 03271+3013 & Y & N & A,B & I & 1.60 $\pm$ 0.10 & 99.0 $\pm$ 13.0\\ 
Per-emb-35 & 03:28:37.090 & +31:13:30.788 & NGC 1333 IRAS1 & Y & Y & A,B & I & 9.10 $\pm$ 0.30 & 103.0 $\pm$ 26.0\\ 
Per-emb-36 & 03:28:57.374 & +31:14:15.772 & NGC 1333 IRAS2B & Y & Y & A,B & I & 5.30 $\pm$ 1.00 & 106.0 $\pm$ 12.0\\ 
Per-emb-37 & 03:29:18.965 & +31:23:14.304 &            & Y & Y & A,B & 0 & 0.50 $\pm$ 0.10 & 22.0 $\pm$ 1.0\\ 
Per-emb-38 & 03:32:29.197 & +31:02:40.759 &            & Y & N & A,B & I & 0.54 $\pm$ 0.01 & 115.0 $\pm$ 21.0\\ 
Per-emb-39 & 03:33:13.781 & +31:20:05.204 &            & N & N & B & I & 0.04 $\pm$ 0.08 & 125.0 $\pm$ 47.0\\ 
Per-emb-40 & 03:33:16.669 & +31:07:54.901 & B1-a & Y & Y & A,B & I & 1.50 $\pm$ 1.00 & 132.0 $\pm$ 25.0\\ 
Per-emb-41 & 03:33:20.341 & +31:07:21.355 & B1-b & Y & Y & A,B & I & 0.17 $\pm$ 0.36 & 157.0 $\pm$ 72.0\\ 
Per-emb-42 & 03:25:39.135 & +30:43:57.909 & L1448C-S & Y & Y & A,B & I & 0.68 $\pm$ 0.85 & 163.0 $\pm$ 51.0\\ 
Per-emb-43 & 03:42:02.160 & +31:48:02.081 &            & N & N & B & I & 0.07 $\pm$ 0.06 & 176.0 $\pm$ 42.0\\ 
Per-emb-44 & 03:29:03.764 & +31:16:03.808 & SVS13A & Y & Y & A,B & 0/I & 32.50 $\pm$ 7.10 & 188.0 $\pm$ 9.0\\ 
Per-emb-45 & 03:33:09.569 & +31:05:31.193 &            & N & N & B & I & 0.05 $\pm$ 0.06 & 197.0 $\pm$ 93.0\\ 
Per-emb-46 & 03:28:00.415 & +30:08:01.013 &            & Y & N & A,B & I & 0.30 $\pm$ 0.07 & 221.0 $\pm$ 7.0\\ 
Per-emb-47 & 03:28:34.507 & +31:00:50.990 & IRAS 03254+3050 & Y & N & A,B & I & 1.20 $\pm$ 0.10 & 230.0 $\pm$ 17.0\\ 
Per-emb-48 & 03:27:38.268 & +30:13:58.448 & L1455-FIR2 & Y & Y & A,B & I & 0.87 $\pm$ 0.04 & 238.0 $\pm$ 14.0\\ 
Per-emb-49 & 03:29:12.957 & +31:18:14.307 &            & Y & Y & A,B & I & 1.10 $\pm$ 0.70 & 239.0 $\pm$ 68.0\\ 
Per-emb-50 & 03:29:07.768 & +31:21:57.128 &            & Y & N & A,B & I & 23.20 $\pm$ 3.00 & 128.0 $\pm$ 23.0\\ 
Per-emb-51 & 03:28:34.536 & +31:07:05.520 &            & N & N & B & I & 0.07 $\pm$ 0.10 & 263.0 $\pm$ 115.0\\ 
Per-emb-52 & 03:28:39.699 & +31:17:31.882 &            & Y & N & A,B & I & 0.16 $\pm$ 0.21 & 278.0 $\pm$ 119.0\\ 
Per-emb-53 & 03:47:41.591 & +32:51:43.672 & B5-IRS1 & Y & N & A,B & I & 4.70 $\pm$ 0.90 & 287.0 $\pm$ 8.0\\ 
Per-emb-54 & 03:29:01.549 & +31:20:20.497 & NGC 1333 IRAS6 & Y & N & A,B & I & 16.80 $\pm$ 2.60 & 131.0 $\pm$ 63.0\\ 
Per-emb-55 & 03:44:43.298 & +32:01:31.236 & IRAS 03415+3152 & Y & Y & A,B & I & 1.80 $\pm$ 0.80 & 309.0 $\pm$ 64.0\\ 
Per-emb-56 & 03:47:05.450 & +32:43:08.240 & IRAS 03439+3233 & Y & N & A,B & I & 0.54 $\pm$ 0.09 & 312.0 $\pm$ 1.0\\ 
Per-emb-57 & 03:29:03.331 & +31:23:14.573 &            & Y & N & A,B & I & 0.09 $\pm$ 0.45 & 313.0 $\pm$ 200.0\\ 
Per-emb-58 & 03:28:58.422 & +31:22:17.481 &            & Y & N & A,B & I & 0.63 $\pm$ 0.47 & 322.0 $\pm$ 88.0\\ 
Per-emb-59 & 03:28:35.040 & +30:20:09.884 &            & N & N & B & I & 0.04 $\pm$ 0.06 & 341.0 $\pm$ 179.0\\ 
Per-emb-60 & 03:29:20.068 & +31:24:07.488 &            & N & N & B & I & 0.28 $\pm$ 1.05 & 363.0 $\pm$ 240.0\\ 
Per-emb-61 & 03:44:21.357 & +31:59:32.514 &            & Y & N & A,B & I & 0.24 $\pm$ 0.16 & 371.0 $\pm$ 107.0\\ 
Per-emb-62 & 03:44:12.977 & +32:01:35.419 &            & Y & N & A,B & I & 1.80 $\pm$ 0.40 & 378.0 $\pm$ 29.0\\ 
Per-emb-63 & 03:28:43.271 & +31:17:32.931 &            & Y & N & A,B & I & 1.90 $\pm$ 0.40 & 436.0 $\pm$ 9.0\\ 
Per-emb-64 & 03:33:12.852 & +31:21:24.020 &            & Y & N & A,B & I & 3.20 $\pm$ 0.60 & 438.0 $\pm$ 8.0\\ 
Per-emb-65 & 03:28:56.316 & +31:22:27.798 &            & Y & N & A,B & I & 0.16 $\pm$ 0.16 & 440.0 $\pm$ 191.0\\ 
Per-emb-66 & 03:43:45.150 & +32:03:58.608 &            & Y & N & A,B & I & 0.69 $\pm$ 0.22 & 542.0 $\pm$ 110.0\\ 
Per-bolo-58 & 03:29:25.464 & +31:28:14.880 & Per-Bolo-58 & N & N & B & 0 & 0.05 $\pm$ 0.50 & 15.0 $\pm$ -99.0\\ 
Per-bolo-45 & 03:29:07.700 & +31:17:16.800 & Per-Bolo-45 & N & N & B & 0 & 0.05 $\pm$ 0.05 & 15.0 $\pm$ -99.0\\ 
L1451-MMS & 03:25:10.245 & +30:23:55.059 & L1451-MMS & Y & N & A,B & 0 & 0.05 $\pm$ 0.05 & 15.0 $\pm$ -99.0\\ 
L1448IRS2E & 03:25:25.660 & +30:44:56.695 & L1448IRS2E & N & N & B & 0 & 0.05 $\pm$ 0.05 & 15.0 $\pm$ -99.0\\ 
B1-bN & 03:33:21.209 & +31:07:43.665 & B1-bN & Y & Y & A,B & 0 & 0.32 $\pm$ 0.10 & 14.7 $\pm$ 1.0\\ 
B1-bS & 03:33:21.355 & +31:07:26.372 & B1-bS & Y & Y & A,B & 0 & 0.70 $\pm$ 0.07 & 17.7 $\pm$ 1.0\\ 
L1448IRS1 & 03:25:09.449 & +30:46:21.933 & L1448IRS1 & Y & Y & A,B & I & -99.90 $\pm$ -99.90 & -99.9 $\pm$ -99.9\\ 
L1448NW & 03:25:35.671 & +30:45:34.193 & L1448IRS3C & Y & Y & A,B & 0 & 1.40 $\pm$ 0.10 & 22.0 $\pm$ 1.0\\ 
L1448IRS3A & 03:25:36.499 & +30:45:21.880 &            & Y & Y & A,B & I & 9.20 $\pm$ 1.30 & 47.0 $\pm$ 2.0\\ 
SVS13C & 03:29:01.970 & +31:15:38.053 & SVS13C & Y & Y & A,B & 0 & 1.50 $\pm$ 0.20 & 21.0 $\pm$ 1.0\\ 
SVS13B & 03:29:03.078 & +31:15:51.740 & SVS13B & Y & Y & A,B & 0 & 1.00 $\pm$ 1.00 & 20.0 $\pm$ 20.0\\ 
IRAS 03363+3207 & 03:39:25.547 & +32:17:07.089 &            & Y & N & A,B & I? & -99.90 $\pm$ -99.90 & -99.9 $\pm$ -99.9\\ 
EDJ2009-161 & 03:28:51.480 & +30:45:00.360 &            & N & N & B & II & 1.10 $\pm$ 0.10 & 1200.0 $\pm$ -99.0\\ 
EDJ2009-263 & 03:30:27.161 & +30:28:29.613 &            & Y & N & A,B & Flat & 0.23 $\pm$ 0.10 & 340.0 $\pm$ -99.0\\ 
EDJ2009-285 & 03:32:46.942 & +30:59:17.797 &            & N & N & B & II & 0.47 $\pm$ 0.10 & 920.0 $\pm$ -99.0\\ 
IRAS 03295+3050 & 03:32:34.066 & +31:00:55.621 & EDJ2009-282 & Y & N & A,B & II & 0.25 $\pm$ 0.10 & 1300.0 $\pm$ -99.0\\ 
L1455IRS2 & 03:27:47.690 & +30:12:04.314 & EDJ2009-133 & Y & N & A,B & Flat & 2.50 $\pm$ 0.10 & 740.0 $\pm$ -99.0\\ 
EDJ2009-333 & 03:42:55.772 & +31:58:44.386 &            & N & N & B & II & 1.30 $\pm$ 0.10 & 980.0 $\pm$ -99.0\\ 
EDJ2009-385 & 03:44:18.168 & +32:04:56.907 &            & Y & N & A,B & II & 0.38 $\pm$ 0.10 & 1200.0 $\pm$ -99.0\\ 
EDJ2009-366 & 03:43:59.651 & +32:01:54.008 &            & Y & N & A,B & II & 1.80 $\pm$ 0.10 & 620.0 $\pm$ -99.0\\ 
EDJ2009-269 & 03:30:44.014 & +30:32:46.812 &            & Y & Y & A,B & II & 1.30 $\pm$ 0.10 & 1200.0 $\pm$ -99.0\\ 
EDJ2009-268 & 03:30:38.231 & +30:32:11.666 &            & N & N & A & Flat & 0.04 $\pm$ 0.10 & 320.0 $\pm$ -99.0\\ 
EDJ2009-183 & 03:28:59.297 & +31:15:48.410 & ASR 106 & Y & Y & A,B & Flat & 3.20 $\pm$ 0.10 & 100.0 $\pm$ -99.0\\ 
EDJ2009-164 & 03:28:53.961 & +31:18:09.349 & ASR 40 & N & N & A,B & II & 0.14 $\pm$ 0.10 & 890.0 $\pm$ -99.0\\ 
EDJ2009-156 & 03:28:51.029 & +31:18:18.409 & ASR 122 & Y & Y & A,B & II & 0.04 $\pm$ 0.10 & 740.0 $\pm$ -99.0\\ 
EDJ2009-172 & 03:28:56.650 & +31:18:35.449 & ASR 120 & Y & N & A,B & II & 0.39 $\pm$ 0.10 & 1100.0 $\pm$ -99.0\\ 
IRAS4B' & 03:29:12.842 & +31:13:06.893 & NGC 1333 IRAS4B' & Y & Y & A,B & 0 & 0.10 $\pm$ 0.10 & 15.0 $\pm$ -99.0\\ 
RAC1999-VLA20 & 03:29:04.255 & +31:16:09.138 &            & Y & N & A,B & Extragalactic & -99.00 $\pm$ -99.00 & -99.0 $\pm$ -99.0\\ 
EDJ2009-233 & 03:29:17.675 & +31:22:44.922 &            & Y & Y & A,B & II & 1.40 $\pm$ 0.10 & 1300.0 $\pm$ -99.0\\ 
EDJ2009-235 & 03:29:18.259 & +31:23:19.758 &            & Y & Y & A,B & II & 0.54 $\pm$ 0.10 & -99.0 $\pm$ -99.0\\ 
SVS3 & 03:29:10.420 & +31:21:59.072 &            & Y & N & A,B & II & 0.54 $\pm$ 0.10 & -99.0 $\pm$ -99.0\\ 
EDJ2009-173 & 03:28:56.964 & +31:16:22.199 & ASR 118, SVS15 & Y & N & A,B & II & 0.10 $\pm$ 0.10 & 1100.0 $\pm$ -99.0\\ 
RAC97-VLA1 & 03:28:57.657 & +31:15:31.450 &            & Y & N & A,B & Extragalactic & -99.00 $\pm$ -99.00 & -99.0 $\pm$ -99.0\\ 
\enddata
\tablecomments{The sources denoted Per-emb-XX correspond to the sources published in \citet{enoch2009},
EDJ2009 sources are from \citet{evans2009}, ASR refers to \citet{aspin1994}, SVS refers to \citep{strom1976} and RAC1999-VLA20 is from \citet{rodriguez1999}.
IRAS4B' was observed in the same field Per-emb-13, EDJ2009-233/235 were observed in the same field, and RAC1999-VLA20 was observed
in the same field as Per-emb-44.}
\tablenotetext{a}{Refers to whether or not the associated source was detected above 5$\sigma$ in the observations (at either 8 mm or 1 cm).} 
\tablenotetext{b}{Denotes whether or not the object comprises a multiple system or is a component of a multiple system based on our data.} 
\tablenotetext{c}{VLA Configuration(s) in which observations were taken.} 
\tablenotetext{d}{Young Stellar Object Classification based on $T_{bol}$ from \citet{enoch2009}, 
\citet{sadavoy2014}, and \citet{young2015}.} 

\end{deluxetable}

\begin{deluxetable}{llllcccccccc}
\tablewidth{0pt}
\rotate
\tabletypesize{\scriptsize}
\tablecaption{Source Properties}
\tablehead{
  \colhead{Source}  & \colhead{RA} & \colhead{Dec}      & 
\colhead{F$_{\nu,\rm int}$} & \colhead{F$_{\nu,\rm peak}$} & \colhead{RMS} &
\colhead{F$_{\nu,\rm int}$} & \colhead{F$_{\nu,\rm peak}$} & \colhead{RMS} & \colhead{Sp. Index} &\colhead{Sp. Index} \\
                    &  &     &   \colhead{(8 mm)}&   \colhead{(8 mm)}&  \colhead{(8 mm)}&  \colhead{(10 mm)}&  \colhead{(10 mm)} &  \colhead{(10 mm)}            \\
                    & \colhead{(J2000)} &  \colhead{(J2000)}    &  \colhead{(mJy)}&   \colhead{(mJy)}&  \colhead{(mJy)}&  \colhead{(mJy)}&  \colhead{(mJy)} &  \colhead{(mJy)} & \colhead{(Int.)} & \colhead{(Peak)}           \\

}
\startdata
Per-emb-1 & 03:43:56.806 & +32:00:50.202 & 0.845 $\pm$ 0.022 & 0.673 & 0.010 & 0.567 $\pm$ 0.019 & 0.464 & 0.009 & 1.653 $\pm$ 0.031 & 1.54 $\pm$ 0.010\\ 
Per-emb-2\\ 
Per-emb-2-comb & 03:32:17.928 & +30:49:47.825 & 2.482 $\pm$ 0.103 & 0.504 & 0.012 & 1.251 $\pm$ 0.068 & 0.355 & 0.010 & 2.843 $\pm$ 0.080 & 1.46 $\pm$ 0.025\\ 
Per-emb-2-A & 03:32:17.932 & +30:49:47.705 & 0.296 $\pm$ 0.040 & 0.210 & 0.017 & 0.185 $\pm$ 0.039 & 0.131 & 0.017 & 1.95 $\pm$ 5.663 & 1.97 $\pm$ 2.139\\ 
Per-emb-2-B & 03:32:17.927 & +30:49:47.753 & 0.137 $\pm$ 0.016 & 0.124 & 0.017 & 0.124 $\pm$ 0.014 & 0.133 & 0.017 & 0.42 $\pm$ 2.468 & -0.28 $\pm$ 3.260\\ 
Per-emb-3 & 03:29:00.575 & +31:12:00.204 & 0.402 $\pm$ 0.022 & 0.363 & 0.012 & 0.291 $\pm$ 0.020 & 0.250 & 0.010 & 1.337 $\pm$ 0.138 & 1.55 $\pm$ 0.047\\ 
Per-emb-4 & 03:28:39.101 & +31:06:01.800 & $<$0.035 $\pm$ 0.012 & $<$0.035 & 0.012 & $<$0.030 $\pm$ 0.010 & $<$0.030 & 0.010 & -99.9 $\pm$ -99.9 & -99.9 $\pm$ -99.9\\ 
Per-emb-5\\ 
Per-emb-5-comb & 03:31:20.939 & +30:45:30.273 & 1.357 $\pm$ 0.035 & 0.814 & 0.013 & 0.702 $\pm$ 0.026 & 0.482 & 0.011 & 2.739 $\pm$ 0.035 & 2.17 $\pm$ 0.013\\ 
Per-emb-5-A & 03:31:20.942 & +30:45:30.263 & 0.449 $\pm$ 0.030 & 0.162 & 0.013 & 0.211 $\pm$ 0.018 & 0.112 & 0.011 & 3.13 $\pm$ 1.101 & 1.55 $\pm$ 1.511\\ 
Per-emb-5-B & 03:31:20.935 & +30:45:30.247 & 0.177 $\pm$ 0.012 & 0.122 & 0.013 & 0.135 $\pm$ 0.010 & 0.117 & 0.011 & 1.12 $\pm$ 0.934 & 0.18 $\pm$ 1.857\\ 
Per-emb-6 & 03:33:14.404 & +31:07:10.715 & 0.280 $\pm$ 0.023 & 0.270 & 0.012 & 0.211 $\pm$ 0.019 & 0.195 & 0.010 & 1.173 $\pm$ 0.259 & 1.35 $\pm$ 0.086\\ 
Per-emb-7 & 03:30:32.681 & +30:26:26.480 & $<$0.037 $\pm$ 0.013 & $<$0.037 & 0.013 & $<$0.032 $\pm$ 0.011 & $<$0.032 & 0.011 & -99.9 $\pm$ -99.9 & -99.9 $\pm$ -99.9\\ 
Per-emb-8 & 03:44:43.982 & +32:01:35.210 & 1.115 $\pm$ 0.029 & 0.835 & 0.013 & 0.739 $\pm$ 0.026 & 0.574 & 0.011 & 1.710 $\pm$ 0.033 & 1.56 $\pm$ 0.011\\ 
Per-emb-9 & 03:29:51.832 & +31:39:05.905 & 0.078 $\pm$ 0.029 & 0.056 & 0.013 & 0.041 $\pm$ 0.017 & 0.051 & 0.011 & 2.677 $\pm$ 5.624 & 0.38 $\pm$ 1.700\\ 
Per-emb-10 & 03:33:16.424 & +31:06:52.063 & 0.369 $\pm$ 0.023 & 0.356 & 0.012 & 0.265 $\pm$ 0.020 & 0.258 & 0.011 & 1.380 $\pm$ 0.159 & 1.33 $\pm$ 0.053\\ 
Per-emb-11\\ 
Per-emb-11-A & 03:43:57.065 & +32:03:04.788 & 1.064 $\pm$ 0.032 & 0.717 & 0.013 & 0.583 $\pm$ 0.025 & 0.435 & 0.011 & 2.50 $\pm$ 0.255 & 2.07 $\pm$ 0.087\\ 
Per-emb-11-B & 03:43:56.881 & +32:03:02.977 & 0.097 $\pm$ 0.021 & 0.105 & 0.013 & 0.079 $\pm$ 0.023 & 0.072 & 0.011 & 0.85 $\pm$ 11.726 & 1.60 $\pm$ 3.494\\ 
Per-emb-11-C & 03:43:57.688 & +32:03:09.975 & 0.086 $\pm$ 0.028 & 0.073 & 0.013 & 0.040 $\pm$ 0.014 & 0.050 & 0.011 & 3.12 $\pm$ 21.052 & 1.62 $\pm$ 7.219\\ 
Per-emb-12\\ 
Per-emb-12-A & 03:29:10.537 & +31:13:30.925 & 9.858 $\pm$ 0.123 & 1.810 & 0.019 & 5.025 $\pm$ 0.076 & 1.292 & 0.014 & 2.80 $\pm$ 0.035 & 1.40 $\pm$ 0.021\\ 
Per-emb-12-B & 03:29:10.427 & +31:13:32.099 & 0.805 $\pm$ 0.057 & 0.459 & 0.019 & 0.534 $\pm$ 0.034 & 0.386 & 0.014 & 1.70 $\pm$ 0.818 & 0.72 $\pm$ 0.281\\ 
Per-emb-13 & 03:29:12.016 & +31:13:08.031 & 3.480 $\pm$ 0.057 & 1.300 & 0.013 & 1.830 $\pm$ 0.041 & 0.823 & 0.011 & 2.668 $\pm$ 0.013 & 1.90 $\pm$ 0.005\\ 
Per-emb-14 & 03:29:13.548 & +31:13:58.150 & 0.865 $\pm$ 0.031 & 0.622 & 0.013 & 0.507 $\pm$ 0.025 & 0.402 & 0.012 & 2.222 $\pm$ 0.063 & 1.81 $\pm$ 0.023\\ 
Per-emb-15 & 03:29:04.055 & +31:14:46.237 & 0.067 $\pm$ 0.025 & 0.059 & 0.012 & 0.070 $\pm$ 0.024 & 0.052 & 0.011 & -0.220 $\pm$ 4.560 & 0.48 $\pm$ 1.486\\ 
Per-emb-16 & 03:43:50.978 & +32:03:24.101 & 0.066 $\pm$ 0.026 & 0.048 & 0.012 & 0.069 $\pm$ 0.027 & 0.055 & 0.011 & -0.200 $\pm$ 5.215 & -0.50 $\pm$ 1.798\\ 
Per-emb-17\\ 
Per-emb-17-A & 03:27:39.104 & +30:13:03.078 & 0.524 $\pm$ 0.019 & 0.468 & 0.010 & 0.364 $\pm$ 0.018 & 0.318 & 0.009 & 1.50 $\pm$ 0.340 & 1.60 $\pm$ 0.113\\ 
Per-emb-17-B & 03:27:39.115 & +30:13:02.840 & 0.105 $\pm$ 0.013 & 0.081 & 0.010 & 0.056 $\pm$ 0.007 & 0.055 & 0.009 & 2.56 $\pm$ 2.940 & 1.59 $\pm$ 3.755\\ 
Per-emb-18\\ 
Per-emb-18-comb & 03:29:11.258 & +31:18:31.073 & 0.733 $\pm$ 0.035 & 0.458 & 0.012 & 0.573 $\pm$ 0.028 & 0.404 & 0.011 & 1.019 $\pm$ 0.079 & 0.52 $\pm$ 0.025\\ 
Per-emb-18-A & 03:29:11.255 & +31:18:31.059 & 0.161 $\pm$ 0.035 & 0.175 & 0.021 & 0.191 $\pm$ 0.041 & 0.150 & 0.019 & -0.73 $\pm$ 8.456 & 0.64 $\pm$ 2.762\\ 
Per-emb-18-B & 03:29:11.261 & +31:18:31.073 & 0.280 $\pm$ 0.054 & 0.169 & 0.021 & 0.161 $\pm$ 0.037 & 0.131 & 0.019 & 2.29 $\pm$ 8.141 & 1.07 $\pm$ 3.313\\ 
Per-emb-19 & 03:29:23.498 & +31:33:29.173 & 0.260 $\pm$ 0.026 & 0.210 & 0.012 & 0.201 $\pm$ 0.022 & 0.179 & 0.011 & 1.064 $\pm$ 0.376 & 0.66 $\pm$ 0.122\\ 
Per-emb-20 & 03:27:43.276 & +30:12:28.781 & 0.172 $\pm$ 0.023 & 0.165 & 0.013 & 0.187 $\pm$ 0.021 & 0.168 & 0.011 & -0.345 $\pm$ 0.522 & -0.09 $\pm$ 0.178\\ 
Per-emb-21 & 03:29:10.668 & +31:18:20.191 & 0.555 $\pm$ 0.024 & 0.496 & 0.012 & 0.300 $\pm$ 0.020 & 0.292 & 0.011 & 2.553 $\pm$ 0.106 & 2.20 $\pm$ 0.035\\ 
Per-emb-22\\ 
Per-emb-22-A & 03:25:22.410 & +30:45:13.254 & 0.532 $\pm$ 0.022 & 0.454 & 0.011 & 0.448 $\pm$ 0.020 & 0.377 & 0.010 & 0.72 $\pm$ 0.338 & 0.77 $\pm$ 0.115\\ 
Per-emb-22-B & 03:25:22.352 & +30:45:13.151 & 0.191 $\pm$ 0.024 & 0.151 & 0.011 & 0.136 $\pm$ 0.019 & 0.122 & 0.010 & 1.39 $\pm$ 3.115 & 0.90 $\pm$ 1.068\\ 
Per-emb-23 & 03:29:17.211 & +31:27:46.302 & 0.157 $\pm$ 0.022 & 0.160 & 0.013 & 0.090 $\pm$ 0.016 & 0.104 & 0.011 & 2.311 $\pm$ 0.904 & 1.79 $\pm$ 0.312\\ 
Per-emb-24 & 03:28:45.297 & +31:05:41.693 & 0.083 $\pm$ 0.014 & 0.079 & 0.010 & 0.052 $\pm$ 0.007 & 0.064 & 0.009 & 1.977 $\pm$ 0.830 & 0.90 $\pm$ 0.661\\ 
Per-emb-25 & 03:26:37.511 & +30:15:27.813 & 0.749 $\pm$ 0.023 & 0.600 & 0.011 & 0.491 $\pm$ 0.019 & 0.413 & 0.009 & 1.750 $\pm$ 0.041 & 1.55 $\pm$ 0.014\\ 
Per-emb-26 & 03:25:38.875 & +30:44:05.283 & 2.065 $\pm$ 0.024 & 1.813 & 0.012 & 1.294 $\pm$ 0.020 & 1.186 & 0.011 & 1.940 $\pm$ 0.006 & 1.76 $\pm$ 0.002\\ 
Per-emb-27\\ 
Per-emb-27-A & 03:28:55.569 & +31:14:37.022 & 1.711 $\pm$ 0.019 & 1.510 & 0.010 & 1.135 $\pm$ 0.016 & 1.018 & 0.008 & 1.70 $\pm$ 0.029 & 1.63 $\pm$ 0.010\\ 
Per-emb-27-B & 03:28:55.563 & +31:14:36.408 & 0.324 $\pm$ 0.020 & 0.286 & 0.010 & 0.215 $\pm$ 0.015 & 0.209 & 0.008 & 1.70 $\pm$ 0.749 & 1.30 $\pm$ 0.248\\ 
Per-emb-28 & 03:43:51.008 & +32:03:08.042 & 0.103 $\pm$ 0.037 & 0.058 & 0.013 & 0.053 $\pm$ 0.025 & 0.050 & 0.012 & 2.729 $\pm$ 5.967 & 0.66 $\pm$ 1.892\\ 
Per-emb-29 & 03:33:17.877 & +31:09:31.817 & 0.565 $\pm$ 0.022 & 0.486 & 0.011 & 0.365 $\pm$ 0.019 & 0.314 & 0.009 & 1.819 $\pm$ 0.073 & 1.82 $\pm$ 0.024\\ 
Per-emb-30 & 03:33:27.303 & +31:07:10.160 & 0.968 $\pm$ 0.022 & 0.817 & 0.011 & 0.701 $\pm$ 0.019 & 0.600 & 0.009 & 1.336 $\pm$ 0.022 & 1.28 $\pm$ 0.007\\ 
Per-emb-31 & 03:28:32.547 & +31:11:05.151 & 0.049 $\pm$ 0.024 & 0.038 & 0.013 & 0.075 $\pm$ 0.023 & 0.054 & 0.012 & -1.812 $\pm$ 5.933 & -1.38 $\pm$ 2.718\\ 
Per-emb-32\\ 
Per-emb-32-A & 03:44:02.403 & +32:02:04.751 & 0.055 $\pm$ 0.022 & 0.045 & 0.011 & 0.028 $\pm$ 0.011 & 0.035 & 0.008 & 2.85 $\pm$ 27.780 & 1.09 $\pm$ 9.967\\ 
Per-emb-32-B & 03:44:02.633 & +32:01:59.431 & 0.096 $\pm$ 0.033 & 0.052 & 0.011 & 0.029 $\pm$ 0.010 & 0.041 & 0.008 & 4.92 $\pm$ 22.100 & 0.98 $\pm$ 7.468\\ 
Per-emb-33\\ 
Per-emb-33-A & 03:25:36.380 & +30:45:14.723 & 0.820 $\pm$ 0.034 & 0.453 & 0.011 & 0.484 $\pm$ 0.031 & 0.257 & 0.010 & 2.19 $\pm$ 0.539 & 2.35 $\pm$ 0.183\\ 
Per-emb-33-B & 03:25:36.312 & +30:45:15.154 & 0.424 $\pm$ 0.034 & 0.258 & 0.011 & 0.317 $\pm$ 0.030 & 0.181 & 0.010 & 1.20 $\pm$ 1.427 & 1.48 $\pm$ 0.427\\ 
Per-emb-33-C & 03:25:36.321 & +30:45:14.914 & 0.312 $\pm$ 0.028 & 0.212 & 0.011 & 0.247 $\pm$ 0.020 & 0.208 & 0.010 & 0.96 $\pm$ 1.320 & 0.09 $\pm$ 0.443\\ 
Per-emb-34 & 03:30:15.163 & +30:23:49.233 & 0.225 $\pm$ 0.021 & 0.241 & 0.013 & 0.175 $\pm$ 0.019 & 0.184 & 0.012 & 1.039 $\pm$ 0.365 & 1.12 $\pm$ 0.119\\ 
Per-emb-35\\ 
Per-emb-35-A & 03:28:37.091 & +31:13:30.788 & 0.430 $\pm$ 0.020 & 0.340 & 0.009 & 0.307 $\pm$ 0.017 & 0.258 & 0.009 & 1.40 $\pm$ 0.480 & 1.15 $\pm$ 0.168\\ 
Per-emb-35-B & 03:28:37.219 & +31:13:31.751 & 0.256 $\pm$ 0.018 & 0.231 & 0.009 & 0.201 $\pm$ 0.018 & 0.158 & 0.009 & 1.01 $\pm$ 1.200 & 1.58 $\pm$ 0.412\\ 
Per-emb-36\\ 
Per-emb-36-A & 03:28:57.374 & +31:14:15.765 & 1.893 $\pm$ 0.026 & 1.545 & 0.013 & 1.190 $\pm$ 0.023 & 0.988 & 0.011 & 1.93 $\pm$ 0.053 & 1.86 $\pm$ 0.018\\ 
Per-emb-36-B & 03:28:57.370 & +31:14:16.073 & 0.308 $\pm$ 0.021 & 0.226 & 0.013 & 0.180 $\pm$ 0.011 & 0.167 & 0.011 & 2.22 $\pm$ 0.767 & 1.26 $\pm$ 0.711\\ 
Per-emb-37 & 03:29:18.965 & +31:23:14.304 & 0.256 $\pm$ 0.024 & 0.232 & 0.012 & 0.172 $\pm$ 0.021 & 0.167 & 0.012 & 1.645 $\pm$ 0.411 & 1.37 $\pm$ 0.136\\ 
Per-emb-38 & 03:32:29.197 & +31:02:40.759 & 0.199 $\pm$ 0.028 & 0.154 & 0.013 & 0.127 $\pm$ 0.023 & 0.112 & 0.011 & 1.873 $\pm$ 0.926 & 1.32 $\pm$ 0.296\\ 
Per-emb-39 & 03:33:13.781 & +31:20:05.204 & $<$0.038 $\pm$ 0.013 & $<$0.038 & 0.013 & $<$0.030 $\pm$ 0.010 & $<$0.030 & 0.010 & -99.9 $\pm$ -99.9 & -99.9 $\pm$ -99.9\\ 
Per-emb-40\\ 
Per-emb-40-A & 03:33:16.669 & +31:07:54.902 & 0.417 $\pm$ 0.018 & 0.356 & 0.009 & 0.312 $\pm$ 0.014 & 0.280 & 0.007 & 1.21 $\pm$ 0.352 & 1.00 $\pm$ 0.118\\ 
Per-emb-40-B & 03:33:16.680 & +31:07:55.269 & 0.054 $\pm$ 0.019 & 0.042 & 0.009 & 0.062 $\pm$ 0.015 & 0.052 & 0.007 & -0.56 $\pm$ 16.636 & -0.86 $\pm$ 5.822\\ 
Per-emb-41 & 03:33:20.341 & +31:07:21.355 & 0.101 $\pm$ 0.024 & 0.102 & 0.013 & 0.081 $\pm$ 0.025 & 0.057 & 0.011 & 0.929 $\pm$ 2.633 & 2.40 $\pm$ 0.982\\ 
Per-emb-42 & 03:25:39.135 & +30:43:57.909 & 0.273 $\pm$ 0.024 & 0.245 & 0.012 & 0.182 $\pm$ 0.019 & 0.180 & 0.011 & 1.677 $\pm$ 0.321 & 1.28 $\pm$ 0.103\\ 
Per-emb-43 & 03:42:02.160 & +31:48:02.081 & $<$0.038 $\pm$ 0.013 & $<$0.038 & 0.013 & $<$0.030 $\pm$ 0.010 & $<$0.030 & 0.010 & -99.9 $\pm$ -99.9 & -99.9 $\pm$ -99.9\\ 
Per-emb-44\\ 
Per-emb-44-A & 03:29:03.766 & +31:16:03.810 & 1.763 $\pm$ 0.022 & 1.501 & 0.011 & 1.115 $\pm$ 0.019 & 0.988 & 0.010 & 1.90 $\pm$ 0.042 & 1.74 $\pm$ 0.015\\ 
Per-emb-44-B & 03:29:03.743 & +31:16:03.790 & 0.508 $\pm$ 0.034 & 0.320 & 0.011 & 0.375 $\pm$ 0.026 & 0.265 & 0.010 & 1.25 $\pm$ 0.841 & 0.78 $\pm$ 0.244\\ 
Per-emb-45 & 03:33:09.569 & +31:05:31.193 & $<$0.038 $\pm$ 0.013 & $<$0.038 & 0.013 & $<$0.030 $\pm$ 0.010 & $<$0.030 & 0.010 & -99.9 $\pm$ -99.9 & -99.9 $\pm$ -99.9\\ 
Per-emb-46 & 03:28:00.415 & +30:08:01.013 & 0.141 $\pm$ 0.042 & 0.089 & 0.015 & 0.118 $\pm$ 0.034 & 0.082 & 0.013 & 0.749 $\pm$ 2.925 & 0.31 $\pm$ 0.897\\ 
Per-emb-47 & 03:28:34.507 & +31:00:50.990 & 0.304 $\pm$ 0.029 & 0.268 & 0.014 & 0.193 $\pm$ 0.022 & 0.177 & 0.012 & 1.874 $\pm$ 0.385 & 1.72 $\pm$ 0.124\\ 
Per-emb-48\\ 
Per-emb-48-A & 03:27:38.277 & +30:13:58.559 & 0.085 $\pm$ 0.022 & 0.077 & 0.011 & 0.072 $\pm$ 0.024 & 0.053 & 0.010 & 0.66 $\pm$ 16.174 & 1.52 $\pm$ 5.375\\ 
Per-emb-48-B & 03:27:38.258 & +30:13:58.320 & 0.064 $\pm$ 0.023 & 0.052 & 0.011 & 0.072 $\pm$ 0.019 & 0.063 & 0.010 & -0.51 $\pm$ 18.333 & -0.79 $\pm$ 6.809\\ 
Per-emb-49\\ 
Per-emb-49-A & 03:29:12.953 & +31:18:14.289 & 0.347 $\pm$ 0.020 & 0.286 & 0.010 & 0.305 $\pm$ 0.021 & 0.217 & 0.009 & 0.53 $\pm$ 0.730 & 1.14 $\pm$ 0.258\\ 
Per-emb-49-B & 03:29:12.976 & +31:18:14.397 & 0.095 $\pm$ 0.017 & 0.089 & 0.010 & 0.117 $\pm$ 0.015 & 0.113 & 0.009 & -0.84 $\pm$ 4.512 & -1.03 $\pm$ 1.685\\ 
Per-emb-50 & 03:29:07.768 & +31:21:57.128 & 1.620 $\pm$ 0.031 & 1.274 & 0.014 & 1.063 $\pm$ 0.024 & 0.918 & 0.012 & 1.749 $\pm$ 0.015 & 1.36 $\pm$ 0.005\\ 
Per-emb-51 & 03:28:34.536 & +31:07:05.520 & $<$0.039 $\pm$ 0.013 & $<$0.039 & 0.013 & $<$0.035 $\pm$ 0.012 & $<$0.035 & 0.012 & -99.9 $\pm$ -99.9 & -99.9 $\pm$ -99.9\\ 
Per-emb-52 & 03:28:39.699 & +31:17:31.882 & 0.083 $\pm$ 0.020 & 0.090 & 0.013 & 0.067 $\pm$ 0.029 & 0.036 & 0.011 & 0.854 $\pm$ 4.118 & 3.81 $\pm$ 2.116\\ 
Per-emb-53 & 03:47:41.591 & +32:51:43.672 & 0.241 $\pm$ 0.034 & 0.182 & 0.015 & 0.190 $\pm$ 0.027 & 0.146 & 0.012 & 0.976 $\pm$ 0.687 & 0.93 $\pm$ 0.219\\ 
Per-emb-54 & 03:29:01.549 & +31:20:20.497 & 0.356 $\pm$ 0.031 & 0.292 & 0.015 & 0.294 $\pm$ 0.021 & 0.292 & 0.012 & 0.791 $\pm$ 0.214 & 0.01 $\pm$ 0.073\\ 
Per-emb-55\\ 
Per-emb-55-A & 03:44:43.298 & +32:01:31.223 & 0.140 $\pm$ 0.012 & 0.114 & 0.010 & 0.058 $\pm$ 0.007 & 0.056 & 0.009 & 3.65 $\pm$ 2.066 & 2.96 $\pm$ 2.977\\ 
Per-emb-55-B & 03:44:43.334 & +32:01:31.636 & 0.078 $\pm$ 0.012 & 0.063 & 0.010 & 0.063 $\pm$ 0.007 & 0.060 & 0.009 & 0.90 $\pm$ 3.377 & 0.17 $\pm$ 4.241\\ 
Per-emb-56 & 03:47:05.450 & +32:43:08.240 & 0.137 $\pm$ 0.023 & 0.134 & 0.014 & 0.054 $\pm$ 0.018 & 0.066 & 0.013 & 3.824 $\pm$ 2.297 & 2.91 $\pm$ 0.859\\ 
Per-emb-57 & 03:29:03.331 & +31:23:14.573 & 0.176 $\pm$ 0.025 & 0.164 & 0.014 & 0.159 $\pm$ 0.032 & 0.119 & 0.013 & 0.429 $\pm$ 1.045 & 1.34 $\pm$ 0.335\\ 
Per-emb-58 & 03:28:58.422 & +31:22:17.481 & 0.071 $\pm$ 0.018 & 0.093 & 0.014 & 0.093 $\pm$ 0.037 & 0.055 & 0.013 & -1.120 $\pm$ 3.830 & 2.16 $\pm$ 1.368\\ 
Per-emb-59 & 03:28:35.040 & +30:20:09.884 & $<$0.042 $\pm$ 0.014 & $<$0.042 & 0.014 & $<$0.037 $\pm$ 0.012 & $<$0.037 & 0.012 & -99.9 $\pm$ -99.9 & -99.9 $\pm$ -99.9\\ 
Per-emb-60 & 03:29:20.068 & +31:24:07.488 & $<$0.041 $\pm$ 0.014 & $<$0.041 & 0.014 & $<$0.036 $\pm$ 0.012 & $<$0.036 & 0.012 & -99.9 $\pm$ -99.9 & -99.9 $\pm$ -99.9\\ 
Per-emb-61 & 03:44:21.357 & +31:59:32.514 & 0.116 $\pm$ 0.027 & 0.101 & 0.014 & 0.056 $\pm$ 0.024 & 0.051 & 0.012 & 2.997 $\pm$ 3.970 & 2.87 $\pm$ 1.297\\ 
Per-emb-62 & 03:44:12.977 & +32:01:35.419 & 0.735 $\pm$ 0.029 & 0.616 & 0.015 & 0.488 $\pm$ 0.024 & 0.417 & 0.012 & 1.703 $\pm$ 0.070 & 1.62 $\pm$ 0.024\\ 
Per-emb-63 & 03:28:43.271 & +31:17:32.931 & 0.378 $\pm$ 0.029 & 0.306 & 0.014 & 0.270 $\pm$ 0.020 & 0.264 & 0.011 & 1.387 $\pm$ 0.193 & 0.61 $\pm$ 0.069\\ 
Per-emb-64 & 03:33:12.852 & +31:21:24.020 & 1.029 $\pm$ 0.029 & 0.891 & 0.015 & 0.810 $\pm$ 0.024 & 0.687 & 0.012 & 0.989 $\pm$ 0.029 & 1.08 $\pm$ 0.010\\ 
Per-emb-65 & 03:28:56.316 & +31:22:27.798 & 0.141 $\pm$ 0.038 & 0.094 & 0.014 & 0.108 $\pm$ 0.032 & 0.070 & 0.013 & 1.105 $\pm$ 2.766 & 1.22 $\pm$ 0.963\\ 
Per-emb-66 & 03:43:45.150 & +32:03:58.608 & $<$0.041 $\pm$ 0.014 & $<$0.041 & 0.014 & $<$0.035 $\pm$ 0.012 & $<$0.035 & 0.012 & -99.9 $\pm$ -99.9 & -99.9 $\pm$ -99.9\\ 
Per-bolo-58 & 03:29:25.464 & +31:28:14.880 & $<$0.037 $\pm$ 0.013 & $<$0.037 & 0.013 & $<$0.036 $\pm$ 0.012 & $<$0.036 & 0.012 & -99.9 $\pm$ -99.9 & -99.9 $\pm$ -99.9\\ 
Per-bolo-45 & 03:29:07.700 & +31:17:16.800 & $<$0.037 $\pm$ 0.012 & $<$0.037 & 0.012 & $<$0.035 $\pm$ 0.012 & $<$0.035 & 0.012 & -99.9 $\pm$ -99.9 & -99.9 $\pm$ -99.9\\ 
L1451-MMS & 03:25:10.245 & +30:23:55.059 & 0.285 $\pm$ 0.023 & 0.276 & 0.013 & 0.173 $\pm$ 0.019 & 0.170 & 0.011 & 2.078 $\pm$ 0.314 & 2.02 $\pm$ 0.106\\ 
L1448IRS2E & 03:25:25.660 & +30:44:56.695 & $<$0.045 $\pm$ 0.015 & $<$0.045 & 0.015 & $<$0.040 $\pm$ 0.013 & $<$0.040 & 0.013 & -99.9 $\pm$ -99.9 & -99.9 $\pm$ -99.9\\ 
B1-bN & 03:33:21.209 & +31:07:43.665 & 1.251 $\pm$ 0.036 & 0.756 & 0.013 & 0.680 $\pm$ 0.028 & 0.458 & 0.011 & 2.529 $\pm$ 0.044 & 2.08 $\pm$ 0.016\\ 
B1-bS & 03:33:21.355 & +31:07:26.372 & 0.906 $\pm$ 0.060 & 0.286 & 0.013 & 0.551 $\pm$ 0.048 & 0.188 & 0.011 & 2.058 $\pm$ 0.206 & 1.74 $\pm$ 0.101\\ 
L1448IRS1\\ 
L1448IRS1-A & 03:25:09.449 & +30:46:21.924 & 0.871 $\pm$ 0.035 & 0.547 & 0.013 & 0.614 $\pm$ 0.031 & 0.398 & 0.012 & 1.45 $\pm$ 0.374 & 1.32 $\pm$ 0.131\\ 
L1448IRS1-B & 03:25:09.413 & +30:46:20.625 & 0.116 $\pm$ 0.022 & 0.080 & 0.013 & 0.063 $\pm$ 0.013 & 0.058 & 0.012 & 2.56 $\pm$ 7.440 & 1.33 $\pm$ 6.129\\ 
L1448NW\\ 
L1448NW-A & 03:25:35.669 & +30:45:34.110 & 0.695 $\pm$ 0.022 & 0.609 & 0.012 & 0.486 $\pm$ 0.020 & 0.429 & 0.010 & 1.49 $\pm$ 0.243 & 1.45 $\pm$ 0.085\\ 
L1448NW-B & 03:25:35.673 & +30:45:34.357 & 0.475 $\pm$ 0.026 & 0.370 & 0.012 & 0.323 $\pm$ 0.023 & 0.254 & 0.010 & 1.61 $\pm$ 0.729 & 1.56 $\pm$ 0.237\\ 
L1448IRS3A & 03:25:36.499 & +30:45:21.880 & 1.067 $\pm$ 0.028 & 0.929 & 0.014 & 0.981 $\pm$ 0.026 & 0.829 & 0.013 & 0.345 $\pm$ 0.024 & 0.47 $\pm$ 0.008\\ 
SVS13C & 03:29:01.970 & +31:15:38.053 & 2.401 $\pm$ 0.031 & 1.717 & 0.013 & 2.134 $\pm$ 0.026 & 1.682 & 0.012 & 0.489 $\pm$ 0.005 & 0.09 $\pm$ 0.002\\ 
SVS13B & 03:29:03.078 & +31:15:51.740 & 1.555 $\pm$ 0.033 & 1.116 & 0.014 & 0.905 $\pm$ 0.027 & 0.723 & 0.013 & 2.245 $\pm$ 0.024 & 1.81 $\pm$ 0.008\\ 
IRAS 03363+3207 & 03:39:25.547 & +32:17:07.089 & 0.898 $\pm$ 0.026 & 0.792 & 0.014 & 0.519 $\pm$ 0.020 & 0.507 & 0.011 & 2.278 $\pm$ 0.041 & 1.85 $\pm$ 0.014\\ 
EDJ2009-161 & 03:28:51.480 & +30:45:00.360 & $<$0.041 $\pm$ 0.014 & $<$0.041 & 0.014 & $<$0.035 $\pm$ 0.012 & $<$0.035 & 0.012 & -99.9 $\pm$ -99.9 & -99.9 $\pm$ -99.9\\ 
EDJ2009-263 & 03:30:27.161 & +30:28:29.613 & 0.088 $\pm$ 0.017 & 0.080 & 0.013 & 0.057 $\pm$ 0.009 & 0.062 & 0.012 & 1.835 $\pm$ 1.058 & 1.01 $\pm$ 1.095\\ 
EDJ2009-285 & 03:32:46.942 & +30:59:17.797 & $<$0.042 $\pm$ 0.014 & $<$0.042 & 0.014 & $<$0.039 $\pm$ 0.013 & $<$0.039 & 0.013 & -99.9 $\pm$ -99.9 & -99.9 $\pm$ -99.9\\ 
IRAS 03295+3050 & 03:32:34.066 & +31:00:55.621 & 0.146 $\pm$ 0.021 & 0.146 & 0.013 & 0.097 $\pm$ 0.020 & 0.095 & 0.011 & 1.717 $\pm$ 1.058 & 1.79 $\pm$ 0.385\\ 
L1455IRS2 & 03:27:47.690 & +30:12:04.314 & 0.044 $\pm$ 0.013 & 0.067 & 0.013 & 0.088 $\pm$ 0.028 & 0.059 & 0.011 & -2.907 $\pm$ 3.303 & 0.53 $\pm$ 1.299\\ 
EDJ2009-333 & 03:42:55.772 & +31:58:44.386 & $<$0.040 $\pm$ 0.013 & $<$0.040 & 0.013 & $<$0.034 $\pm$ 0.011 & $<$0.034 & 0.011 & -99.9 $\pm$ -99.9 & -99.9 $\pm$ -99.9\\ 
EDJ2009-385 & 03:44:18.168 & +32:04:56.907 & 0.247 $\pm$ 0.061 & 0.095 & 0.013 & 0.130 $\pm$ 0.028 & 0.077 & 0.011 & 2.647 $\pm$ 1.881 & 0.87 $\pm$ 0.677\\ 
EDJ2009-366 & 03:43:59.651 & +32:01:54.008 & 0.152 $\pm$ 0.021 & 0.162 & 0.013 & 0.137 $\pm$ 0.021 & 0.128 & 0.011 & 0.454 $\pm$ 0.742 & 0.96 $\pm$ 0.244\\ 
EDJ2009-269\\ 
EDJ2009-269-A & 03:30:44.014 & +30:32:46.811 & 0.258 $\pm$ 0.035 & 0.163 & 0.013 & 0.224 $\pm$ 0.043 & 0.115 & 0.013 & 0.58 $\pm$ 5.130 & 1.46 $\pm$ 1.704\\ 
EDJ2009-269-B & 03:30:43.978 & +30:32:46.576 & 0.210 $\pm$ 0.039 & 0.127 & 0.013 & 0.121 $\pm$ 0.026 & 0.096 & 0.013 & 2.29 $\pm$ 7.484 & 1.15 $\pm$ 2.559\\ 
EDJ2009-268 & 03:30:38.231 & +30:32:11.666 & $<$0.040 $\pm$ 0.013 & $<$0.040 & 0.013 & $<$0.034 $\pm$ 0.011 & $<$0.034 & 0.011 & -99.9 $\pm$ -99.9 & -99.9 $\pm$ -99.9\\ 
EDJ2009-183\\ 
EDJ2009-183-A & 03:28:59.296 & +31:15:48.405 & 0.242 $\pm$ 0.048 & 0.099 & 0.011 & 0.122 $\pm$ 0.018 & 0.114 & 0.011 & 2.84 $\pm$ 5.665 & -0.58 $\pm$ 2.012\\ 
EDJ2009-183-B & 03:28:59.373 & +31:15:48.399 & 0.047 $\pm$ 0.025 & 0.045 & 0.011 & 0.056 $\pm$ 0.023 & 0.037 & 0.011 & -0.75 $\pm$ 40.680 & 0.86 $\pm$ 13.324\\ 
EDJ2009-164 & 03:28:53.961 & +31:18:09.349 & $<$0.055 $\pm$ 0.018 & $<$0.055 & 0.018 & $<$0.052 $\pm$ 0.017 & $<$0.052 & 0.017 & -99.9 $\pm$ -99.9 & -99.9 $\pm$ -99.9\\ 
EDJ2009-156\\ 
EDJ2009-156-A & 03:28:51.031 & +31:18:18.380 & 0.193 $\pm$ 0.032 & 0.139 & 0.014 & 0.107 $\pm$ 0.026 & 0.080 & 0.011 & 2.44 $\pm$ 7.858 & 2.32 $\pm$ 2.511\\ 
EDJ2009-156-B & 03:28:51.111 & +31:18:15.448 & 0.130 $\pm$ 0.026 & 0.107 & 0.014 & 0.079 $\pm$ 0.024 & 0.075 & 0.011 & 2.07 $\pm$ 11.924 & 1.48 $\pm$ 3.341\\ 
EDJ2009-172 & 03:28:56.650 & +31:18:35.449 & 0.329 $\pm$ 0.028 & 0.224 & 0.010 & 0.277 $\pm$ 0.027 & 0.173 & 0.009 & 0.716 $\pm$ 0.293 & 1.07 $\pm$ 0.077\\ 
SVS13A2 & 03:29:03.386 & +31:16:01.622 & 0.336 $\pm$ 0.027 & 0.317 & 0.014 & 0.270 $\pm$ 0.023 & 0.263 & 0.013 & 0.905 $\pm$ 0.239 & 0.78 $\pm$ 0.076\\ 
IRAS4B' & 03:29:12.842 & +31:13:06.893 & 1.556 $\pm$ 0.060 & 0.536 & 0.013 & 0.813 $\pm$ 0.044 & 0.319 & 0.011 & 2.697 $\pm$ 0.077 & 2.15 $\pm$ 0.030\\ 
BD +30 547 & 03:28:57.222 & +31:14:18.991 & 0.075 $\pm$ 0.027 & 0.064 & 0.014 & 0.103 $\pm$ 0.029 & 0.073 & 0.012 & -1.328 $\pm$ 3.539 & -0.56 $\pm$ 1.283\\ 
RAC1999-VLA20 & 03:29:04.255 & +31:16:09.138 & 0.161 $\pm$ 0.039 & 0.100 & 0.014 & 0.168 $\pm$ 0.037 & 0.109 & 0.013 & -0.171 $\pm$ 1.874 & -0.35 $\pm$ 0.592\\ 
EDJ2009-233 & 03:29:17.675 & +31:22:44.922 & 0.454 $\pm$ 0.027 & 0.348 & 0.012 & 0.375 $\pm$ 0.027 & 0.281 & 0.012 & 0.785 $\pm$ 0.148 & 0.89 $\pm$ 0.052\\ 
EDJ2009-235 & 03:29:18.259 & +31:23:19.758 & 0.040 $\pm$ 0.012 & 0.069 & 0.012 & 0.080 $\pm$ 0.030 & 0.061 & 0.012 & -2.820 $\pm$ 3.906 & 0.52 $\pm$ 1.218\\ 
SVS3 & 03:29:10.420 & +31:21:59.072 & 0.210 $\pm$ 0.029 & 0.177 & 0.014 & 0.270 $\pm$ 0.041 & 0.130 & 0.012 & -1.047 $\pm$ 0.725 & 1.27 $\pm$ 0.257\\ 
EDJ2009-173 & 03:28:56.964 & +31:16:22.199 & 0.232 $\pm$ 0.018 & 0.254 & 0.014 & 0.176 $\pm$ 0.009 & 0.176 & 0.012 & 1.147 $\pm$ 0.153 & 1.53 $\pm$ 0.132\\ 
RAC97-VLA1 & 03:28:57.657 & +31:15:31.450 & 0.122 $\pm$ 0.031 & 0.098 & 0.014 & 0.155 $\pm$ 0.020 & 0.158 & 0.012 & -1.011 $\pm$ 1.425 & -1.98 $\pm$ 0.438\\ 
\enddata

\end{deluxetable}

\begin{deluxetable}{lllll}
\tablewidth{0pt}
\tabletypesize{\scriptsize}
\tablecaption{Class 0 Multiple Systems}
\tablehead{
  \colhead{Source}  &   \colhead{Separation} &\colhead{Separation} & \colhead{Flux Difference} & \colhead {Type}  \\
  \colhead{}  &   \colhead{(\arcsec)} &\colhead{(AU)} & \colhead{(Log [F$_1$/F$_2$])} &   \\

}
\startdata
Per-emb-2 & 0.080 $\pm$ 0.006 & 18.4 $\pm$ 1.3 & 0.17 $\pm$ 0.15 & Class 0\\
Per-emb-18 & 0.085 $\pm$ 0.004 & 19.6 $\pm$ 0.9 & -0.02 $\pm$ 0.12 & Class 0\\
Per-emb-5 & 0.097 $\pm$ 0.006 & 22.3 $\pm$ 1.4 & 0.26 $\pm$ 0.07 & Class 0\\
L1448NW & 0.251 $\pm$ 0.004 & 57.7 $\pm$ 1.0 & 0.17 $\pm$ 0.03 & Class 0\\
Per-emb-33 & 0.264 $\pm$ 0.008 & 60.7 $\pm$ 1.8 & 0.12 $\pm$ 0.05\\
Per-emb-17 & 0.278 $\pm$ 0.014 & 63.9 $\pm$ 3.1 & 0.75 $\pm$ 0.08 & Class 0\\
Per-emb-44 & 0.300 $\pm$ 0.003 & 69.0 $\pm$ 0.7 & 0.48 $\pm$ 0.02 & Class 0\\
Per-emb-27 & 0.620 $\pm$ 0.003 & 142.6 $\pm$ 0.7 & 0.75 $\pm$ 0.03 & Class 0\\
Per-emb-22 & 0.751 $\pm$ 0.004 & 172.8 $\pm$ 1.0 & 0.51 $\pm$ 0.05 & Class 0\\
Per-emb-33 & 0.795 $\pm$ 0.004 & 182.8 $\pm$ 1.0 & 0.33 $\pm$ 0.04 & Class 0\\
Per-emb-12 & 1.830 $\pm$ 0.002 & 420.8 $\pm$ 0.4 & 1.04 $\pm$ 0.02 & Class 0\\
Per-emb-11 & 2.951 $\pm$ 0.008 & 678.8 $\pm$ 1.7 & 0.93 $\pm$ 0.08 & Class 0\\
Per-emb-44+SVS13A2 & 5.314 $\pm$ 0.004 & 1222.2 $\pm$ 0.9 & 1.78 $\pm$ 0.03 & Class 0/I\\
Per-emb-32 & 6.066 $\pm$ 0.022 & 1395.3 $\pm$ 5.0 & -0.16 $\pm$ 0.27 & Class 0/I\\
Per-emb-33+L1448IRS3A & 7.317 $\pm$ 0.004 & 1683.0 $\pm$ 0.9 & -0.24 $\pm$ 0.02 & Class 0-Class I\\
Per-emb-26+Per-emb-42 & 8.104 $\pm$ 0.005 & 1864.0 $\pm$ 1.2 & 1.97 $\pm$ 0.03 & Class 0-Class I\\
Per-emb-11 & 9.469 $\pm$ 0.025 & 2177.8 $\pm$ 5.8 & 0.76 $\pm$ 0.13 & Class 0\\
Per-emb-8+Per-emb-55 & 9.557 $\pm$ 0.013 & 2198.2 $\pm$ 2.9 & 2.38 $\pm$ 0.10 & Class 0-Class I\\
Per-emb-37+EDJ2009+235 & 10.556 $\pm$ 0.009 & 2427.8 $\pm$ 2.2 & 1.53 $\pm$ 0.11 & Class 0-Class II\\
Per-emb-13+IRAS4B' & 10.654 $\pm$ 0.005 & 2450.4 $\pm$ 1.2 & 0.81 $\pm$ 0.02 & Class 0-Class 0\\
Per-emb-21+Per-emb-18 & 13.252 $\pm$ 0.004 & 3048.0 $\pm$ 1.0 & -0.44 $\pm$ 0.02 & Class 0-Class 0\\
B1-bS+Per-emb-41 & 13.957 $\pm$ 0.014 & 3210.1 $\pm$ 3.2 & 2.08 $\pm$ 0.09 & Class 0-Class 0/I\\
Per-emb-44+SVS13B & 14.932 $\pm$ 0.002 & 3434.4 $\pm$ 0.5 & 0.40 $\pm$ 0.01 & Class 0/I-Class 0\\
Per-emb-16+Per-emb-28 & 16.063 $\pm$ 0.037 & 3694.5 $\pm$ 8.5 & -0.05 $\pm$ 0.17 & Class 0-Class 0\\
B1-bN+B1-bS & 17.395 $\pm$ 0.009 & 4000.8 $\pm$ 2.0 & 0.31 $\pm$ 0.03 & Class 0-Class 0\\
Per-emb-33+L1448NW & 21.503 $\pm$ 0.004 & 4945.6 $\pm$ 1.0 & -0.23 $\pm$ 0.02 & Class 0-Class 0\\
Per-emb-18+Per-emb-49 & 27.474 $\pm$ 0.007 & 6319.1 $\pm$ 1.5 & 0.33 $\pm$ 0.03 & Class 0-Class I\\
Per-emb-12+Per-emb-13 & 29.739 $\pm$ 0.002 & 6840.0 $\pm$ 0.5 & 1.04 $\pm$ 0.01 & Class 0-Class 0\\
Per-emb-36+Per-emb-27 & 31.420 $\pm$ 0.001 & 7226.6 $\pm$ 0.3 & 0.15 $\pm$ 0.01 & Class 0-Class I\\
Per-emb-6+Per-emb-10 & 31.947 $\pm$ 0.005 & 7347.9 $\pm$ 1.2 & -0.28 $\pm$ 0.03 & Class 0-Class 0\\
Per-emb-37+EDJ2009+233 & 33.704 $\pm$ 0.006 & 7752.0 $\pm$ 1.3 & -0.62 $\pm$ 0.04 & Class 0-Class II\\
Per-emb-44+SVS13C & 34.528 $\pm$ 0.001 & 7941.5 $\pm$ 0.3 & -0.21 $\pm$ 0.01 & Class 0/I-Class 0\\
Per-emb-32+EDJ2009+366 & 36.605 $\pm$ 0.010 & 8419.2 $\pm$ 2.4 & -1.31 $\pm$ 0.11 & Class 0/I-Class II\\
\enddata
\tablecomments{This table includes Class 0 + Class 0, Class 0 + Class I, and Class 0 + Class II multiple systems. The flux difference
is calculated in the 9 mm band.}
\end{deluxetable}

\begin{deluxetable}{llll}
\tablewidth{0pt}
\tabletypesize{\scriptsize}
\tablecaption{Class I Multiple Systems}
\tablehead{
  \colhead{Source}  &   \colhead{Separation} &\colhead{Separation} & \colhead{Flux Differencepe}  \\
  \colhead{}  &   \colhead{(\arcsec)} & \colhead{(AU)} & \colhead{(Log [F$_1$/F$_2$])} \\

}
\startdata
Per-emb-36 & 0.311 $\pm$ 0.005 & 71.6 $\pm$ 1.2 & 0.80 $\pm$ 0.02\\
Per-emb-49 & 0.313 $\pm$ 0.009 & 71.9 $\pm$ 2.0 & 0.46 $\pm$ 0.11\\
Per-emb-48 & 0.346 $\pm$ 0.019 & 79.5 $\pm$ 4.4 & 0.06 $\pm$ 0.17\\
Per-emb-40 & 0.391 $\pm$ 0.022 & 90.0 $\pm$ 5.1 & 0.94 $\pm$ 0.19\\
Per-emb-55 & 0.618 $\pm$ 0.009 & 142.1 $\pm$ 2.0 & 0.14 $\pm$ 0.07\\
EDJ2009-183 & 1.025 $\pm$ 0.028 & 235.8 $\pm$ 6.4 & 0.45 $\pm$ 0.13\\
L1448IRS1 & 1.424 $\pm$ 0.015 & 327.4 $\pm$ 3.5 & 1.02 $\pm$ 0.09\\
Per-emb-35 & 1.908 $\pm$ 0.003 & 438.8 $\pm$ 0.7 & 0.23 $\pm$ 0.04\\
EDJ2009-156 & 3.107 $\pm$ 0.011 & 714.6 $\pm$ 2.5 & 0.17 $\pm$ 0.13\\
Per-emb-58+Per-emb-65 & 28.878 $\pm$ 0.023 & 6641.9 $\pm$ 5.3 & -0.44 $\pm$ 0.13\\
\enddata
\tablecomments{This table includes only Class I + Class I multiple systems. The flux difference
is calculated in the 9 mm band.}
\end{deluxetable}

\begin{deluxetable}{lllll}
\tablewidth{0pt}
\tabletypesize{\scriptsize}
\tablecaption{Class II Multiple Systems}
\tablehead{
  \colhead{Source}  &   \colhead{Separation} &\colhead{Separation} & \colhead{Flux Difference}   \\
  \colhead{}  &   \colhead{(\arcsec)} & \colhead{(AU)} & \colhead{(Log [F$_1$/F$_2$])}  \\

}
\startdata
EDJ2009-269 & 0.524 $\pm$ 0.007 & 120.6 $\pm$ 1.6 & 0.12 $\pm$ 0.08\\
EDJ2009-156 & 3.107 $\pm$ 0.011 & 714.6 $\pm$ 2.5 & 0.17 $\pm$ 0.13\\
\enddata
\tablecomments{This table includes only Class II + Class II multiple systems. The flux difference
is calculated in the 9 mm band.}
\end{deluxetable}

\begin{deluxetable}{lllll}
\tablewidth{0pt}
\tabletypesize{\scriptsize}
\tablecaption{Multiplicity and Companion Star Fractions}
\tablehead{
  \colhead{Sample/Sub-sample}  & \colhead{Separation Range} & \colhead{S:B:T:Q:5:6}  &   \colhead{MF} & \colhead{CSF}\\
 }
\startdata
Full Sample & 15 - 10000 AU &
37:15:5:2:2:1 & 0.40 $\pm$ 0.06 & 0.71 $\pm$ 0.06\\
Class 0 &  15 - 10000  AU &
13:7:5:2:2:1 & 0.57 $\pm$ 0.09 & 1.2 $\pm$ 0.20\\

Class I &  15 - 10000  AU &
20:6:0:0:0:0 & 0.23 $\pm$ 0.08 & 0.23 $\pm$ 0.08\\
Class II &  15 - 10000  AU &
4:2:0:0:0:0 & 0.33 $\pm$ 0.19 & 0.33 $\pm$ 0.19\\
\\
Full Sample &  15 - 5000 AU & 
42:19:5:1:1:1 & 0.39 $\pm$ 0.06 & 0.59 $\pm$ 0.06\\
Class 0  & 15 - 5000 AU & 
15:10:5:1:1:1 & 0.55 $\pm$ 0.09 & 0.97 $\pm$ 0.03\\
Class I  & 15 - 5000 AU & 
22:7:0:0:0:0 & 0.24 $\pm$ 0.08 & 0.24 $\pm$ 0.08\\
Class II &  15 - 5000 AU & 
5:2:0:0:0:0 & 0.29 $\pm$ 0.17 & 0.29 $\pm$ 0.17\\
\\
Full Sample  & 50 - 5000 AU & 
45:16:5:1:1:1 & 0.35 $\pm$ 0.06 & 0.55 $\pm$ 0.06\\
Class 0  & 50 - 5000 AU & 
18:7:5:1:1:1 & 0.45 $\pm$ 0.09 & 0.88 $\pm$ 0.06\\
Class I  & 50 - 5000 AU & 
22:7:0:0:0:0 & 0.24 $\pm$ 0.08 & 0.24 $\pm$ 0.08\\
Class II  & 50 - 5000 AU & 
5:2:0:0:0:0 & 0.29 $\pm$ 0.17 & 0.29 $\pm$ 0.17\\
\\

Full Sample  & 15 - 2000 AU & 
51:21:1:1:0:0 & 0.31 $\pm$ 0.05 & 0.35 $\pm$ 0.06\\
Class 0 &  15 - 2000 AU & 
24:11:1:1:0:0 & 0.35 $\pm$ 0.08 & 0.43 $\pm$ 0.08\\
Class I &  15 - 2000 AU & 
21:8:0:0:0:0 & 0.28 $\pm$ 0.08 & 0.28 $\pm$ 0.08\\
Class II &  15 - 2000 AU & 
6:2:0:0:0:0 & 0.25 $\pm$ 0.15 & 0.25 $\pm$ 0.15\\
\\
Full Sample &  50 - 2000 AU & 
54:18:1:1:0:0 & 0.27 $\pm$ 0.05 & 0.31 $\pm$ 0.05\\
Class 0 & 50 - 2000 AU & 
27:8:1:1:0:0 & 0.27 $\pm$ 0.07 & 0.35 $\pm$ 0.08\\
Class I &  50 - 2000 AU & 
21:8:0:0:0:0 & 0.28 $\pm$ 0.08 & 0.28 $\pm$ 0.08\\
Class II  & 50 - 2000 AU & 
6:2:0:0:0:0 & 0.25 $\pm$ 0.15 & 0.25 $\pm$ 0.15\\
\\
Full Sample  & 15 - 1000 AU & 
57:20:1:0:0:0 & 0.27 $\pm$ 0.05 & 0.28 $\pm$ 0.05\\
Class 0  & 15 - 1000 AU & 
29:10:1:0:0:0 & 0.28 $\pm$ 0.07 & 0.30 $\pm$ 0.07\\
Class I  & 15 - 1000 AU & 
22:8:0:0:0:0 & 0.27 $\pm$ 0.08 & 0.27 $\pm$ 0.08\\
Class II  & 15 - 1000 AU & 
6:2:0:0:0:0 & 0.25 $\pm$ 0.15 & 0.25 $\pm$ 0.15\\
\\
Full Sample  & 50 - 1000 AU & 
60:17:1:0:0:0 & 0.23 $\pm$ 0.05 & 0.24 $\pm$ 0.05\\
Class 0 &  50 - 1000 AU & 
32:7:1:0:0:0 & 0.20 $\pm$ 0.06 & 0.22 $\pm$ 0.07\\
Class I  & 50 - 1000 AU & 
22:8:0:0:0:0 & 0.27 $\pm$ 0.08 & 0.27 $\pm$ 0.08\\
Class II &  50 - 1000 AU & 
6:2:0:0:0:0 & 0.25 $\pm$ 0.15 & 0.25 $\pm$ 0.15\\

\\
Full Sample  & 100 - 1000 AU & 
67:12:0:0:0:0 & 0.15 $\pm$ 0.04 & 0.15 $\pm$ 0.04\\
Class 0  & 100 - 1000 AU & 
36:5:0:0:0:0 & 0.12 $\pm$ 0.05 & 0.12 $\pm$ 0.05\\
Class I &  100 - 1000 AU & 
25:5:0:0:0:0 & 0.17 $\pm$ 0.07 & 0.17 $\pm$ 0.07\\
Class II &  100 - 1000 AU & 
6:2:0:0:0:0 & 0.25 $\pm$ 0.15 & 0.25 $\pm$ 0.15\\
\enddata
\tablecomments{Note that the uncertainties throughout the text
are caculated assuming binomial statistics, $\sigma_{CSF}$ = (N$_{comp}$(1-N$_{comp}$/N$_{sys}$)$^{-0.5}$ 
$\times$ 1/N$_{sys}$ where N$_{comp}$ is the number of companions and N$_{sys}$ is the number of systems.
$\sigma_{MF}$ is calculated similarly, but by substituting N$_{mult}$ (number of multiple systems) for N$_{comp}$.
Poisson statistics are not used because the criteria of N$_{comp}$ $>>$ N$_{sys}$ is not met. However, we note
that the variance calculated assuming binomial statistics is only slightly smaller than that of 
Poisson statistics. For the case of CSF $>$ 1.0, $\sigma_{CSF}$ is not a real number and we revert to 
Poisson statistics in this case.}
\end{deluxetable}

\begin{deluxetable}{lllll}
\tablewidth{0pt}
\tabletypesize{\scriptsize}
\tablecaption{Position Angles of Outflows and Close Multiples}
\tablehead{
  \colhead{Source}  &   \colhead{Outflow PA} &\colhead{Companion PA} & \colhead{Relative PA} & \colhead{References}\\
  \colhead{}  &   \colhead{(\degr)} & \colhead{(\degr)} & \colhead{(\degr)}  & \\

}
\startdata
   Per-emb-2 & 127 & 302.8 & 4.2 & 1\\
  Per-emb-33 & 285 & 292.1 & 7.1 & 2, 3\\
  Per-emb-27 & 24 & 187.6 & 16.4 & 4\\
  Per-emb-36 & 24 & 352.2 & 31.8 & 4\\
  Per-emb-44 & 130 & 266.1 & 43.9 & 4\\
   Per-emb-5 & 125 & 256.1 & 48.9 & 5 \\
  Per-emb-35 & 290 & 59.7 & 50.3 & 5\\
  Per-emb-22 & 318 & 262.1 & 55.9 & 3\\
 Per-emb-107 & 308 & 10.5 & 62.5 & 3, 5\\
  Per-emb-12 & 200 & 309.9 & 70.1 & 4\\
  Per-emb-17 & 250 & 148.2 & 78.2 & 5\\
  Per-emb-40 & 280 & 20.2 & 79.8 & 5\\ 
  Per-emb-18 & 345 & 82.3 & 82.7 & 6\\
\enddata
\tablecomments{Only sources with known outflows and separations less than 500 AU are included. 
Outflow position angles (PA) taken from
(1) \citet{schnee2012}, (2) \citet{kwon2006}, (3) \citet{tobin2015}, 
(4) \citet{plunkett2013}, (5) \citet{lee2015}, (6) \citet{davis2008}. The relative PA column, is the absolute value
of the companion position angle relative to the outflow position angle; as defined this angle will not be larger
than 90\degr. }
\end{deluxetable}

\end{document}